\documentclass[reprint,preprintnumbers,
amsmath,amssymb,aps,nofootinbib,superscriptaddress]{revtex4-1}
\usepackage[pdftex]{graphicx}
\usepackage{dcolumn}
\usepackage{bm}
\usepackage{natbib}
\usepackage{cases}
\usepackage[pdftex]{hyperref}
\usepackage{color}
\definecolor{phthaloblue}{rgb}{0.0, 0.06, 0.54}
\definecolor{bluscuro}{rgb}{0.15, 0.2, .85}
\definecolor{rossos}{cmyk}{0,1,1,0.55}
\definecolor{bluchiaro}{cmyk}{1,.3,0.,0.1}
\hypersetup{
    colorlinks=true,
    linkcolor=bluscuro,
    citecolor=phthaloblue,
    filecolor=phthaloblue,
    urlcolor=phthaloblue,
    }
\makeatletter \def\@eqnnum{{\normalsize \normalcolor (\theequation)}}  \makeatother %

\begin{document}
\preprint{\begin{minipage}[b]{1\linewidth}
\begin{flushright} KEK-TH-1965, KEK-Cosmo-201 \end{flushright}
\end{minipage}}
\title{Electroweak Vacuum Instability and Renormalized Vacuum Field Fluctuations in Friedmann-Lemaitre-Robertson-Walker Background}
\author{Kazunori Kohri}
\email{kohri@post.kek.jp}
\affiliation{KEK Theory Center, IPNS, KEK, Tsukuba, Ibaraki 305-0801, Japan}
\affiliation{The Graduate University of Advanced Studies (Sokendai),Tsukuba, Ibaraki 305-0801, Japan}
\affiliation{Rudolf Peierls Centre for Theoretical Physics,
The University of Oxford, 1 Keble Road, Oxford, OX1 3NP, UK}
\author{Hiroki Matsui}
\email{matshiro@post.kek.jp}
\affiliation{KEK Theory Center, IPNS, KEK, Tsukuba, Ibaraki 305-0801, Japan}
\affiliation{The Graduate University of Advanced Studies (Sokendai),Tsukuba, Ibaraki 305-0801, Japan}

\begin{abstract}
The cosmological Higgs vacuum stability has been an attractive research subject and it is crucial to accurately follow the development of the Higgs fluctuations.
In this work, we thoroughly investigate how the vacuum fluctuations of the Higgs field affect the stability of the electroweak vacuum in
Friedmann-Lemaitre-Robertson-Walker (FLRW) background.
Adopting adiabatic (WKB) approximation or adiabatic
regularization methods, we clearly show that vacuum fluctuations of the Higgs field in the FLRW background depend on the curvature and also masses of the Higgs or other scalar fields. 
The Higgs fluctuations can generate true vacuum bubbles and 
trigger off a collapse of the electroweak vacuum.
Furthermore we clearly show that the effective Higgs potential 
in the FLRW background is modified by the Higgs vacuum fluctuations.
The vacuum fluctuations of the standard model fields can stabilize or
destabilize the effective Higgs potential through backreaction effects.
Considering the improved effective Higgs potential 
with the Higgs vacuum fluctuations  $\left< {  \delta \phi  }^{ 2 } \right>$ in various backgrounds,
we provide new cosmological constraints on
the mass of the Higgs-coupled scalar fields
and a quantitative description of the Higgs
stability in the FLRW background.
\end{abstract}
\date{\today}

\maketitle
\flushbottom
\section{Introduction}
\label{sec:intro}
The Large Hadron Collider (LHC) experiments discovered the Higgs boson and established
the Standard Model (SM) of particle physics.
But currently central values of the Higgs boson mass 
$m_{h}=125.09 \pm 0.21 ({\rm stat}) \pm 0.11({\rm syst})\ {\rm GeV}$~\cite{Aad:2015zhl,Aad:2013wqa,
Chatrchyan:2013mxa,Giardino:2013bma} and 
the top quark mass $m_{t}=172.44\pm 0.13 ({\rm stat}) \pm 0.47 ({\rm syst})\ {\rm GeV}$~\cite{Khachatryan:2015hba} suggest that 
the effective Higgs potential develops an instability about the scale $\Lambda_{I} \approx 10^{11}\ {\rm GeV}$.
Therefore, if there are no new physics to stabilize the Higgs field,
the current electroweak vacuum is not stable and finally causes
a vacuum decay through quantum tunneling~\cite{Kobzarev:1974cp,Coleman:1977py,Callan:1977pt}.
Fortunately, the vacuum decay timescale is longer than the age of the Universe~\cite{Degrassi:2012ry,Isidori:2001bm,
Ellis:2009tp,EliasMiro:2011aa}, and therefore,
it has been thought that the metastability of our electroweak vacuum 
does not cause cosmological problems to the observed Universe.

However, the recent investigations~\cite{Espinosa:2007qp,Fairbairn:2014zia,Lebedev:2012sy,Kobakhidze:2013tn,Enqvist:2013kaa,Herranen:2014cua,Kobakhidze:2014xda,Kamada:2014ufa,Enqvist:2014bua,Hook:2014uia,Kearney:2015vba,Espinosa:2015qea,Kohri:2016qqv,Czerwinska:2016fky,East:2016anr} 
reveal that the metastable electroweak vacuum becomes incompatible with large-field inflation models.
It is well-known that the vacuum fluctuations 
$\left<{  \delta \phi   }^{ 2 } \right>$ of the quantum field glow rapidly in the inflationary de Sitter phase.
If the inflationary de Sitter fluctuations of the Higgs field $\left<{  \delta \phi   }^{ 2 } \right>$
overcome the barrier of the effective Higgs potential $V_{\rm eff}\left( \phi  \right)$,
an unwanted vacuum transition to a Planck-scale true vacuum immediately occurs and cause a collapse of the Universe.
Furthermore, even after the inflation, the large vacuum fluctuations of the Higgs field are generated 
via parametric resonance or tachyonic resonance, and can become potentially problematic~\cite{Herranen:2015ima,
Kohri:2016wof,Ema:2016kpf,Enqvist:2016mqj,Postma:2017hbk,Ema:2017loe}.
Besides that, the false vacuum decay of the Higgs 
can be enhanced in Schwarzschild background~\cite{Burda:2015isa,Burda:2015yfa,
Burda:2016mou,Grinstein:2015jda,
Tetradis:2016vqb,Cheung:2013sxa,Canko:2017ebb,Gorbunov:2017fhq,Kohri:2017ybt},
and therefore the existence of the tiny primordial black holes might not favor the metastability of the Higgs vacuum.
The cosmological Higgs vacuum stability has been an attractive research subject and it is crucial to accurately follow the development of the Higgs fluctuations.

In the present paper 
we thoroughly investigate how the vacuum fluctuations of the Higgs field
affect the stability of the electroweak vacuum in Friedmann-Lemaitre-Robertson-Walker (FLRW) background spacetime.
We consider the vacuum fluctuations of the Higgs field described by 
the two-point correlation function $\left<{  \delta \phi   }^{ 2 } \right>$ 
using several methods of the quantum field theory (QFT) in curved spacetime
like adiabatic (WKB) approximation or adiabatic regularization.
Then, we derive the improved effective potential in FRW background
based on Ref.\cite{Ringwald:1987ui}.
Previous works~\cite{Espinosa:2007qp,Fairbairn:2014zia,Lebedev:2012sy,Kobakhidze:2013tn,Enqvist:2013kaa,Herranen:2014cua,Kobakhidze:2014xda,Kamada:2014ufa,Enqvist:2014bua,Hook:2014uia,Kearney:2015vba,Espinosa:2015qea,Kohri:2016qqv,Czerwinska:2016fky,East:2016anr} 
about the Higgs vacuum stability in FRW background
are based on the the standard effective potential in curved spacetime 
like Eq.~(\ref{eq:fhlfgkdg}).
However, the vacuum fluctuations $\left< {  \delta \phi  }^{ 2 } \right>$
of the Higgs field can modify the effective 
potential and we must consider the modified effective potential like
Eqs.~(\ref{eq:gddjkkdg}) or (\ref{eq:fhlgkdg}).
This matter was not pointed out in previous investigation except for 
our paper~\cite{Kohri:2016qqv}, but it is not in a comprehensive manner.
The vacuum fluctuations $\left< {  \delta \phi  }^{ 2 } \right>$ depend on the mass, couplings or the background, and therefore, we consider various situations in Section~\ref{sec:non-adiabatic} and Section~\ref{sec:background}.
Considering the improved effective Higgs potential 
with the Higgs vacuum fluctuations  $\left< {  \delta \phi  }^{ 2 } \right>$ in various backgrounds,
we provide new cosmological constraints on
the mass of the Higgs-coupled scalar fields
and a quantitative description of the Higgs
vacuum stability in the FLRW background
\footnote{
In this paper we focus on the vacuum fluctuations of the Higgs field 
and neglect backreaction effects of other field fluctuations like gauge bosons or fermions.
These backreaction effects would become also crucial for the Higgs vacuum stability in the FLRW background.
We plan to perform a detailed analysis of the Higgs vacuum stability 
including these effects in the future works. }.

This paper is organized as follows. In Section~\ref{sec:potential} 
we derive the standard effective potential in curved spacetime by using the adiabatic (WKB) approximation.
In Section~\ref{sec:adiabatic} we consider 
the renormalized vacuum fluctuations in the FLRW background
where the mass of the quantum field is larger than the curvature scale.
In Section~\ref{sec:non-adiabatic} we discuss
the renormalized vacuum fluctuations of the massless fields in the FLRW background
and provide the detail calculations of the renormalized vacuum fluctuations in the adiabatic regularization methods.
In Section~\ref{sec:background} we consider the renormalized vacuum fluctuations in
the dynamical scalar field background.
In Section~\ref{sec:instability} we discuss how the vacuum fluctuations of the Higgs field
affect the stability of the electroweak vacuum. 
We clearly show that the large Higgs fluctuations in the FLRW background
modify the standard effective Higgs potential as the backreaction effects 
and also generate true vacuum bubbles or domains.
We discuss various cosmological constraints on the metastable electroweak vacuum in the FLRW background.
Finally, in Section~\ref{sec:conclusion} we conclude our work.

\section{Standard effective potential in curved background}
\label{sec:potential}
The cosmological dynamics of the Higgs field can be determined by the effective potential.
The matters of the effective potential in curved background has been thoroughly investigated 
in the literature~\cite{Shore:1979as,Toms:1982af,Toms:1983qr,Hu:1984js,
Buchbinder:1985js,Ringwald:1987ui,
Balakrishnan:1991pm,Muta:1991mw,Kirsten:1993jn,Elizalde:1993ee,Elizalde:1993ew,Elizalde:1994im,
Elizalde:1994ds,Elizalde:1994gv,Gorbar:2002pw,Gorbar:2003yt,Gorbar:2003yp,Maroto:2014oca,Czerwinska:2015xwa,Albareti:2016cvx}
and there are a variety of formulation to derive the effective potential in curved background. 
In this section, we discuss the standard effective potential via the adiabatic (WKB) approximation method
following the literature~\cite{Ringwald:1987ui}.
This formulation can clearly handle the UV divergences of the vacuum field fluctuations and 
simply derive the effective potential in curved background.

In this present paper, we assume the Friedmann-Lemaitre-Robertson-Walker (FLRW) background which is described by 
the FLRW metric 
\begin{align}
g_{\mu\nu}={\rm diag}\left( -1,\frac { { a }^{ 2 }\left( t \right)  }{ 1-K { r }^{ 2 } } ,{ a }^{ 2 }\left( t \right) { r }^{ 2 },{ a }^{ 2 }\left( t \right) { r }^{ 2 }\sin^{2} { \theta  }  \right)\label{eqsdffedg},
\end{align}
where $a = a\left(t\right)$ express the scale factor with the cosmic time $t$ and $K $ is the spatial curvature parameter.
The positive, zero, and negative values of the spatial curvature parameter $K $ 
are related with closed, flat, and hyperbolic spacetime.
For the spatially flat spacetime, we can take $K=0$ and the Ricci scalar is given as
\begin{align}
R=6\left[ 
{ \left( \frac { \dot { a }  }{ a }  \right)  }^{ 2 }+\left( \frac { \ddot { a }  }{ a }  \right)  
\right]=6\left(\frac{a''}{a^{3}}\right)\label{eq:dgdgfdgedg}, 
\end{align}
where $\eta$ is the conformal time and defined by $d\eta=dt/a$.
In the radiation dominated Universe, the scale factor becomes $a\left(t\right) \propto t^{1/2}$ 
and the Ricci scalar is expressed as $R=0$. On the other hand, in the matter dominated Universe,
the scale factor becomes $a\left(t\right) \propto t^{2/3}$ and 
the Ricci scalar is expressed as $R=3H^{2}$. Finally, in the de Sitter Universe,
the scale factor becomes $a\left(t\right) \propto e^{Ht}$ and 
the Ricci scalar is expressed as $R=12H^{2}$.

The bare (unrenormalized) action for the Higgs field with the potential $V\left( \phi  \right)$ in curved background is given by 
\begin{align}
S\left[ \phi  \right] =\int { { d }^{ 4 }x\sqrt { -g } \left( \frac { 1 }{ 2 } { g }^{ \mu \nu  }
{ \partial   }_{ \mu  }\phi {\partial  }_{ \nu  }\phi +V\left( \phi  \right)  \right)  } \label{eq:ddddddgedg},
\end{align}
where we assume the simple form for the Higgs potential with bare parameters as
\begin{align}
V\left( \phi  \right) =\frac{1}{2}\left(m^{2}+\xi R\right)\phi^{2}+\frac{\lambda}{4}\phi^{4} \label{eq:aaaadg}.
\end{align}
Thus, the Klein-Gordon equation for the Higgs field are written as
\begin{align}
\Box \phi-m^{2}\phi-\xi R\phi-\lambda\phi^{3}=0 \label{eq:dsssssdg},
\end{align}
where $\Box$ expresses the generally covaiant d'Alembertian operator, 
$\Box =g^{\mu\nu}{ \nabla  }_{ \mu  }{  \nabla   }_{ \nu  }=1/\sqrt { -g } 
{ \partial  }_{ \mu  }\left( \sqrt { -g } { \partial  }^{ \mu  } \right) $
and $\xi$ is the non-minimal Higgs-gravity coupling constant.

In the QFT, we treat the Higgs field $\phi \left( \eta ,x \right)$ as the field operator acting on the ground states,
then the Higgs field  $\phi \left( \eta ,x \right)$ 
is decomposed into a classic field and a quantum field as
\begin{align}
\phi \left( \eta ,x \right) =\phi_{ c}+\delta \phi \left(\eta ,x \right) \label{eq:kkkkgedg},
\end{align}
where the vacuum expectation value of the Higgs field is 
$\phi_{ c}=\left< 0  \right| { \phi \left(\eta ,x\right) }\left| 0 \right>$ 
and $\left< 0  \right| { \delta\phi \left(\eta ,x\right) }\left| 0 \right>=0$.
By introducing the renormalized parameters and the counterterms as $m^{2}=m^{2}\left(\mu\right)+\delta m^{2}$,
$\xi=\xi\left(\mu\right)+\delta \xi$ and $\lambda=\lambda\left(\mu\right)+\delta \lambda$,
we can obtain the Klein-Gordon equations in the one-loop approximation as 
\begin{align}
&\Box \phi_{ c} -\left(m^{2}\left(\mu\right)+\delta m^{2}\right)\phi_{ c} -\left(\xi\left(\mu\right)
+\delta \xi\right)R\phi_{ c}  \\ &-3\left(\lambda\left(\mu\right)+\delta \lambda \right)
\left< \delta { \phi  }^{ 2 } \right>\phi_{ c} -\left(\lambda\left(\mu\right)+\delta \lambda \right)\phi^{3}_{ c} =0,\nonumber \\
&\left(\Box -m^{2}\left(\mu\right)-\xi\left(\mu\right) R
-3\lambda\left(\mu\right)\phi^{2}_{ c}\right)\delta \phi =0 \label{eq:dppppg},
\end{align}
From here we drop the subscript of the classic field $\phi_{ c}$ for convenience. 
The quantum Higgs field $\delta \phi$ is decomposed into each $k$ modes as,
\begin{align}
\delta \phi \left( \eta ,x \right) =\int { { d }^{ 3 }k\left( { a }_{ k }\delta { \phi  }_{ k }\left( \eta ,x \right) +{ a }_{ k }^{ \dagger  }\delta { \phi  }_{ k }^{ * }\left( \eta ,x \right)  \right)  }  \label{eq:ddfkkfledg},
\end{align}
where
\begin{align}
{ \delta \phi  }_{ k }\left( \eta ,x \right) =\frac { { e }^{ ik\cdot x } }{ { \left( 2\pi  \right)  }^{ 3/2 }
\sqrt { C\left( \eta  \right)  }  } \delta { \chi  }_{ k }\left( \eta  \right)  \label{eq:slkdlkgdg},
\end{align}
with $C\left(\eta \right)=a^{2}\left(\eta \right)$.
Now, we can build a complete set of the mode functions, which are orthonormal 
with respect to the scalar product in curved background
\begin{align}
\left( \delta { \phi  }_{ k },\delta { \phi  }_{ k' } \right) =-i\int _{ \Sigma  }^{  }{ 
 d{ \Sigma  }^{ \mu  } \sqrt { -{ g }_{ \Sigma  } }
\left[ { \delta { \phi  } }_{ k }\left( { \partial  }_{ \mu  }\delta { \phi  }_{ k' }^{ \ast  } \right) 
-\left( { \partial  }_{ \mu  }{ \delta { \phi  } }_{ k } \right) \delta { \phi  }_{ k' }^{ \ast  } \right]  },
\end{align}
where $d{ \Sigma  }^{ \mu  }=n^{\mu}d\Sigma$ is expressed by the a unit timelike vector $n^{\mu}$
and the volume element $d\Sigma$.
These orthonormal mode solutions satisfy 
\begin{align}
\left( \delta { \phi  }_{ k },\delta { \phi  }_{ k' } \right) ={ \delta  }\left( k-k' \right) \label{eq:sldjgfdg}
\end{align}
The creation and annihilation operators of $\delta { \phi  }_{ k }$ are required to satisfy the commutation relations
\begin{align}
\bigl[ { a }_{ k },{ a }_{ k' }\bigr] =\bigl[ { a }_{ k }^{ \dagger  } ,{ a }_{ k' }^{ \dagger  } \bigr] =0  ,\ \
\bigl[ { a }_{ k },{ a }_{ k' }^{ \dagger  } \bigr] ={ \delta  }\left( k-k' \right),
\end{align}
where the in-vacuum state $\left|  0 \right>$ is defined as $a_{k}\left|  0 \right>=0 $
and depends on the boundary conditions of the mode functions $\delta { \phi  }_{ k }$.
Different boundary conditions of $\delta { \phi  }_{ k }$ corresponds to different initial state of the quantum vacuum.
The vacuum field fluctuations $\left<{  \delta \phi   }^{ 2 } \right>$ of the Higgs field can be written as
\begin{align}
\left< 0  \right| { \delta \phi^{2}  }\left| 0 \right>&=\int { { d }^{ 3 }k{ \left| \delta { \phi  }_{ k }\left( \eta ,x \right)  \right|  }^{ 2 } } \label{eq:ddkg;ergedg},\\
&=\frac { 1 }{ 2{ \pi  }^{ 2 }C\left( \eta  \right)  } \int _{ 0 }^{ \infty  } { dk { k }^{ 2 }{ \left| \delta { \chi  }_{ k } \right|  }^{ 2 } }  \label{eq:xkkgdddgedg},
\end{align}
where $\left<{  \delta \phi   }^{ 2 } \right>$ has ultraviolet (quadratic and logarithmic) divergences,
which require a regularization, e.g. cut-off regularization or dimensional regularization,
and must be cancelled by the counterterms of the couplings.

From Eq.~(\ref{eq:dppppg}), the Klein-Gordon equation for the quantum rescaled field $\delta \chi$ is written by
\begin{align}
{ \delta \chi}''_{ k }+{ \Omega  }_{ k }^{ 2 }\left( \eta  \right) { \delta \chi  }_{ k }=0 \label{eq:dlksdsg},
\end{align}
where
\begin{align}
{ \Omega  }_{ k }^{ 2 }\left( \eta  \right) ={ k }^{ 2 }+C\left( \eta  \right) \left( { m }^{ 2 } 
+3\lambda \phi^{2}+\left( \xi -1/6 \right) R \right).
\end{align}
The orthonormal condition of Eq.~(\ref{eq:sldjgfdg}) for the mode functions $\delta \chi$
can be given by
\begin{align}
\delta { \chi  }_{ k }{ { \delta { \chi  } '}_{ k }^{ \ast  }-{ \delta { \chi  } }'_{ k }\delta { \chi  }_{ k }^{ \ast  } }=i,
\end{align}
which is the normalization of the mode function $\delta \chi \left( \eta  \right)$.
Eq.~(\ref{eq:dlksdsg}) is consistent with the differential equation of the harmonic oscillator with time-dependent mass.
Thus, we can rewrite the mode function $\delta \chi \left( \eta  \right)$ by the two complex function ${ \alpha  }_{ k }\left( \eta  \right)$
and ${ \beta  }_{ k }\left( \eta  \right) $ as
\begin{align}
\delta { { \chi  }_{ k } }\left( \eta  \right) =\frac { 1 }{ \sqrt { 2{ \Omega  }_{ k }\left( \eta  \right)  }  } \left\{ { \alpha  }_{ k }\left( \eta  \right) 
{ \delta \varphi }_{k}\left( \eta  \right) +{ \beta  }_{ k }\left( \eta  \right){ \delta \varphi }_{k}^{*}\left( \eta  \right)  \right\} \label{eq:dldfghdhg},
\end{align}
where ${ \delta \varphi }_{k}\left( \eta  \right) $ are given by
\begin{align}
{ \delta \varphi }_{k}\left( \eta  \right) ={\rm exp}\left\{ -i\int _{  }^{ \eta  }{ \Omega  }_{ k }\left( \eta_{1} \right) { d\eta_{1}}   \right\}.
\end{align}
From Eq.~(\ref{eq:dlksdsg}), we can obtain the relations for ${ \alpha  }_{ k }\left( \eta  \right)$
and ${ \beta  }_{ k }\left( \eta  \right) $ as the following
\begin{align}
{ \alpha  }_{ k }'=\frac{1}{2}\frac{{\Omega  }_{ k }'}{{\Omega  }_{ k }} { \beta  }_{ k }{ \delta \varphi^{*}_{k} }^{2}\left( \eta  \right), \ \
{ \beta  }_{ k }'=\frac{1}{2}\frac{{\Omega  }_{ k }'}{{\Omega  }_{ k }} { \alpha  }_{ k }{ \delta \varphi }_{k}^{2}\left( \eta  \right) \label{eq:dgjjgdhg}.
\end{align}
The Wronskian condition can be written by
\begin{align}
{ \left| { \alpha  }_{ k }\left( \eta  \right)  \right|  }^{ 2 }-{ \left| { \beta  }_{ k }\left( \eta  \right)  \right|  }^{ 2 }=1.
\end{align}
The initial conditions for ${ \alpha  }_{ k }\left( \eta_{0}  \right) $ and ${ \beta  }_{ k }\left( \eta_{0}  \right) $ 
corresponds to the choice of the in-vacuum. From Eq.~(\ref{eq:dldfghdhg}),
the vacuum field fluctuations $\left<{  \delta \phi   }^{ 2 } \right>$ of the Higgs field can be given by
\begin{align} 
\left<{  \delta \phi   }^{ 2 } \right>=\frac { 1 }{ 4{ \pi  }^{ 2 }C\left( \eta  \right)  }
\int _{ 0 }^{ \infty  }dk{ k }^{ 2 }
{ \Omega  }_{ k }^{ -1 }\biggl\{ 1+2{ \left| { \beta  }_{ k } \right|  }^{ 2 }  \nonumber \\
+{ \alpha  }_{ k }{ \beta  }_{ k }^{ * }{ \delta \varphi }_{k}^{2}+{ \alpha  }_{ k }^{ * }{ \beta  }_{ k }{ \delta \varphi^{*}_{k} }^{2} \biggr\}  ,
\end{align}
where the number density of the created particles and the corresponding energy density 
are given by
\begin{align}
N&=\frac { 1 }{2\pi^{2} a^{3}\left( \eta  \right)} \int_{ 0 }^{ \infty  } { { dk }k^{2}{ \left| { \beta  }_{ k } \right|  }^{ 2 } } ,\\
\rho&=\frac { 1 }{ 2\pi^{2} a^{4}\left( \eta  \right)  } \int_{ 0 }^{ \infty  }
 { { dk }k^{2}{ { \Omega  }_{ k }
\left| { \beta  }_{ k } \right|  }^{ 2 } } .
\end{align}
For simplicity, we define $n_{k}$ and $z_{k}$ as the following
\begin{align}
n_{k}={ \left| { \beta  }_{ k } \right|  }^{ 2 }, \quad z_{k}={ \alpha  }_{ k }{ \beta  }_{ k }^{ * }{ \delta \varphi }_{k}^{2}.
\end{align}
From Eq.~(\ref{eq:dgjjgdhg}), $n_{k}$ and $z_{k}$ satisfy the following differential equations
\begin{align}
{ n  }_{ k }'=\frac{{\Omega  }_{ k }'}{{\Omega  }_{ k }}{\rm Re}z_{k}, \ \
{ z  }_{ k }'=\frac{{\Omega  }_{ k }'}{{\Omega  }_{ k }}\left(n_{k}+\frac{1}{2}\right)-2i{\Omega  }_{ k }z_{k}
\label{eq:dgjdasahg}.
\end{align}
To solve Eq.~(\ref{eq:dgjdasahg}), we must take adequately the initial conditions.
For simplicity, we choose the following condition 
\begin{align}
n_{k}\left(\eta_{0}\right)=z_{k}\left(\eta_{0}\right)=0,
\end{align} 
which is equivalent to $\alpha_{k}\left(\eta_{0}\right)=1$, $\beta_{k}\left(\eta_{0}\right)=0$ and
corresponds to the Minkowski vacuum state which has no excited particles
\footnote{Note that this state is not identified as the Bunch-Davies vacuum
which fixes the Bogoliubov coefficients with the subhorizon limit ${\left| k\eta  \right|} \gg 1$ }.
The quantity $n_{k}={ \left| { \beta  }_{ k }\left( \eta  \right) \right|  }^{ 2 }$ can be interpreted as 
the number density created in the curved background.
By using $n_{k}$ and $z_{k}$, we obtain the following expression 
of the vacuum field fluctuations as
\begin{align}
\left<{  \delta \phi   }^{ 2 } \right>=\frac { 1 }{ 4{ \pi  }^{ 2 }C\left( \eta  \right)  }
\int _{ 0 }^{ \infty  }{ dk{ k }^{ 2 }{ \Omega  }_{ k }^{ -1 }\left\{ 1+2n_{k}+2{\rm Re}z_{k} \right\}  } \label{eq:dgsfoufdhg},
\end{align}
where we must solve adequately  Eq.~(\ref{eq:dgjdasahg}) and insert $n_{k}$ and $z_{k}$ 
into Eq.~(\ref{eq:dgsfoufdhg}) in order to obtain the vacuum field fluctuations $\left<{  \delta \phi   }^{ 2 } \right>$ of the Higgs field.
It is difficult to solve analytically Eq.~(\ref{eq:dgjdasahg}), and therefore, 
we generally use the adiabatic (WKB) approximation method, which is valid in 
large mass, large momentum mode or slowly-varying background as follows 
\begin{align}
\left| { \Omega ' }_{ k }/{ \Omega  }_{ k }^{ 2 } \right| \ll 1  \label{eq:adgkdhjhf}. 
\end{align}
By using the adiabatic approximation method, 
$n_{k}$ and $z_{k}$ can be approximated as follows~\cite{Ringwald:1987ui}:
\begin{align}
{ n }_{ k }&={ n }_{ k }^{ (2) }+{ n }_{ k }^{ (4) }+\cdots,  \label{eq:dsfufdhg} \\ 
{\rm Re}z_{k}&={\rm Re}z_{k}^{(2)}+{\rm Re}z_{k}^{(4)}+\cdots,  \label{eq:dgsfskkjkg}
\end{align}
where superscripts $(i)$ express the adiabatic order and the second order expressions are given by
\begin{align}
{ n }_{ k }^{ (2) }&=\frac{1}{16}\frac{{ \Omega ' }_{ k }^{ 2 }}{{ \Omega  }_{ k }^{ 4}},\\ 
{\rm Re}z_{k}^{(2)}&=\frac{1}{8}\frac{{ \Omega ''  }_{ k }}{{ \Omega  }_{ k }^{ 3}}
-\frac{1}{4}\frac{{ \Omega ' }_{ k }^{ 2 }}{{ \Omega  }_{ k }^{ 4}}.
\end{align}
The forth order adiabatic expressions are given by
\begin{align}
{ n }_{ k }^{ (4) }=&
-\frac{{ \Omega ' }_{ k }{ \Omega '''}_{ k }}{32{ \Omega  }_{ k }^{6}}
+\frac{{ \Omega '' }_{ k }^{2}}{64{ \Omega  }_{ k }^{6}}
+\frac{5{ \Omega ' }_{ k }^{2}{ \Omega ''}_{ k }}{32{ \Omega  }_{ k }^{7}}
-\frac{45{ \Omega ' }_{ k }^{4}}{256{ \Omega  }_{ k }^{8}},\\ 
{\rm Re}z_{k}^{(4)}=&
-\frac{{ \Omega ''''}_{ k }}{32{ \Omega  }_{ k }^{5}}
+\frac{11{ \Omega ' }_{ k }{ \Omega ''' }_{ k }  }{32{ \Omega  }_{ k }^{6}}
-\frac{115{ \Omega ' }_{ k }^{2}{ \Omega ''}_{ k }}{64{ \Omega  }_{ k }^{7}}\nonumber  \\& 
+\frac{7{ \Omega '' }_{ k }^{2}}{32{ \Omega  }_{ k }^{6}}
+\frac{45{ \Omega ' }_{ k }^{4}}{32{ \Omega  }_{ k }^{8}}.
\end{align}
By using the adiabatic (WKB) approximation method, 
we can obtain the following approximation of the vacuum field fluctuations of the Higgs field as
\begin{align}
\left<{  \delta \phi   }^{ 2 } \right>=\left<{  \delta \phi   }^{ 2 } \right>^{(0)}+\left<{  \delta \phi   }^{ 2 } \right>^{(2)}
+\left<{  \delta \phi   }^{ 2 } \right>^{(4)}+\cdots  \label{eq:adbatigudf} ,
\end{align}
where
\begin{align}
\left<{  \delta \phi   }^{ 2 } \right>^{(0)}&=\frac { 1 }{ 4{ \pi  }^{ 2 }C\left( \eta  \right)  }
\int _{ 0 }^{ \infty  }{ dk{ k }^{ 2 }{ \Omega  }_{ k }^{ -1 }  }  \label{eq:adbatikdjhf} , \\
\left<{  \delta \phi   }^{ 2 } \right>^{(2n)}&=\frac { 1 }{ 4{ \pi  }^{ 2 }C\left( \eta  \right)  }
\int _{ 0 }^{ \infty  }{ dk{ k }^{ 2 }{ \Omega  }_{ k }^{ -1 }\left\{ 2n^{(2n)}_{k}+2{\rm Re}z^{(2n)}_{k} \right\}  }  \label{eq:adsdldjhf}.
\end{align}
Although the higher order approximation can become finite,
the lowest order approximation has UV (quadratic and logarithmic) divergences.
However, the divergences in the lowest order expression are the same as the
divergences in the flat background.
Thus, we can regularize the divergence integral via the cut-off regularization or 
the dimensional regularization and offset the divergences by the counterterms of the couplings.

By using the dimensional regularization, we obtain the following lowest order expression as
\begin{align}
\left<{  \delta \phi   }^{ 2 } \right>^{(0)}
=\frac { { M }^{ 2 }\left( \phi  \right)  }{ 16{ \pi  }^{ 2 } } \left[ \ln { \left( \frac { { M }^{ 2 }\left( \phi  \right)  }{ { \mu  }^{ 2 } }  \right)  }
 -\frac { 1 }{ \epsilon  } -\log { 4\pi }-\gamma - 1  \right]  \label{eq:dklgkhldg},
\end{align}
with
\begin{align}
{ M }^{ 2 }\left( \phi  \right)={ m }^{ 2 }\left(\mu\right) +3\lambda\left(\mu\right) \phi^{2}+\left( \xi\left(\mu\right) -1/6 \right) R,
\end{align}
where $\mu$ is the renormalization scale and $\gamma$ is the Euler-Mascheroni constant.
The counterterms $\delta { m }^{ 2 }$, $\delta \xi$ and $\delta \lambda$ must cancel these divergences 
and are given by
\begin{align}
\delta { m }^{ 2 }&=\frac {3 \lambda\left(\mu\right) m^{2}\left(\mu\right)  }{ 16{ \pi  }^{ 2 } }\left(\frac { 1 }{ \epsilon  } +\log { 4\pi }
+\gamma  \right)  , \\
\delta \xi &=\frac {3 \lambda\left(\mu\right) }{ 16{ \pi  }^{ 2 } } \left( \xi\left(\mu\right) -\frac { 1 }{ 6 }  \right)\left(\frac { 1 }{ \epsilon  } +\log { 4\pi }+\gamma  \right), \\ 
\delta \lambda &=\frac { 9{ \lambda\left(\mu\right) } }{ 16{ \pi  }^{ 2 } } \left(\frac { 1 }{ \epsilon  } +\log { 4\pi }+\gamma  \right) .
\end{align}
Thus, the renormalized vacuum field fluctuations of the Higgs field of the lowest order can be given by
\begin{align}
\left<{  \delta \phi   }^{ 2 } \right>^{(0)}_{\rm ren}
=\frac { { M }^{ 2 }\left( \phi  \right)  }{ 16{ \pi  }^{ 2 } } \left[ \ln { \left( \frac { {M}^{ 2 }\left( \phi  \right)  }{ { \mu  }^{ 2 } }  \right)  } 
-1   \right]  \label{eq:sfhdhkhldg},
\end{align}
where the above expression corresponds to the renormalized vacuum fluctuations in flat background.
From the renormalized expression of Eq.~(\ref{eq:sfhdhkhldg}),
we can construct the one-loop evolution equation as follows:
\begin{align}
\ddot { \phi  } +3H\dot { \phi  } +\frac { \partial { V }_{ \rm eff }\left( \phi  \right)  }{ \partial \phi  } =0,\label{eq:ewhghdldg}
\end{align}
where the one-loop effective potential
in curved background is give by
\begin{align}
V_{\rm eff}\left( \phi  \right) &=\frac{1}{2}m^{2}\left(\mu\right)\phi^{2}+\frac{1}{2}\xi \left(\mu\right)R\phi^{2}
+\frac{\lambda\left(\mu\right)}{4}\phi^{4} \nonumber \\& + \frac { { M }^{ 4}\left( \phi  \right)  }{ 64{ \pi  }^{ 2 } } 
\left[ \ln { \left( \frac { {M}^{ 2 }\left( \phi  \right)  }{ { \mu  }^{ 2 } }  \right)  } -\frac { 3 }{ 2 }    \right] \label{eq:fjdjsfdg},
\end{align}
From the one-loop effective potential of Eq.~(\ref{eq:fjdjsfdg}), the one-loop $\beta$ functions are given by 
\begin{align}
\beta_{\lambda } &\equiv \frac { d\lambda  }{ d\ln { \mu  }  } =\frac { 18{ \lambda  }^{ 2 } }{ { \left( 4\pi  \right)  }^{ 2 } } 
 \label{eq:dkldgedg},\\
\beta_{\xi } &\equiv \frac { d\xi  }{ d\ln { \mu  }  } =\frac { 6{ \lambda  }  }{ { \left( 4\pi  \right)  }^{ 2 } }\left( \xi -1/6 \right) \label{eq:dgeddlkgg}, \\
\beta_{ m^{2} }&\equiv \frac { d{ { m }^{ 2 } }  }{ d\ln { \mu  }  } =\frac { 6{ \lambda  }m^{2} }{ { \left( 4\pi  \right)  }^{ 2 } }  \label{eq:ddakdlgg}.
\end{align}
Although Eq.~(\ref{eq:ewhghdldg}) is the standard expression
to describe the cosmological dynamics of the Higgs field $\phi\left(t\right)$, 
this expression does not include the high-order vacuum fluctuations $\left<{  \delta \phi   }^{ 2 } \right>^{(2n)}$ which corresponds to the gravitational particle productions in curved background.
Therefore, the correct effective-evolution equation is given as follows~\cite{Ringwald:1987ui}:
\begin{align}
\ddot { \phi  } +3H\dot { \phi  } +\frac { \partial { V }_{ \rm eff }\left( \phi  \right)  }{ \partial \phi  } +
3\lambda(\mu) \left<{  \delta \phi   }^{ 2 } \right>^{(2n)}\phi =0,\label{eq:bdhldg}
\end{align}
which require the modification of the standard effective potential.
The redefined/modified effective potential in curved background 
is given as follows: 
\begin{align}
V_{\rm eff}\left( \phi  \right) &= \frac{1}{2}m^{2}(\mu)\phi^{2}+\frac{1}{2}\xi (\mu)R\phi^{2}
+\frac{3\lambda(\mu)}{2}\left< {\delta \phi  }^{ 2 } \right>^{(2n)}\phi^{2} \nonumber \\
&+\frac{\lambda(\mu)}{4}\phi^{4} + \frac { { M }^{ 4}\left( \phi  \right)  }{ 64{ \pi  }^{ 2 } } 
\left[ \ln { \left( \frac { {M}^{ 2 }\left( \phi  \right)  }{ { \mu  }^{ 2 } }  \right)  } -\frac { 3 }{ 2 }    \right] \label{eq:fsddsdsjsfdg}.
\end{align}
which properly include gravitational vacuum effects.
The additional term originates from the particle production 
in curved spacetime and depends on the vacuum state.
Note that in flat spacetime the vacuum state is unique 
and the effective potential has no additional terms~\cite{Maroto:2014oca,Albareti:2016cvx}. 
Here we considered the modification of the effective potential 
from the particle production effects in curved spacetime.
On the other hand, the particle production can also affect the spacetime~\cite{Ford:1984hs,Mottola:1984ar,Mottola:1985qt,Antoniadis:1985pj}
as the back-reaction effects.

\section{Renormalized vacuum fluctuations from adiabatic (WKB) approximation method}
\label{sec:adiabatic}
In previous section we show that the lowest-order (Minkowskian) vacuum field fluctuations 
contract the one-loop effective potential.
However, the higher-order adiabatic vacuum field fluctuations 
appear as a result of the particle production effects in curved background, and therefore, 
provide a significant contribution to dynamical evolutions of the Higgs field.
To obtain the exact one-loop evolution equation in curved background,
we must count up the higher order of the adiabatic approximation.
From Eq.~(\ref{eq:dsfufdhg}), (\ref{eq:dgsfskkjkg}), and (\ref{eq:adsdldjhf}), 
the second (adiabatic) order expressions of the vacuum field fluctuations are given by~\cite{Ringwald:1987ui}
\begin{align}
\left<{  \delta \phi   }^{ 2 } \right>^{(2)}=\frac { 1 }{ 16{ \pi  }^{ 2 }C\left( \eta  \right)  }
\int _{ 0 }^{ \infty  }{ dk{ k }^{ 2 }{ \Omega  }_{ k }^{ -1 }\left\{ 
\frac{{ \Omega''  }_{ k }}{{ \Omega }_{ k }^3}-\frac{3}{2}\frac{{ \Omega' }_{ k }^2}{{ \Omega }_{ k }^4}\right\}  } 
\label{eq:dgsghfhg},
\end{align}
with
\begin{align}
{ \Omega  }_{ k }^{ 2 }={ k }^{ 2 }+C\left( \eta  \right) \left( { m }^{ 2 } 
+3\lambda \phi^{2}+\left( \xi -1/6 \right) R \right).
\end{align}
Thus, we can obtain the following expression as
\begin{align}
\left<{  \delta \phi   }^{ 2 } \right>^{(2)}&=\frac { 1 }{ 16{ \pi  }^{ 2 }C\left( \eta  \right)  }
\int _{ 0 }^{ \infty  }{ dk{ k }^{ 2 }\Omega  }_{ k }^{ -1 }\biggl\{ 
\frac{ \left(\bar { M } \bar { M } ''+\bar { M } '^{2} \right)     }{{ \Omega }_{ k }^4}
\nonumber \\  &-\frac{5}{2}\frac{\bar { M } ^{2}\bar { M }'^2}{{ \Omega }_{ k }^6}\biggr\} 
\label{eq:dgsgjjgmng},
\end{align}
with
\begin{align}
{\bar { M }  }^{ 2 }=C\left( \eta  \right){ M }^{ 2 }\left( \phi  \right).
\end{align}
Now, we must perform the integral of Eq.~(\ref{eq:dgsgjjgmng}).
As already pointed out, the high-order adiabatic expressions like $\left<{  \delta \phi   }^{ 2 } \right>^{(2)}$
are UV finite and therefore there is no need to renormalize the high-order vacuum field fluctuations.
The corresponding integrals converge to the finite values by
\begin{align}
I\left( \alpha  \right) &\equiv \int _{ 0 }^{ \infty  }{ dk{ k }^{ 2 } } { \left( { k }^{ 2 }+{\bar { M } }^{ 2 } \right)  }^{ -\alpha  } \\
&=\frac { {\bar { M }  }^{ 3-2\alpha  } }{ 2 } \frac { \Gamma \left( 3/2 \right) \Gamma \left( \alpha -3/2 \right)  }{ \Gamma \left( \alpha  \right)  } 
\label{eq:eqgmng},
\end{align}
where the above expression is valid for $\alpha>3/2$.
By using Eq.~(\ref{eq:eqgmng}), the second (adiabatic) order of the vacuum field fluctuations $\left<{  \delta \phi   }^{ 2 } \right>^{(2)}$
are given as follows
\begin{align}
\left<{  \delta \phi   }^{ 2 } \right>^{(2)}&=\frac { 1 }{ 16{ \pi  }^{ 2 }C\left( \eta  \right)  }
\biggl\{ \left(\bar { M }\bar { M }''+\bar { M }'^{2} \right)I\left( \frac{5}{2}  \right)  \nonumber \\&
-\frac{5}{2}\bar { M }^{2}\bar { M }'^2I\left( \frac{7}{2}  \right)\biggr\} \nonumber \\
&=\frac { 1 }{ 48{ \pi  }^{ 2 }C\left( \eta  \right)  }\frac{\bar { M }''}{\bar { M }} 
\label{eq:dggdfgg}.
\end{align}
Thus, the renormalized vacuum fluctuations in curved background via the adiabatic (WKB) approximation method 
are given by
\begin{align*}
\left<{  \delta \phi   }^{ 2 } \right>_{\rm ren}
&= \left<{  \delta \phi   }^{ 2 } \right>^{(0)}+ \left<{  \delta \phi   }^{ 2 } \right>^{(2)}+\cdots \nonumber \\
&= \frac { { M }^{ 2 }  }{ 16{ \pi  }^{ 2 } } \left[ \ln { \left( \frac { {M}^{ 2 }  }{ { \mu  }^{ 2 } }  \right)  } 
-1    \right]  + \frac { 1 }{ 48{ \pi  }^{ 2 }C\left( \eta  \right)  }\frac{\bar { M }''}{\bar { M } }  +\cdots ,
\end{align*}
where the first term express the Minkowskian renormalized vacuum field fluctuations,
and the second term describes the dynamical contribution of the renormalized vacuum fluctuations,
which corresponds to the particle production effects. 
Next let us consider the second (adiabatic) order expression at 
proper time $t$ as
\begin{align}
\left<{  \delta \phi   }^{ 2 } \right>^{(2)}
&= \frac { 1 }{ 48{ \pi  }^{ 2 }  }\left\{ \frac{a''}{a^3}+2\frac{a'}{a^2}\frac{M'}{M}+\frac{1}{a^2}\frac{M''}{M}\right\} \nonumber \\
&= \frac { 1 }{ 48{ \pi  }^{ 2 }  }\left\{ \frac{ \ddot { a } }{a}+\frac { { \dot { a }  }^{ 2 } }{ { a }^{ 2 } } 
+3\frac{\dot { a } }{a}\frac{\dot { M}}{M}+\frac{\ddot { M}}{M}\right\} \nonumber \\
&= \frac { 1 }{ 48{ \pi  }^{ 2 }  } \Biggl\{
\frac { \ddot { a }  }{ a } +\frac { { \dot { a }  }^{ 2 } }{ { a }^{ 2 } } +\frac { 3 }{ 2 } \frac { \dot { a }  }{ a } \frac { \left( \xi -1/6 \right) \dot { R } 
+6\lambda \phi \dot { \phi  }  }{ { m }^{ 2 }+\left( \xi -1/6 \right) R+3\lambda { \phi  }^{ 2 } } \nonumber \\
&-\frac { 1 }{ 4 } \frac { { \left( \left( \xi -1/6 \right) \dot { R } +6\lambda \phi \dot { \phi  }  \right)  }^{ 2 } }{ { \left( { m }^{ 2 }+\left( \xi -1/6 \right) R+3\lambda { \phi  }^{ 2 } \right)  }^{ 2 } }  \nonumber \\
&+\frac { 1 }{ 2 } \frac { \left( \xi -1/6 \right) \ddot { R } +6\lambda \left( \phi \ddot { \phi  } 
+{ \dot { \phi  }  }^{ 2 } \right)  }{ { m }^{ 2 }+\left( \xi -1/6 \right) R+3\lambda { \phi  }^{ 2 } } \Biggr\} .
\label{eq:dggdhfhgg}
\end{align}
If we consider the near-conformal coupling case $\xi \simeq1/6$,
we can obtain the following expression 
\footnote{Note that if the non-minimal coupling is very small $\xi \ll 1$, 
one can safely neglect the curvature mass, and the perturbation calculation
breaks down in the large coupling case $\xi \gg 1$. Thus, we focus on the 
near-conformal coupling case $\xi \simeq1/6$.}, 
\begin{align}
\left<{  \delta \phi   }^{ 2 } \right>^{(2)}
&= \frac { 1 }{ 48{ \pi  }^{ 2 }  } \Biggl\{
\frac { \ddot { a }  }{ a } +\frac { { \dot { a }  }^{ 2 } }{ { a }^{ 2 } } +\frac { 3 }{ 2 } \frac { \dot { a }  }{ a } 
\frac {6\lambda \phi \dot { \phi  }  }{ { m }^{ 2 }+3\lambda { \phi  }^{ 2 } } \nonumber \\
&-\frac { 1 }{ 4 } \left(\frac { 6\lambda \phi \dot { \phi  } }{ { m }^{ 2 }+3\lambda { \phi  }^{ 2 } }\right)^{2} 
+\frac { 1 }{ 2 } \frac {6\lambda \phi \ddot { \phi  } 
+6\lambda{ \dot { \phi  }  }^{ 2 }  }{ { m }^{ 2 }+3\lambda { \phi  }^{ 2 } } \Biggr\} \nonumber \\
&= \frac { 1 }{ 48{ \pi  }^{ 2 }  } \Biggl\{
\frac { R }{ 6 }+\frac { 3H }{ 2 }
\frac {6\lambda \phi \dot { \phi  }  }{ { m }^{ 2 }+3\lambda { \phi  }^{ 2 } } \nonumber \\
&-\frac { 1 }{ 4 } \left(\frac { 6\lambda \phi \dot { \phi  } }{ { m }^{ 2 }+3\lambda { \phi  }^{ 2 } }\right)^{2} 
+\frac { 1 }{ 2 } \frac { 6\lambda \phi \ddot { \phi  } 
+6\lambda{ \dot { \phi  }  }^{ 2 }  }{ { m }^{ 2 }+3\lambda { \phi  }^{ 2 } } \Biggr\} 
\label{eq:dljkjfjjlg}.
\end{align}
For nearly constant Higgs field, the time-derivative terms of $\dot { \phi  }$ and $\ddot { \phi  }$ are negligible
and the second (adiabatic) order expressions of the vacuum fluctuations
are simplified as
\begin{align}
\left<{  \delta \phi   }^{ 2 } \right>^{(2)}
\simeq \frac { R }{ 288{ \pi  }^{ 2 } } 
\label{eq:dlsdufjjlg}.
\end{align}
Therefore, in the near-conformal coupling case 
($\xi\simeq1/6$ and $m\lesssim  H$), 
we have the high-order vacuum fluctuations corresponding to the particle production effects as follows:
\begin{align}
{ \left< { \delta \phi   }^{ 2 } \right>  }_{ \rm ren }
\simeq\frac { R }{ 288{ \pi  }^{ 2 } }+\mathcal{O}\left(R^{2}\right)+\cdots \label{eq:oesdhg}.
\end{align}
In the radiation dominated Universe, the Ricci scalar becomes $R=0$, and
in the matter dominated Universe, the Ricci scalar becomes $R=3H^{2}$.
On the other hand, in the de Sitter Universe, the Ricci scalar becomes $R=12H^{2}$.
Thus, we summarize the renormalized vacuum field fluctuations 
in the massive conformal coupling case ($\xi\simeq1/6$ and $m\lesssim   H$) as follows:
\begin{align}
\left< {  \delta \phi  }^{ 2 } \right>_{\rm ren}\simeq\begin{cases} 0 ,\quad\quad\quad\quad\quad
\left( {\rm radiation\  dominated\ Universe} \right) 
\\  H^{2}/96\pi^{2},\quad\ \ \left( {\rm matter\  dominated\ Universe} \right)  \\  
H^{2}/24\pi^{2}. \quad\ \ \left( {\rm de\ Sitter\ Universe} \right) \end{cases} \label{eq:klkdjghjhsg}
\end{align}
Note that the massive vacuum field fluctuations in curved background are described by Eq.~(\ref{eq:oesdhg}).
However, the massless vacuum field fluctuations cannot satisfy the adiabatic (WKB) condition of Eq.~(\ref{eq:adgkdhjhf})
to be
\begin{align}
\frac { { \Omega ' }_{ k } }{ { \Omega  }_{ k }^{ 2 } } \simeq \frac { 2H }{ m } \ll 1\label{eq:fddkghgg},
\end{align}
where we assume $m={\rm const}$,
and therefore, the adiabatic expansion method does not provide exact expressions in the massless case. 
In a small mass or rapid varying background, the vacuum field fluctuations are generally enlarged to be
\begin{align}
{ \left< { \delta \phi   }^{ 2 } \right>  }_{ \rm ren }
\gg \mathcal{O}\left(H^{2}\right)\label{eq:fdgsdhg}.
\end{align}
where the vacuum field fluctuations in the non-adiabatic case
are generally larger than the adiabatic one.
This situation cosmologically occurs during inflation for massless scalar fields
or during preheating stage of the parametric resonance (see, e.g. Ref.\cite{Kofman:1997yn}).
In next section, we discuss vacuum field fluctuations in the non-anabatic case.

\section{Renormalized vacuum fluctuations from adiabatic regularization method}
\label{sec:non-adiabatic}
In the non-anabatic case, e.g., a small mass or rapid varying background,
we must usually solve the following equation with the suitable in-vacuum,
\begin{align}
\left<{  \delta \phi   }^{ 2 } \right>_{\rm ren}=\frac { 1 }{ 4{ \pi  }^{ 2 }C\left( \eta  \right)  }
\int _{ 0 }^{ \infty  }{ dk{ k }^{ 2 }{ \Omega  }_{ k }^{ -1 }\left\{2n_{k}+2{\rm Re}z_{k} \right\}  } \label{eq:dfsufsdhg}
\end{align}
where
\begin{align}
{ n  }_{ k }'=\frac{{\Omega  }_{ k }'}{{\Omega  }_{ k }}{\rm Re}z_{k}, \ \
{ z  }_{ k }'=\frac{{\Omega  }_{ k }'}{{\Omega  }_{ k }}\left(n_{k}+\frac{1}{2}\right)-2i{\Omega  }_{ k }z_{k}\label{eq:fufudsdhg}.
\end{align}
However, it is a hard task to calculate 
the above equations in the non-anabatic case.
If we assume unspecified initial conditions or any initial vacuum,
we obtain the following expression of ${ z  }_{ k }$~\cite{Ringwald:1987ui},
\begin{align}
{ z }_{ k }\left( \eta  \right) =& \int _{ { \eta  }_{ 0 } }^{ \eta  }{ d{ \eta  }_{ 1 } } 
\frac { { \Omega ' }_{ k }\left( { \eta  }_{ 1 } \right)  }{ { \Omega  }_{ k }\left( { \eta  }_{ 1 } \right)  } 
\left( { n }_{ k }\left( { \eta  }_{ 1 } \right) +\frac { 1 }{ 2 }  \right) \nonumber  \\ &
\times {\rm exp}\left\{ -2i\int _{ { \eta  }_{ 1 } }^{ \eta  }{ d{ \eta  }_{ 2 }{ \Omega  }_{ k }\left( { \eta  }_{ 2 } \right)  }  \right\}    \nonumber  \\ &
+{ z }_{ k }\left( { \eta  }_{ 0 } \right)  {\rm exp}\left\{ -2i\int _{ { \eta  }_{ 0 } }^{ \eta  }
{ d{ \eta  }_{ 2 }{ \Omega  }_{ k }\left( { \eta  }_{ 2 } \right)  }  \right\},
\end{align} 
where we must adequately solve Eq.~(\ref{eq:fufudsdhg}) and inset into Eq.~(\ref{eq:dfsufsdhg}),
and therefore, there is usually no other way except numerical calculations in the non-anabatic regime.
However, if we analytically calculate the exact mode function of $\delta \chi$
from the Klein-Gordon equation of Eq.~(\ref{eq:dfsufsdhg}) with the suitable in-vacuum,
we can obtain the renormalized vacuum fluctuations ${ \left< { \delta \phi   }^{ 2 } \right>  }_{ \rm ren }$
by removing the UV divergences of ${ \left< { \delta \phi   }^{ 2 } \right>  }$
via adiabatic regularization or point-splitting regularization.

Next, let us review the adiabatic regularization~\cite{bunch1980adiabatic,Parker:1974qw,Fulling:1974pu,Fulling:1974zr,
Anderson:1987yt,Haro:2010zz,Haro:2010mx,birrell1978application,Kohri:2016lsj} 
which is the extremely powerful method
to obtain the renormalized vacuum fluctuations even in the non-adiabatic regime.
The adiabatic regularization is not the mathematical method of regularizing divergent integrals like a kind of 
dimensional regularization or cut-off regularization.
As previously discussed, the divergences of ${ \left< { \delta \phi   }^{ 2 } \right>  }$ come from 
the lowest-order adiabatic mode, and therefore,
we can remove these divergences by subtracting the lowest-order adiabatic (Minkowskian) vacuum field fluctuations 
${ \left< { \delta \phi   }^{ 2 } \right>^{(0)} }$ from  ${ \left< { \delta \phi   }^{ 2 } \right>}$. Thus, we can obtain
the renormalized expression of the adiabatic or the non-adiabatic vacuum fluctuations as follows,
\begin{align}
{ \left< { \delta \phi   }^{ 2 } \right>}_{\rm ren}
&={ \left< { \delta \phi   }^{ 2 } \right>}-{ \left< { \delta \phi   }^{ 2 } \right>^{(0)} }
\label{eq:adshdkhf}  \\
&=\frac { 1 }{ 4{ \pi  }^{ 2 }C\left( \eta  \right)  }
\int _{ 0 }^{ \infty  }{ dk{ k }^{ 2 }{ \Omega  }_{ k }^{ -1 }\left\{2n_{k}+2{\rm Re}z_{k} \right\}  }  \nonumber \\
&=\frac { 1 }{ 4{ \pi  }^{ 2 }C\left( \eta  \right)  } \left[\int _{ 0 }^{ \infty  }{ dk2{ k }^{ 2 }{ \left| \delta { \chi  }_{ k } \right|  }^{ 2 } }-
\int _{ 0 }^{ \infty  }{ dk{ k }^{ 2 }{ \Omega  }_{ k }^{ -1 }  } \right]  \nonumber,
\end{align}
where we must obtain the exact mode function of $\delta { \chi  }_{ k }$ with appropriate in-vacuum.
Note that the above formulation is improved in comparison with the literature~\cite{bunch1980adiabatic,Parker:1974qw,Fulling:1974pu,Fulling:1974zr,
Anderson:1987yt,Haro:2010zz,Haro:2010mx,birrell1978application,Haro:2010zz,Haro:2010mx,Kohri:2016lsj}.
This method is equivalent to the point-splitting regularization which regularizes divergences 
via the point separation in the two-point function.

As a concrete example, how to use the adiabatic regularization, 
we consider the vacuum field fluctuations of the massless minimally-coupled
scalar field ($\xi=0$ and $m=0$) in de Sitter background
(for the detailed discussions see Refs.~\cite{Haro:2010zz,Haro:2010mx}).
In this case, the mode function $\delta { \chi  }_{ k }\left( \eta  \right) $ can be exactly given by 
\begin{align}
\delta { \chi  }_{ k }\left( \eta  \right) =\frac{1}{\sqrt{2k}}\left\{ { \alpha }_{ k }\delta { \varphi  }_{ k }\left( \eta  \right) 
+{ \beta }_{ k }\delta { \varphi }^{*}_{ k }\left( \eta  \right) \right\} \label{eq:ohjijgedg},
\end{align}
where
\begin{align}
\delta { \varphi  }_{ k }\left( \eta  \right) ={ e }^{ -ik\eta  }\left( 1+\frac { 1 }{ ik\eta  }  \right)\label{eq:fhghgedg}. 
\end{align}
In the massless minimally coupled case, the vacuum field fluctuations $\left< { \delta \phi   }^{ 2 } \right>$ 
have not only ultraviolet divergences but also infrared divergences. 
Thus, we assume that the Universe changes from the radiation-dominated 
Universe to the de Sitter Universe in order to avoid the infrared divergences
\begin{align}
a\left( \eta  \right) =\begin{cases} 2-\frac { \eta  }{ { \eta  }_{ 0 } } ,\quad \left( \eta <{ \eta  }_{ 0 } \right)  \\ 
\frac { \eta  }{ { \eta  }_{ 0 } } ,\quad\quad\ \left( \eta >{ \eta  }_{ 0 } \right)  \end{cases}\label{eq:fhsdljedg}
\end{align}
where ${ \eta  }_{ 0 }=-1/H$, and 
we choose the mode function as the in-vacuum state 
\begin{align}
\delta { \chi  }_{ k }={ e }^{ -ik\eta  }/\sqrt { 2k }\label{eq:fhsdsedg}.
\end{align}
during the radiation-dominated region $\left( \eta <{ \eta  }_{ 0 } \right)$.
By requiring the conditions $\delta { \chi   }_{ k }\left( \eta  \right)$ 
and $\delta { \chi   }_{ k }'\left( \eta  \right)$ at the matching time ${ \eta  }={ \eta  }_{ 0 }$,  
we obtain the corresponding coefficients of the mode function as follows
\begin{align}
{ \alpha }_{ k }&=1+\frac { H }{ ik } -\frac { { H }^{ 2 } }{ 2{ k }^{ 2 } }, \\
{ \beta }_{ k }&=-\frac{H^{2}}{2k^{2}}e^{\frac{2ik}{H}} 
={ \alpha }_{ k }+\frac { 2ik }{ 3H } +\mathcal{O}\left( \frac { { k }^{ 2 } }{ { H }^{ 2 } }  \right) 
\label{eq:fkhkdedg}.
\end{align}
By using the above coefficients of ${ \alpha }_{ k }$ and ${ \beta }_{ k }$,
we obtain the suitable mode function of $\delta { \chi  }_{ k }$.
For small $k$ modes in the de Sitter Universe $\left( \eta  >{ \eta  }_{ 0 } \right)$, 
we can approximate the mode function to be
\begin{align}
{ \left| \delta { \chi  }_{ k } \right|  }^{ 2 }=\frac { 1 }{ 2k } 
\left[ { \left( \frac { 2 }{ 3H\eta  } +2+\frac { { H }^{ 2 }{ \eta  }^{ 2 } }{ 6 }  \right)  }^{ 2 }
+\mathcal{O}\left( \frac { { k }^{ 2 } }{ { H }^{ 2 } }  \right) +\cdots  \right] \label{eq:joooegedg}.
\end{align}
Here it is notable that one has no infrared divergences because
$k^{2}{ \left| \delta { \chi  }_{ k } \right|  }^{ 2 }\approx \mathcal{O}\left(k\right)$.
For large $k$ modes,  we can obtain the following expression 
\begin{align}
{ \left| \delta { \chi  }_{ k } \right|  }^{ 2 }=&\frac { 1 }{ 2k } \Biggl[ 1+\frac { 1 }{ { k }^{ 2 }{ \eta  }^{ 2 } } 
-\frac { { H }^{ 2 } }{ { k }^{ 2 } } \cos { \left( 2k\left( 1/H+\eta  \right)  \right)  }\nonumber \\
&+\mathcal{O}\left( \frac { { H }^{ 3 } }{ { k }^{ 3 } }  \right) +\cdots  \Biggr] \label{eq:ogfhghedg}.
\end{align}
Here, we must require the cut-off of $k$ mode form the 
the adiabatic (WKB) condition ${ \Omega  }_{ k }^{ 2 }>0$ to be $k>\sqrt{2}/\left| \eta \right|=\sqrt{2}aH$.
Therefore, we can obtain the renormalized vacuum fluctuations 
form Eq.~(\ref{eq:adshdkhf}) as follows:
\begin{align}
{ \left< { \delta \phi   }^{ 2 } \right>  }_{ \rm ren }
=&\lim _{ \Lambda \rightarrow \infty }\frac { 1 }{ 4{ \pi  }^{ 2 }C\left( \eta  \right)  }  
\Biggl[ \int _{ 0 }^{ \Lambda  }{ 2k^{2}{ \left| \delta { \chi  }_{ k } \right|  }^{ 2 }dk } 
-\int _{ \sqrt{2}/\left| \eta \right| }^{ \Lambda  }{ dk{ k }^{ 2 }{ \Omega  }_{ k }^{ -1 }} \Biggr]\nonumber  \\
=&\lim _{ \Lambda \rightarrow \infty }\frac { 1 }{ 4{ \pi  }^{ 2 }C\left( \eta  \right)  }
\Biggl[ \int _{ 0 }^{ \Lambda  }{ 2k^{2}{ \left| \delta { \chi  }_{ k } \right|  }^{ 2 }dk } \nonumber \\
&-\int _{\sqrt{2}/\left| \eta \right| }^{ \Lambda  }{ \frac { { k }^{ 2 } }{ \sqrt{k^{2}-2/\eta^{2}}}  dk} \Biggr]  \\
=&\lim _{ \Lambda \rightarrow \infty  }\frac { 1 }{ 4{ \pi  }^{ 2 }C\left( \eta  \right)  }
\Biggl[ \int _{ 0 }^{ \Lambda  }{ 
2k^{2}{ \left| \delta { \chi  }_{ k } \right|  }^{ 2 }dk } \nonumber \\
&-\int _{\sqrt{2}/\left| \eta \right| }^{ \Lambda  }
{ \left( k + \frac{1}{k\eta^{2}}+\cdots \right) dk} \Biggr] 
\label{eq:fhgegedg}.
\end{align}
For large $k$ modes, we can use Eq.~(\ref{eq:ogfhghedg}) 
and subtract the UV divergences as the following 
\begin{align}
&\lim _{ \Lambda \rightarrow \infty  }\frac { 1 }{ 4{ \pi  }^{ 2 }C\left( \eta  \right)  }
\Biggl[ \int _{\sqrt{2}/\left| \eta \right| }^{ \Lambda  }{ 
 \left(k+\frac { 1 }{ { k }{ \eta  }^{ 2 } }  \right) dk } \nonumber \\
&-\int _{\sqrt{2}/\left| \eta \right| }^{ \Lambda  }
{ \left( k + \frac{1}{k\eta^{2}} \right) dk} \Biggr] =0 \label{eq:ohffddg}.
\end{align}
Thus, we obtain the following expression of the renormalized vacuum fluctuations as
\begin{align}
{ \left< { \delta \phi   }^{ 2 } \right>  }_{ \rm ren }
=&\frac { 1 }{ 2{ \pi  }^{ 2 }C\left( \eta  \right)  } \int _{ 0 }^{ \sqrt{2}/\left| \eta \right|  }
{ k^{2}{ \left| \delta { \chi  }_{ k } \right|  }^{ 2 }dk } \nonumber \\ &
+\frac { { \eta  }^{ 2 }{ H }^{ 2 } }{ 4{ \pi  }^{ 2 } } \int _{ \sqrt { 2 } /\left| \eta  \right|  }^{ \infty  }
\Biggl( -\frac { { H }^{ 2 } }{ { k }^{ 2 } } \cos { \left( 2k\left( 1/H+\eta  \right)  \right)  } \nonumber \\
&+\mathcal{O}\left( \frac { { H }^{ 3 } }{ { k }^{ 3 } }  \right) +\cdots  \Biggr)  kdk
\label{eq:oehf:lhg}.
\end{align}
At the late cosmic-time ($\eta\simeq0$ corresponds to $N_{\rm tot}=Ht\gg1$),
we have the following approximation 
\begin{align}
{ \left< { \delta \phi   }^{ 2 } \right>  }_{ \rm ren }
&\simeq  \frac { { \eta  }^{ 2 }{ H }^{ 2 } }{ 2{ \pi  }^{ 2 } } \int _{ 0 }^{ \sqrt{2}/\left| \eta \right|  }
{ k^{2}{ \left| \delta { \chi  }_{ k } \right|  }^{ 2 }dk }, \nonumber \\ 
&\simeq \frac { 1 }{ 9{ \pi  }^{ 2 } } \int _{ 0 }^{ H }{ kdk } 
+\frac { { H }^{ 2 } }{ 4{ \pi  }^{ 2 } } \int _{ H }^{ \sqrt { 2 } /\left| \eta  \right|  }{ \frac { 1 }{ k } dk } \label{eq:oflhkegedg},
\end{align}
where we approximate the mode function $\delta { \chi  }_{ k  }\left( \eta \right)$ 
from Eq.~(\ref{eq:joooegedg}) and Eq.~(\ref{eq:ogfhghedg}) as the following
\begin{align}
{ \left| \delta { \chi  }_{ k } \right|  }^{ 2 } \simeq  \begin{cases}  
\frac{1}{2k}{ \left( \frac { 2 }{ 3H\eta  } +2+\frac { { H }^{ 2 }{ \eta  }^{ 2 } }{ 6 }  \right)  }^{ 2 }
 \quad\quad \left( 0\le  k \le  H \right)  \\  
 \frac{1}{2k}\left(1+\frac { 1 }{ { k }^{ 2 }{ \eta  }^{ 2 } } \right)
  \quad\quad\quad\quad  \left( H\le  k \le \sqrt { 2 } /\left| \eta  \right|  \right) \end{cases} 
\label{eq:klgffsg}
\end{align}
Therefore, we can finally obtain the well-know expression as follows
\begin{align}
{ \left< { \delta \phi   }^{ 2 } \right>  }_{ \rm ren }
&\simeq \frac { { H }^{ 2 } }{ 18{ \pi  }^{ 2 } } +\frac { { H }^{ 2 } }{ 4{ \pi  }^{ 2 } } 
\left( \frac { 1 }{ 2 } \log { 2 } +Ht \right) \nonumber \\
&\simeq \frac { { H }^{ 3 } }{ 4{ \pi  }^{ 2 } } t\label{eq:odhsdhedg},
\end{align}
which grows as cosmic-time proceeds.

Next, let us consider the massive minimally coupled scalar field 
($\xi \ll 1$ and $m \ll H$) in de Sitter background.
This situation is cosmologically important in order to understand 
the origin of the primordial perturbations or the self-backreaction of the inflaton field in inflationary Universe
(see, e.g. Ref.\cite{Linde:1982uu,Vilenkin:1983xq}).
In this case, the mode function $\delta { \chi  }_{ k }\left( \eta  \right) $ can be given by 
\begin{align}
\delta { \chi  }_{ k }\left( \eta  \right) =\sqrt{\frac{\pi}{4}}\eta^{1/2}\left\{ { \alpha }_{ k }{ H}_{ \nu }^{(2)}\left( k\eta  \right) 
+{ \beta }_{ k }{ H}_{ \nu }^{(1)}\left( k\eta  \right)  \right\} \label{eq:ohghggedg},
\end{align}
with
\begin{align}
\nu \equiv \sqrt{\frac{9}{4}-\frac{m^{2}}{H^{2}}}\simeq \frac{3}{2}-\frac{m^{2}}{3H^{2}},
\end{align}
where ${ H}_{ \nu }^{(1, 2 )}\left( k\eta  \right)$ are the Hankel functions.
For simplicity we assume the spacetime transition from the radiation-dominated Universe to the de Sitter Universe
and require the matching conditions at $\eta =\eta_{0}$ to determine the Bogoliubov coefficients
\begin{align}
\alpha_{k}=&\frac { 1 }{ 2i } \sqrt { \frac { \pi k{ \eta  }_{ 0 } }{ 2 }  } \biggl( \left( -i+\frac { H }{ 2k }  \right) { H }_{ \nu  }^{ (1)}\left( k{ \eta  }_{ 0 } \right) 
\nonumber  \\  &\quad -{ H }_{ \nu  }^{ (1)' }\left( k{ \eta  }_{ 0 } \right)  \biggr) { e }^{ ik/H } \label{eq:ksfdedg}, \\
\beta_{k}=&-\frac { 1 }{ 2i } \sqrt { \frac { \pi k{ \eta  }_{ 0 } }{ 2 }  } \biggl( \left( -i+\frac { H }{ 2k }  \right) { H }_{ \nu  }^{ (2) }\left( k{ \eta  }_{ 0 } \right) 
\nonumber \\  &\quad -{ H }_{ \nu  }^{ (2)' }\left( k{ \eta  }_{ 0 } \right)  \biggr) { e }^{ ik/H } \label{eq:kgdggedg}.
\end{align}
The renormalized vacuum fluctuations from Eq.~(\ref{eq:adshdkhf}) are given as follows:
\begin{align}
{ \left< { \delta \phi   }^{ 2 } \right>  }_{ \rm ren }
=&\lim _{ \Lambda \rightarrow \infty }\frac { 1 }{ 4{ \pi  }^{ 2 }C\left( \eta  \right)  }  
\Biggl[ \int _{ 0 }^{ \Lambda  }{ 2k^{2}{ \left| \delta { \chi  }_{ k } \right|  }^{ 2 }dk } \nonumber \\
&\quad -\int _{ \sqrt{2}/\left| \eta \right| }^{ \Lambda  }{ dk{ k }^{ 2 }{ \Omega  }_{ k }^{ -1 }} \Biggr] \nonumber \\
=&\frac { \eta^{2}H^{2} }{ 2{ \pi  }^{ 2 } } \int _{ 0 }^{ H }{ k^{2}{ \left| \delta { \chi  }_{ k } \right|  }^{ 2 }dk }  \nonumber \\
& \quad +\frac { \eta^{2}H^{2} }{ 2{ \pi  }^{ 2 }} \int _{ H }^{ \sqrt{2}/\left| \eta \right| }{ k^{2}{ \left| \delta { \chi  }_{ k } \right|  }^{ 2 }dk }  
\label{eq:kdddegedg}.
\end{align}
The divergence parts exactly cancel as previously discussed,
\begin{align}
\lim _{ \Lambda \rightarrow \infty }\frac { 1 }{ 4{ \pi  }^{ 2 }C\left( \eta  \right)  }  
\Biggl[ \int _{ \sqrt{2}/\left| \eta \right|}^{ \Lambda  }{ 2k^{2}{ \left| \delta { \chi  }_{ k } \right|  }^{ 2 }dk } 
-\int _{ \sqrt{2}/\left| \eta \right| }^{ \Lambda  }{ dk{ k }^{ 2 }{ \Omega  }_{ k }^{ -1 }} \Biggr] ,
\end{align}
where we must take the adiabatic mode cut-off as $k> \sqrt{2-m^{2}/H^{2}}/\left|\eta \right| \simeq \sqrt{2}/\left|\eta \right|$.
It is more difficult task in the massive case than in the massless case to obtain exactly the 
renormalized vacuum fluctuations from Eq.~(\ref{eq:ksfdedg}) and Eq.~(\ref{eq:kgdggedg}).
However, by using the asymptotic behavior of the Hankel functions,
we can easily get the renormalized vacuum fluctuations of ${ \left< { \delta \phi   }^{ 2 } \right>  }_{ \rm ren }$
via the adiabatic regularization method (for the details, see Ref.\cite{Haro:2010zz,Haro:2010mx}).

By using the following formula of the Hankel functions 
\begin{align}
{ H }_{ \nu  }^{ (1,2)' }\left( k{ \eta  }_{ 0 } \right)=
{ H }_{ \nu-1  }^{ (1,2) }\left( k{ \eta  }_{ 0 } \right)-\frac{\nu}{k{ \eta  }_{ 0 }}{ H }_{ \nu  }^{ (1,2) }\left( k{ \eta  }_{ 0 } \right),
\end{align}
and the Bessel function of the first kind defined by $J_{\nu}=( { H }_{ \nu  }^{ (1) }+{ H }_{ \nu  }^{ (2) } )/2$,
we obtain the following expression 
\begin{align}
\left| { \alpha  }_{ k }-{ \beta  }_{ k } \right| =\sqrt { \frac { \pi k }{ 2H }  } \left| { J }_{ \nu -1 }\left( k{ \eta  }_{ 0 } \right) +\left( i-\frac { H }{ 2k } +\frac { \nu H }{ k }  \right) { J }_{ \nu  }\left( k{ \eta  }_{ 0 } \right)  \right| .
\end{align}
For small $k$ modes, the the Bessel function and the Hankel function asymptotically behave as
\begin{align}
J_{\nu}\left(k\eta_{0}\right)&\simeq \frac { 1 }{ \Gamma \left( \nu +1 \right)  } { \left( \frac { k{ \eta  }_{ 0 } }{ 2 }  \right)  }^{ \nu  },\\
{ H }_{ \nu }^{ (2) }\left( k{ \eta  }_{ 0 } \right)&\simeq -{ H }_{ \nu }^{ (1) }\left( k{ \eta  }_{ 0 } \right)
\simeq \frac { i }{ \pi  } \Gamma \left( \nu  \right) { \left( \frac { k{ \eta  }_{ 0 } }{ 2 }  \right)  }^{ -\nu  }.
\end{align}
Thus, we can obtain the following expression of the mode function,
\begin{align}
{ \left| \delta { \chi  }_{ k } \right|  }^{ 2 } &\simeq
\frac { \pi  }{ 4 } \left| \eta  \right| { \left| { \alpha  }_{ k }-{ \beta  }_{ k } \right|  }^{ 2 }{ { \left| { H }_{ \nu }^{ (2) }\left( k{ \eta  }\right) \right|  }^{ 2 } }\nonumber \\ &\simeq
\frac{2}{9k}\left(H\left| \eta \right|\right)^{1-2\nu}\quad\ \left( 0\le  k \le  H \right)\label{eq:kddffsedg}.
\end{align}
For large $k$ modes, we can approximate the Bogoliubov coefficients as $\alpha_{k} \simeq1$ and $\beta_{k} \simeq 0$
and evaluate the mode function 
\begin{align}
\delta { \chi  }_{ k }\left( \eta  \right) \simeq \sqrt{\frac{\pi}{4}}\eta^{1/2}{ H}_{ \nu }^{(2)}\left( k\eta  \right) .
\end{align}
Thus, we obtain the following expression 
\begin{align}
{ \left| \delta { \chi  }_{ k } \right|  }^{ 2 } \simeq \frac{\left| \eta \right|}{16}
\left(\frac{k\left| \eta \right|}{2}\right)^{-2\nu}\quad  \left( H\le  k \le \sqrt { 2 } /\left| \eta  \right|  \right).\label{eq:kdyydg}
\end{align}
From Eq.~(\ref{eq:kddffsedg}) and Eq.~(\ref{eq:kdyydg}),
the renormalized vacuum fluctuations are given by
\begin{align}
{ \left< { \delta \phi   }^{ 2 } \right>  }_{ \rm ren }
&\simeq  \frac { \left(H\left| \eta \right|\right)^{3-2\nu} }{ 9{ \pi  }^{ 2 } } \int _{ 0 }^{ H }{ k dk }  \nonumber \\
&+\frac { H^{2}\left| \eta \right|^{3-2\nu} }{ 4{ \pi  }^{ 2 }\cdot2^{3-2\nu}} \int _{ H }^{ \sqrt{2}/\left| \eta \right| }{ k^{2-2\nu}dk }\\
&\simeq \frac { H^{2} }{ 18{ \pi  }^{ 2 } }e^{-\frac{2m^{2}t}{3H}}
+\frac { 3H^{4} }{ 8{ \pi  }^{ 2 }m^{2}} \left(1-e^{-\frac{2m^{2}t}{3H}}\right)
\label{eq:kfhshggdg}.
\end{align}
For late cosmic-time ($N_{\rm tot}=Ht\gg H^{2}/m^{2}$),
the renormalized vacuum fluctuations ${ \left< { \delta \phi   }^{ 2 } \right>  }_{ \rm ren }$ 
in de Sitter background are approximately written as
\begin{align}
{ \left< { \delta \phi   }^{ 2 } \right>  }_{ \rm ren }
\simeq \frac { 3H^{4} }{ 8{ \pi  }^{ 2 }m^{2}} 
\label{eq:klkddussg}.
\end{align}
These vacuum field fluctuations as described by Eq.~(\ref{eq:klkdjghjhsg}) and Eq.~(\ref{eq:klkddussg})
are corresponding to the quantum particle creation from the curved background, 
and therefore, once generated vacuum fluctuations remains on the cosmological timescale.
However, if the created particles can decay into other particles,
the created vacuum field fluctuations would disappear on the particle decay timescale.

\section{Renormalized vacuum field fluctuations in dynamical scalar field background}
\label{sec:background}
In the general cosmological situations, the background Higgs field dynamically changes and 
does not stagnate for all times.
The dynamical variation of the Higgs field or other scalar field coupled with the Higgs field
provide a varying effective mass and 
leads to real particle productions or the vacuum fluctuations of the Higgs field.
Even in the slowly varying scalar field background, the generated vacuum field fluctuations are non-negligible.
In this section, we consider the vacuum filed fluctuations in the slowly varying scalar field background
following the literature~\cite{Ringwald:1987ui}.

\subsection{The Higgs field background}
\label{sec:higgs-background}
For convenience, we rewrite Eq.~(\ref{eq:dfsufsdhg}) in order to 
obtain the renormalized vacuum field fluctuations on the dynamical Higgs field background,
\begin{align}
\left<{  \delta \phi   }^{ 2 } \right>_{\rm ren}=\frac { 1 }{ 4{ \pi  }^{ 2 }C\left( \eta  \right)  }
\int _{ 0 }^{ \infty  }{ dk{ k }^{ 2 }{ \Omega  }_{ k }^{ -1 }\left\{2n_{k}+2{\rm Re}z_{k} \right\}  } \label{eq:dfsudfgg},
\end{align}
where ${ n  }_{ k }$ and ${ z  }_{ k }$ are determined by the differential equations of Eq.~(\ref{eq:fufudsdhg}) as follows
\begin{align}
{ n  }_{ k }'=\frac{{\Omega  }_{ k }'}{{\Omega  }_{ k }}{\rm Re}z_{k}, \ \
{ z  }_{ k }'=\frac{{\Omega  }_{ k }'}{{\Omega  }_{ k }}\left(n_{k}+\frac{1}{2}\right)-2i{\Omega  }_{ k }z_{k}\label{eq:fufudfg}.
\end{align}
For simplicity we assume the initial conditions to be $ { n  }_{ k }\left(\eta_{0}\right)={ z }_{ k }\left(\eta_{0}\right)=0$,
and obtain the following equations,
\begin{align}
{ n  }_{ k }\left(\eta\right)&= \int _{ { \eta  }_{ 0 } }^{ \eta  }{ d{ \eta  }_{ 1 } } \int _{ { \eta  }_{ 0 } }^{ \eta_{1}  }{ d{ \eta  }_{ 2 } } 
\frac{{\Omega'  }_{ k }\left(\eta_{1}\right)}{{\Omega }_{ k }\left(\eta_{1}\right)}
\frac{{\Omega'  }_{ k }\left(\eta_{2}\right)}{{\Omega }_{ k }\left(\eta_{2}\right)} \nonumber \\ & \times 
\cos \bigl\{ 2\int _{ { \eta  }_{ 2 } }^{ \eta_{1}  }{ d{ \eta  }_{ 3 } } {\Omega }_{ k }\left(\eta_{3}\right)\bigr\}
\left( \frac{1}{2}+ { n  }_{ k }\left(\eta_{2}\right)  \right)   \label{eq:fnlg},  \\
{\rm Re}z_{k}\left(\eta\right)&=\int _{ { \eta  }_{ 0 } }^{ \eta  }{ d{ \eta  }_{ 1 } }
\frac{{\Omega'  }_{ k }\left(\eta_{1}\right)}{{\Omega }_{ k }\left(\eta_{1}\right)}
\cos \bigl\{ 2\int _{ { \eta  }_{ 1 } }^{ \eta  }{ d{ \eta  }_{ 2 } } {\Omega }_{ k }\left(\eta_{2}\right)\bigr\} \nonumber \\ & \times
\left( \frac{1}{2}+\int _{ { \eta  }_{ 0 } }^{ \eta_{1}  }{ d{ \eta  }_{ 3 } }
\frac{{\Omega'  }_{ k }\left(\eta_{3}\right)}{{\Omega }_{ k }\left(\eta_{3}\right)}{\rm Re}z_{k}\left(\eta_{3}\right)  \right) \label{eq:fnlsdsg}.
\end{align}
Let us consider the following condition as
\begin{align}
\left|  \int _{ { \eta  }_{ 0 } }^{ \eta  }{ d{ \eta  }_{ 1 } }
\frac{{\Omega'  }_{ k }\left(\eta_{1}\right)}{{\Omega }_{ k }\left(\eta_{1}\right)} \right| \ll 1,
\end{align}
which corresponds to the small time-difference of ${\bar { M } }^{ 2 }\left( \eta \right)$ as follows
\begin{align}
\left|  { \bar { M } }^{ 2 }\left( { \eta  } \right) -{ \bar { M } }^{ 2 }\left( { \eta_{0}  } \right)\right| 
\ll 2{ \bar { M } }^{ 2 }\left( { \eta  } \right) {\rm or}\ 2{ \bar { M } }^{ 2 }\left( { \eta_{0}  } \right).
\end{align}
In this assumption, we can approximate these equations of Eq.~(\ref{eq:fnlg}) and Eq.~(\ref{eq:fnlsdsg}) as follows:
\begin{align}
{ n  }_{ k }\left(\eta\right)\simeq & \  0 \label{eq:fnlweerg}, \\
{\rm Re}z_{k}\left(\eta\right)\simeq &\ \frac{1}{2}\int _{ { \eta  }_{ 0 } }^{ \eta  }{ d{ \eta  }_{ 1 } }
\frac{{\Omega'  }_{ k }\left(\eta_{1}\right)}{{\Omega }_{ k }\left(\eta_{1}\right)}
\cos \bigl\{ 2\int _{ { \eta  }_{ 1 } }^{ \eta  }{ d{ \eta  }_{ 2 } } {\Omega }_{ k }\left(\eta_{2}\right)\bigr\}. \nonumber
\end{align}
Furthermore, we can approximate Eq.~(\ref{eq:fnlweerg}) as the following 
\begin{align}
{\rm Re}z_{k}\left(\eta\right) &\simeq \frac{1}{2}\int _{ { \eta  }_{ 0 } }^{ \eta  }{ d{ \eta  }_{ 1 } }
\frac{\bar{ M }\left(\eta_{1}\right)\bar{ M }'\left(\eta_{1}\right)  }{{\Omega }_{ k }^{2}\left(\eta_{1}\right)}
\cos \bigl\{ 2\int _{ { \eta  }_{ 1 } }^{ \eta  }{ d{ \eta  }_{ 2 } } {\Omega }_{ k }\left(\eta_{2}\right)\bigr\} \nonumber  \\
&\simeq  \frac{1}{2{\Omega }_{ k }^{2}\left(\eta \right)}\int _{ { \eta  }_{ 0 } }^{ \eta  }{ d{ \eta  }_{ 1 } }
\bar{ M }\left(\eta_{1}\right)\bar{ M }'\left(\eta_{1}\right)
\cos \bigl\{2 {\Omega }_{ k }\left(\eta \right)\left(\eta - \eta_{1}\right)\bigr\}.
\end{align}
From Eq.~(\ref{eq:dfsudfgg}), we obtain the renormalized vacuum field fluctuations,
\begin{align}
\left<{  \delta \phi   }^{ 2 } \right>_{\rm ren}&=\frac { 1 }{ 2{ \pi  }^{ 2 }C\left( \eta  \right)  }
\int _{ 0 }^{ \infty  }{ dk{ k }^{ 2 }{ \Omega  }_{ k }^{ -1 }\left\{n_{k}+{\rm Re}z_{k} \right\}  } \nonumber \\
&\simeq \frac { 1 }{ 2{ \pi  }^{ 2 }C\left( \eta  \right)  }\int _{ 0 }^{ \infty  } dk{ k }^{ 2 }{ \Omega  }_{ k }^{ -3 }
\int _{ { \eta  }_{ 0 } }^{ \eta  }{ d{ \eta  }_{ 1 } } \bar{ M }\left(\eta_{1}\right)\bar{ M }'\left(\eta_{1}\right)
\nonumber \\ & \times
\cos \bigl\{2 {\Omega }_{ k }\left(\eta \right)\left(\eta - \eta_{1}\right)\bigr\}
\label{eq:errudfgg}.
\end{align}
By integration by parts, we have the following expression,
\begin{align}
\left<{  \delta \phi   }^{ 2 } \right>_{\rm ren} &\simeq 
\frac { \bar{ M }^{2}\left(\eta \right)}{ 8{ \pi  }^{ 2 }C\left( \eta  \right)  }\left(\bar{ M }^{2}\left(\eta_{0} \right)-\bar{ M }^{2}\left(\eta \right)\right) 
\int _{ 0 }^{ \infty  } dk{ \Omega  }_{ k }^{ -3 } \nonumber \\  &+
\frac { 1 }{ 4{ \pi  }^{ 2 }C\left( \eta  \right)  }\int _{ 0 }^{ \infty  } dk { \Omega  }_{ k }^{ -1 }
\int _{ { \eta  }_{ 0 } }^{ \eta  }{ d{ \eta  }_{ 1 } } \nonumber \\ & \times
\bar{ M }\left(\eta_{1}\right)\bar{ M }'\left(\eta_{1}\right)
\cos \bigl\{2 {\Omega }_{ k }\left(\eta \right)\left(\eta - \eta_{1}\right)\bigr\} \nonumber \\  &+
\frac { \bar{ M }^{2}\left(\eta \right)}{ 4{ \pi  }^{ 2 }C\left( \eta  \right)  }\int _{ 0 }^{ \infty  } dk { \Omega  }_{ k }^{ -2 }
\int _{ { \eta  }_{ 0 } }^{ \eta  }{ d{ \eta  }_{ 1 } } \label{eq:errsdlfjsd} \\ & \times
\left( { \bar{ M } }^{ 2 }\left( { \eta_{1}  } \right) -{ \bar{ M } }^{ 2 }\left( { \eta_{0}  } \right) \right)
\sin \bigl\{2 {\Omega }_{ k }\left(\eta \right)\left(\eta - \eta_{1}\right)\bigr\} \nonumber
\end{align}
which is equivalent to the result 
by using the perturbation technique~\cite{Davies:1979se}.
By performing the integration, we obtain the following expression,
\begin{align}
\left<{  \delta \phi   }^{ 2 } \right>_{\rm ren} &\simeq 
\frac { 1}{ 8{ \pi  }^{ 2 }a^{2}\left(\eta \right)  }\left(\bar{ M }^{2}\left(\eta_{0} \right)-\bar{ M }^{2}\left(\eta \right)\right)  \nonumber \\  
&-\frac { 1 }{ 8{ \pi  }^{ 2 }a^{2}\left( \eta  \right)  }
\int _{ { \eta  }_{ 0 } }^{ \eta  }{ d{ \eta  }_{ 1 } }
\bar{ M }\left(\eta_{1}\right)\bar{ M }'\left(\eta_{1}\right) \nonumber \\ & \times
N_{0} \left( 2\bar{ M }\left(\eta - \eta_{1}\right)\right) \nonumber \\  &+
\frac { \bar{ M }^{2}\left(\eta \right)}{ 8{ \pi  }^{ 2 }a^{2}\left( \eta  \right)  }
\int _{ { \eta  }_{ 0 } }^{ \eta  }{ d{ \eta  }_{ 1 } } 
\left( \eta-\eta_{1} \right) \label{eq:erreljfjsd} \\ & \times
\left( { \bar{ M } }^{ 2 }\left( { \eta_{1}  } \right) -{ \bar{ M } }^{ 2 }\left( { \eta_{0}  } \right) \right)
F \left( 2\bar{ M }\left(\eta - \eta_{1}\right)\right) \nonumber
\end{align}
where $N_{0}\left(x\right)$ is the Bessel function, $F\left(x\right)$ is
combination of Bessel function $N_{\alpha}\left(x\right)$
and Struve functions $\bm{H}_{\alpha}\left(x\right)$ defined by 
$F\left(x\right)\equiv \bm{H}_{0}\left(x\right)N_{1}\left(x\right)
+ N_{0}\left(x\right)\bm{H}_{-1}\left(x\right)$ and $\bar{ M }\left(\eta \right)$ is described by $\bar{ M }\left(\eta \right) \simeq 3\lambda a\left(\eta \right) \phi^{2} \left(\eta \right)$.

When the expansion of the Universe is slow and the background Higgs field $\phi \left(\eta \right)$
evolves quickly on the cosmological timescale,
the vacuum field fluctuations evolve in proportion to $\bar{ M }\left(\eta \right)$. 
The vacuum field fluctuations given by Eq.~(\ref{eq:erreljfjsd}) would be approximately equal to
the first-order adiabatic approximation of Eq.~(\ref{eq:adbatigudf})
where the odd-order adiabatic number density is zero as $n^{(2n+1)}_{k}=0$.
As previously discussed, the second-order approximation of 
the vacuum field fluctuations are given by Eq.~(\ref{eq:dggdfgg}),
\begin{align}
\left<{  \delta \phi   }^{ 2 } \right>_{\rm ren}
&=\frac { 1 }{ 48{ \pi  }^{ 2 }a^{2}\left( \eta  \right)  }\frac{\bar{ M }''\left(\eta \right) }{\bar{ M }\left(\eta \right) }\label{eq:dggdkf}, \\
{ \bar{ M }  }^{ 2 }\left( \eta \right) &=a^2\left( \eta \right)\left(m^{2}+3\lambda \phi^{2}+\left(\xi-1/6\right)R\right) \nonumber.
\end{align}
Thus, we obtain the following expression of 
the second-order adiabatic vacuum fluctuations to be 
\begin{align}
\left<{  \delta \phi   }^{ 2 } \right>_{\rm ren}
=&\frac { 1 }{ 48{ \pi  }^{ 2 }  } \Biggl\{
\frac { \ddot { a }  }{ a } +\frac { { \dot { a }  }^{ 2 } }{ { a }^{ 2 } } +\frac { 3 }{ 2 } \frac { \dot { a }  }{ a } \frac { \left( \xi -1/6 \right) \dot { R } 
+6\lambda \phi \dot { \phi  }  }{ { m }^{ 2 }+\left( \xi -1/6 \right) R+3\lambda { \phi  }^{ 2 } } \nonumber \\
&-\frac { 1 }{ 4 } \frac { { \left( \left( \xi -1/6 \right) \dot { R } +6\lambda \phi \dot { \phi  }  \right)  }^{ 2 } }{ { \left( { m }^{ 2 }+\left( \xi -1/6 \right) R+3\lambda { \phi  }^{ 2 } \right)  }^{ 2 } }  \nonumber \\
&+\frac { 1 }{ 2 } \frac { \left( \xi -1/6 \right) \ddot { R } +6\lambda \left( \phi \ddot { \phi  } 
+{ \dot { \phi  }  }^{ 2 } \right)  }{ { m }^{ 2 }+\left( \xi -1/6 \right) R+3\lambda { \phi  }^{ 2 } } \Biggr\} .
\label{eq:dgferedhfg}
\end{align}
If the large background Higgs field $\phi \left(t \right)$ exists, and 
we can safely neglect the mass terms or the non-minimal curvature terms, 
the second-order adiabatic expression of the vacuum field fluctuations are written as
\begin{align}
\left<{  \delta \phi   }^{ 2 } \right>_{\rm ren}
\simeq &\frac { 1 }{ 48{ \pi  }^{ 2 }  } \Biggl\{
\frac{1}{6}R+\frac {3H\dot { \phi  }  }{  { \phi  } } +\frac{\ddot{\phi}}{\phi}\Biggr\}.
\label{eq:dgfeuhfg}
\end{align}
From Eqs.~(\ref{eq:dgferedhfg}) and ~(\ref{eq:dgfeuhfg}),
when the curvature effects are negligible, and the Higgs background field 
evolves quickly as $\phi \left(t \right) \approx e^{-{ M  }\left( \phi \right)t}$ or
$\phi \left(t \right) \approx \sin \left({ M  }\left( \phi \right)t\right)$,
the renormalized vacuum fluctuations on the dynamical Higgs field background can be approximated by
\begin{align}
\left<{  \delta \phi   }^{ 2 } \right>_{\rm ren}
\simeq \frac { M^{2}\left(\phi \right) }{ 48{ \pi  }^{ 2 }  } .
\label{eq:dgfskdhfg}
\end{align}
If the Higgs field has the large effective mass ${ M  }\left( \phi \right)$, 
the  Higgs background field develops rapidly on the cosmological timescale
and the vacuum field fluctuations of the Higgs field glow in proportional to the Higgs mass $ { M  }\left( \phi \right)$.

\subsection{The scalar (inflaton) field background}
\label{sec:scalar-background}
When there are other coherent or classical scalar fields $S$ like the inflaton field which 
couple the Higgs field with $\lambda_{\phi S}$,
the effective mass of the Higgs field can be generated as
$m^{2}_{\phi S}=\lambda_{\phi S}S^{2}$. 
The effective Higgs mass becomes ${ \bar{M}  }^{ 2 }\left( \eta \right) =a^2\left( \eta \right)\left(m^{2}+3\lambda \phi^{2}
+\lambda_{\phi S}S^{2}+\left(\xi-1/6\right)R\right)$
and the second-order adiabatic vacuum fluctuations are given as follows:
\begin{align}
&\left<{  \delta \phi   }^{ 2 } \right>_{\rm ren}=\nonumber \\
&\frac { 1 }{ 48{ \pi  }^{ 2 }  } \Biggl\{
\frac { \ddot { a }  }{ a } +\frac { { \dot { a }  }^{ 2 } }{ { a }^{ 2 } }
+\frac { 3 }{ 2 } \frac { \dot { a }  }{ a } \frac { \left( \xi -1/6 \right) \dot { R } 
+6\lambda \phi \dot { \phi  }+2\lambda_{\phi S}S\dot{S}  }{ { m }^{ 2 }+\left( \xi -1/6 \right) R+3\lambda { \phi  }^{ 2 } +\lambda_{\phi S}S^{2}} \nonumber \\
&-\frac { 1 }{ 4 } \frac { { \left( \left( \xi -1/6 \right) \dot { R } +6\lambda \phi \dot { \phi  }+2\lambda_{\phi S}S\dot{S}   \right)  }^{ 2 } }{ { \left( { m }^{ 2 }
+\left( \xi -1/6 \right) R+3\lambda { \phi  }^{ 2 }+\lambda_{\phi S}S^{2} \right) }^{ 2 }}  \nonumber \\
&+\frac { 1 }{ 2 } \frac { \left( \xi -1/6 \right) \ddot { R } +6\lambda \left( \phi \ddot { \phi  } 
+{ \dot { \phi  }  }^{ 2 } \right) +2\lambda_{\phi S} \left( S \ddot { S  }+\dot{S}^2 \right) }{ { m }^{ 2 }+\left( \xi -1/6 \right) R+3\lambda { \phi  }^{ 2 } +\lambda_{\phi S}S^{2}} \Biggr\} .
\label{eq:dgffsfdhfg}
\end{align}
For large background scalar field $S \left(t \right)$, 
the second-order adiabatic vacuum fluctuations are given by
\begin{align}
\left<{  \delta \phi   }^{ 2 } \right>_{\rm ren}
\simeq &\frac { 1 }{ 48{ \pi  }^{ 2 }  } \Biggl\{
\frac{1}{6}R+\frac {3H\dot { S }  }{  { S } } +\frac{\ddot{S}}{S}\Biggr\}.
\label{eq:dgfdsdfdhfg}
\end{align}
The evolution of the background scalar field $S \left(t \right)$ is determined by the effective scalar potential ${ V }_{\rm eff}\left( S  \right)$. 
Thus, the renormalized vacuum fluctuations of the Higgs field on the dynamical background scalar field are given by
\begin{align}
\left<{  \delta \phi   }^{ 2 } \right>_{\rm ren}
\simeq \frac { m^{2}_{S} }{ 48{ \pi  }^{ 2 }  }  ,
\label{eq:dgdfsdhfg}
\end{align}
where $m_{S}$ is defined by ${ V }_{\rm eff}''\left( S  \right)=m_{S}^{2}$.
The vacuum fluctuations of the Higgs field expand in proportion to the curvature scale $R$, 
the mass of the Higgs field $\phi$ or the scalar field $S$ in the FLRW background.

\section{Electroweak vacuum instability in FLRW background}
\label{sec:instability}
So far we have discussed the vacuum field fluctuations of the Higgs field in various situations.
In this section, we investigate the electroweak vacuum instability in the FLRW background
by using the results of Section~\ref{sec:adiabatic}, Section~\ref{sec:non-adiabatic}
and Section~\ref{sec:background}.

The stability of the electroweak vacuum is determined by 
the dynamics of the background Higgs field and 
the vacuum fluctuations of the Higgs field.
As previous discussed in Section~\ref{sec:potential},
the one-loop effective evolution equation of the Higgs field is written as follows:
\begin{align}
\ddot { \phi  } +3H\dot { \phi  } +\frac { \partial { V }_{ \rm eff }\left( \phi  \right)  }{ \partial \phi  } =0,\label{eq:ewhghldg}
\end{align}
where the one-loop standard model effective Higgs potential in curved background can written as~\cite{Shore:1979as,Herranen:2014cua}
\begin{align}
V_{\rm eff}\left( \phi  \right) &=\frac{1}{2}m^{2}(\mu)\phi^{2}+\frac{1}{2}\xi (\mu)R\phi^{2}
+\frac{\lambda(\mu)}{4}\phi^{4}\label{eq:fhlfgkdg} \\ &+\sum _{ i=1 }^{ 9 }{ \frac { { n }_{ i } }{ 64{ \pi  }^{ 2 } } { M }_{ i }^{ 4 }\left( \phi  \right) \left[ \log{ \frac { { M }_{ i }^{ 2 }\left( \phi  \right)  }{ { \mu  }^{ 2 } }  } -{ C }_{ i } \right]  } \nonumber ,\\
{ M }_{ i }^{ 2 }\left( \phi  \right) &={ \kappa  }_{ i }{ \phi  }^{ 2 }+{ \kappa }'_{ i }+\theta_{i} R\label{eq:gkljkkdg},
\end{align}
where the coefficients $n_{i}$, ${ \kappa  }_{ i }$, ${ \kappa }'_{ i }$, $\theta_{i}$ and $C_{i}$
are given by Table I of Ref.\cite{Herranen:2014cua}. 
The effective evolution equation and the one-loop effective potential in curved 
background
has been well-known in the literature~\cite{Shore:1979as,Toms:1982af,Toms:1983qr,Hu:1984js,Buchbinder:1985js,Ringwald:1987ui,
Balakrishnan:1991pm,Muta:1991mw,Kirsten:1993jn,Elizalde:1993ee,Elizalde:1993ew,Elizalde:1994im,
Elizalde:1994ds,Elizalde:1994gv,Gorbar:2002pw,Gorbar:2003yt,Gorbar:2003yp,Maroto:2014oca,Czerwinska:2015xwa,Albareti:2016cvx}

As the previous discussed, however, the additional contribution from the gravitational vacuum fluctuations of the Higgs field 
change the effective evolution equation of the Higgs field as follow:
\begin{align}
\ddot { \phi  } +3H\dot { \phi  } +\frac { \partial { V }_{ \rm eff }\left( \phi  \right)  }{ \partial \phi  } +
3\lambda(\mu) \left<{  \delta \phi   }^{ 2 } \right>_{\rm ren}\phi =0,\label{eq:bdf,fhldg}
\end{align}
where the vacuum fluctuations term provides the effective mass
and this formulation was first discussed by the literature~\cite{Ringwald:1987ui}. This expression can be obtained even by replacing the Higgs field
$\phi^{2}\rightarrow \phi^{2}+\left< {  \delta \phi  }^{ 2 } \right>_{\rm ren} $
so as to include the backreaction terms from the Higgs vacuum 
fluctuations~\cite{Kohri:2016qqv}.
Thus, the standard model Higgs potential 
in curved background should be modified as follows: 
\begin{align}
V_{\rm eff}\left( \phi  \right) &= \frac{1}{2}m^{2}(\mu)\phi^{2}+\frac{1}{2}\xi (\mu)R\phi^{2}
+\frac{3\lambda(\mu)}{2}\left< {\delta \phi  }^{ 2 } \right>_{\rm ren}\phi^{2} \nonumber \\
&+\frac{\lambda(\mu)}{4}\phi^{4} +\sum _{ i=1 }^{ 9 }{ \frac { { n }_{ i } }{ 64{ \pi  }^{ 2 } } 
{ M }_{ i }^{ 4 }\left( \phi  \right) \left[ \log{ \frac { { M }_{ i }^{ 2 }\left( \phi  \right)  }
{ { \mu  }^{ 2 } }  } -{ C }_{ i } \right]  }, \nonumber \\ 
{ \left< { \delta \phi   }^{ 2 } \right>}_{\rm ren}
&=\frac { 1 }{ 4{ \pi  }^{ 2 }C\left( \eta  \right)  }
\int _{ 0 }^{ \infty  }{ dk{ k }^{ 2 }{ \Omega  }_{ k }^{ -1 }\left\{2n_{k}+2{\rm Re}z_{k} \right\}  } \label{eq:gddjkkdg},
\end{align}
which includes the backreaction of the Higgs fluctuation.

Next let us discuss some issues of the renormalization scale $\mu$.
Generally speaking, we take the renormalization scale $\mu$
so as to suppress the high order log-corrections about $\log{ ( { { M }_{i}^{ 2 }\left( \phi  \right)  }/{ { \mu  }^{ 2 } } ) }$.
In Minkowski spacetime as $R=0$, 
we usually take the renormalization scale to be $\mu \approx \phi$.
The renormalization scale $\mu$ corresponds to the phenomenological/cosmological energy scale described as the effective mass of the scalar field.
Although the log-correction in Eq.~(\ref{eq:gkljkkdg}) does not include the vacuum fluctuation terms,
the high-order expressions would have these terms and therefore
the renormalization scale should be taken as 
$\mu^{2} \approx \phi^{2}+\left< {\delta \phi  }^{ 2 } \right>_{\rm ren}+R$.

The running couplings $m^{2}(\mu)$, $\xi (\mu)$ and $\lambda(\mu)$ change depending on the renormalization scale $\mu$.
The running Higgs self-coupling $\lambda(\mu)$ becomes negative at the high-energy scale $\Lambda_{I}$
\footnote{
The instability scale $\Lambda_{I}$ can be approximately determined by
the value of the Higgs boson mass and the top quark mass.
The current measurements of the Higgs boson mass
$m_{h}=125.09 \pm 0.21 ({\rm stat}) \pm 0.11({\rm syst})\ {\rm GeV}$~\cite{Aad:2015zhl,Aad:2013wqa,
Chatrchyan:2013mxa,Giardino:2013bma} and 
the top quark mass $m_{t}=172.44\pm 0.13 ({\rm stat}) \pm 0.47 ({\rm syst})\ {\rm GeV}$~\cite{Khachatryan:2015hba} 
show the instability scale to be  $\Lambda_{I} \approx 10^{11}\ {\rm GeV}$~\cite{Buttazzo:2013uya}
although this instability scale $\Lambda_{I}$ depends on the gauge (see~\cite{DiLuzio:2014bua,Andreassen:2014eha,Andreassen:2014gha,Lalak:2016zlv,
Espinosa:2016uaw,Espinosa:2016nld} for the detail discussions).}.
If the renormalization scale becomes larger than the instability scale to be
$\mu^{2} \simeq R+\left< {\delta \phi  }^{ 2 } \right>_{\rm ren} \gtrsim \Lambda_{I}^{2}$,
the running Higgs self-coupling $\lambda(\mu)$ becomes negative and 
the backreaction term of the Higgs fluctuation 
destabilizes the Higgs potential~\cite{Kohri:2016qqv}.
On the other hand, 
the vacuum fluctuations of the $W/Z$ bosons and the top quark
expressed by
$\left< {  \delta W }^{ 2 } \right>_{\rm ren}$, $\left< {  \delta Z }^{ 2 } \right>_{\rm ren}$ and $\left< {  \delta t }^{ 2 } \right>_{\rm ren}$
can stabilize the effective Higgs potential.
The modified effective Higgs potential 
including the vacuum fluctuation of the various SM fields 
can be written as follows:
\begin{align}
V_{\rm eff}\left( \phi  \right) &=\frac{1}{2}m^{2}( \mu )\phi^{2}
+\frac{1}{2}\xi(\mu )R\phi^{2} 
+\frac{3\lambda(\mu)}{2}\left< {\delta \phi  }^{ 2 } \right>_{\rm ren}\phi^{2} \nonumber \\
&+\frac{\lambda (\mu )}{4}\phi^{4}+\frac{g^{2}(\mu)}{8}\left< {  \delta W }^{ 2 } \right>_{\rm ren}\phi^{2} \label{eq:fhlgkdg} \\  &
+ \frac{\left[ g^{2}(\mu) + g'^{2}(\mu)\right]}{8}
\left< {  \delta Z }^{ 2 } \right>_{\rm ren}\phi^{2} 
 +\frac{y^{2}_{t}(\mu)}{4}\left< {  \delta t }^{ 2 } \right>_{\rm ren}\phi^{2} \nonumber \\
&+\sum _{  i=1 }^{9}{ \frac { { n }_{ i } }{ 64{ \pi  }^{ 2 } } { M }_{ i }^{ 4 }\left( \phi  \right) 
\left[ \log{ \frac { { M }_{ i }^{ 2 }\left( \phi  \right)  }{ \mu^{ 2 } }  } -{ C }_{ i } \right]  } \nonumber,
\end{align}
where the vacuum fluctuations of the Higgs, W/Z bosons and the top quark 
strongly depend on their masses.
Especially, these backreaction effects of the W/Z bosons and top quark
would become also crucial factors of the Higgs vacuum stability 
in the FLRW background. 
In this present paper, however, we focus on only the backreaction of 
the Higgs fluctuation and leaves detailed discussion of the Higgs vacuum stability with the backreaction of the SM particles 
for a forthcoming work.

The Higgs field can cosmologically acquire various effective masses from various couplings.
The non-minimal curvature coupling $\xi \left( \mu  \right) $ provides an extra contribution to the Higgs field mass.
Furthermore, if there are coherent scalar fields 
to couple the Higgs field with $\lambda_{\phi S}$,
the dynamical mass of the Higgs field can be generated by the interaction
$\lambda_{\phi S}S^{2}$ where $S$ is the Higgs-coupled scalar field. 
In this section, let us consider the curvature mass $\xi \left( \mu  \right) R $
and the dynamical mass $\lambda_{\phi S}S^{2}$.
The magnitude relation of the effective mass $m^{2}_{\rm eff} \simeq \xi \left( \mu  \right) R +\lambda_{\phi S}S^{2}$
and the renormalized vacuum fluctuations of the Higgs field 
determine the stability of the effective Higgs potential.
If the effective Higgs potential is destabilized by the vacuum field fluctuations,
the Higgs effective potential becomes negative as ${\partial  V_{\rm eff}\left( \phi  \right)} / { \partial \phi  }\lesssim 0$, and therefore, 
the coherent Higgs field $\phi\left(t\right)$ on the entire Universe 
rolls down to the Planck-scale true vacuum.

For $ \xi (\mu)R \gg \lambda_{\phi S}S^{2}$
the renormalized vacuum fluctuations of the Higgs field are summarize as
\begin{align}
\left< {  \delta \phi  }^{ 2 } \right>_{\rm ren}\begin{cases}\simeq
 R/288\pi^{2} \quad\ \left( \xi(\mu)\gtrsim   \mathcal{O}\left(10^{-1}\right) \right) \\  
\gtrsim \mathcal{O}\left( R \right) \quad\quad\ \ \left(  \xi(\mu) \lesssim   \mathcal{O}\left(10^{-1}\right) \right)
\end{cases} 
\label{eq:sffdusg}
\end{align}@article{Shore:1979as,
      author         = "Shore, Graham M.",
      title          = "{Radiatively Induced Spontaneous Symmetry Breaking and
                        Phase Transitions in Curved Space-Time}",
      journal        = "Annals Phys.",
      volume         = "128",
      year           = "1980",
      pages          = "376",
      doi            = "10.1016/0003-4916(80)90326-7",
      reportNumber   = "HUTP-79-A070",
      SLACcitation   = "
}
For de-Sitter background with $R=12H^{2}$,
the renormalized Higgs fluctuations are given by
\begin{align}
\left< {  \delta \phi  }^{ 2 } \right>_{\rm ren}\simeq\begin{cases}
 H^{2}/24\pi^{2} \quad \quad\quad\  \left( \xi(\mu)\gtrsim   \mathcal{O}\left(10^{-1}\right) \right) \\
{ H }^{ 2} / 32{ \pi  }^{ 2 }\xi(\mu) \quad\ \left(  \xi(\mu) \lesssim   \mathcal{O}\left(10^{-1}\right) \right)
\end{cases} 
\label{eq:kfsdfhdsg}
\end{align}
where the above expressions are valid during the inflation.
However, after inflation, the non-minimal curvature term $\xi\left(\mu\right) R$ can generate 
the enormous Higgs vacuum fluctuations via tachyonic resonance as 
${ \left< { \delta \phi   }^{ 2 } \right>  }_{ \rm ren} \gg  \mathcal{O}\left( R \right)$
where the non-minimal curvature term $\xi\left(\mu\right)$ is relatively large.
If we assume the simple chaotic inflation model, 
we can numerically obtain the constraint of the tachyonic resonance not to generate the large Higgs vacuum fluctuations
as  $\xi(\mu)\lesssim  \mathcal{O}\left(10\right)$ (see Ref.\cite{Herranen:2015ima,
Kohri:2016wof,Ema:2016kpf,Enqvist:2016mqj,Postma:2017hbk,Ema:2017loe} for the detailed discussions).

For $\mu^{2} \simeq R+\left< {\delta \phi  }^{ 2 } \right>_{\rm ren} \gtrsim \Lambda_{I}^{2}$,
the Higgs self-coupling $\lambda\left(\mu\right)$ becomes negative
\footnote{
For $\mu^{2} \simeq R+\left< {\delta \phi  }^{ 2 } \right>_{\rm ren} \lesssim \Lambda_{I}^{2}$,
the running Higgs self-coupling $\lambda\left(\mu\right)$ becomes positive unless $\phi \gtrsim  \Lambda_{I}$. 
Thus, the homogeneous Higgs field $\phi\left(t\right)$ can not classically roll down into the Planck-scale true vacuum. 
However, the large vacuum fluctuations of the Higgs field can generate  AdS domains or bubbles 
as shown in Eq.~(\ref{eq:hswedg}).}
and the destabilization of the effective Higgs potential 
can be determined by the following relation 
$\xi (\mu)R \lesssim  \left| {\lambda(\mu)} \right|\left< {\delta \phi  }^{ 2 } \right>_{\rm ren} $
where we can assume $\lambda(\mu)\simeq -0.01$.
In de-Sitter background
\footnote{
During inflation, the curvature mass $\xi (\mu)R$ stabilize the effective Higgs potential 
and suppress  AdS domains/bubbles. 
Thus, the electroweak vacuum decay can be avoided 
if the relatively large non-minimal curvature coupling $\xi (\mu)$ is introduced.},
we obtain the condition of the non-minimal coupling to be ${ \xi (\mu) }\lesssim \mathcal{O}\left(10^{-3}\right)$
not to destabilize the effective Higgs potential~\cite{Kohri:2016qqv}.
In the radiation/matter dominated Universe, we can expect the same constraint of the non-minimal coupling.
If ${ \xi (\mu) }$ does not satisfy this condition, the effective Higgs potential $V_{\rm eff}\left( \phi  \right)$ is destabilized,
the coherent Higgs field goes out to the negative Planck vacuum and
leads to the collapse of the Universe.

For $ \xi (\mu)R \ll \lambda_{\phi S}S^{2}$,
the renormalized vacuum fluctuations of the Higgs field are given by
\begin{align}
\left< {  \delta \phi  }^{ 2 } \right>_{\rm ren}\simeq\begin{cases}
 M^{2}\left(\phi\right)/48\pi^{2} \quad\quad  \left( \lambda_{\phi S}S^{2} \lesssim  \lambda\phi^{2}  \right) \\
{ m }^{ 2}_{S} / 48{ \pi  }^{ 2 }  \quad\quad\quad\ \  \left(  \lambda_{\phi S}S^{2} \gtrsim   \lambda\phi^{2}  \right)
\end{cases} 
\label{eq:ssddsdsg}
\end{align}
where the above expressions are valid for the slowly varying scalar field.
In the rapid varying case, the Higgs fluctuations
become generally larger than the above expressions.
As well-known facts, in the parametric/tachyonic resonance
during preheating stage, 
the vacuum fluctuations exponentially grow 
where a complicated numerical analysis is required.
If we assume the simple $m^{2}_{S}S^{2}$ chaotic inflation model,
we can numerically obtain the restriction of the parametric resonance
not to generate the large Higgs fluctuations
as $\lambda_{\phi S}\lesssim   \mathcal{O}\left(10^{-8}\right)$ (see Ref.\cite{Herranen:2015ima,
Kohri:2016wof,Ema:2016kpf,Enqvist:2016mqj,Postma:2017hbk,Ema:2017loe} 
for the detail).

For $\mu^{2} \simeq R+\left< {\delta \phi  }^{ 2 } \right>_{\rm ren}\gtrsim \Lambda_{I}^{2}$,
the effective Higgs potential is destabilized in $\lambda_{\phi S}S^{2}\lesssim  \left| {\lambda(\mu)} \right|\left< {\delta \phi  }^{ 2 } \right>_{\rm ren} $.
Considering $m^{2}_{S}S^{2}$ chaotic inflation where 
the inflaton field $S$ has the Planck-field value $S \approx M_{\rm Pl}\approx 10^{19}\ {\rm GeV}$,
the stabilization condition during inflation is $\lambda_{\phi S}\gtrsim \mathcal{O}\left(10^{-13}\right)$
\footnote{
For $m^{2}_{S}S^{2}$ chaotic inflation where $m_{S}\approx H\approx 10^{14}\ {\rm GeV}$, 
the renormalized vacuum fluctuations of the Higgs field are written as 
\begin{align*}
\left< {  \delta \phi  }^{ 2 } \right>_{\rm ren}\simeq \frac{3H^4}{8\pi^2\lambda_{\phi S}S^{2}}+\frac{m^{2}_{S}}{48\pi^2},
\end{align*}
where we ignore the curvature mass term $\xi (\mu)12H^{2}$.}.
Thus, the inflaton-Higgs coupling $\lambda_{\phi S}$ can 
stabilize the Higgs potential during inflation.
After inflation, however, the parametric/tachyonic resonance 
via the coherent oscillation of $S$
can generate the enormous Higgs fluctuations with the relatively large coupling $\lambda_{\phi S}$.
Moreover, if the inflaton/other scalar field $S$ satisfy the following relations
$\lambda(\mu) \left< {  \delta \phi  }^{ 2 } \right>_{\rm ren}\simeq \lambda(\mu) m_{S}^{2}/48\pi^{2} \gtrsim  \Lambda_{I}^2$ 
and $\lambda(\mu) m_{S}^{2}/48\pi^{2} \gtrsim   \lambda_{\phi S}S^{2}$, 
the Higgs fluctuations destabilize the effective Higgs potential.
This situation could easily happen after inflation.
If we take $\Lambda_{I} \approx 10^{11}\ {\rm GeV}$ and $\lambda(\mu) \simeq -0.01$,
we obtain new constraint of the mass of the inflaton/scalar field to be $m_{S}  \lesssim   10^{13}\ {\rm GeV}$.

On the other hand, the vacuum fluctuations of the Higgs field expressed as $\left< {  \delta \phi  }^{ 2 } \right>_{\rm ren}$
can cause directly the vacuum transition to the true vacuum~\cite{Espinosa:2007qp,Fairbairn:2014zia,Lebedev:2012sy,Kobakhidze:2013tn,Enqvist:2013kaa,Herranen:2014cua,Kobakhidze:2014xda,Kohri:2016wof,Kamada:2014ufa,Enqvist:2014bua,Hook:2014uia,Kearney:2015vba,Espinosa:2015qea}. 
This situation is essentially different from the phenomenon discussed previously. 
If the inhomogeneous Higgs fields overcome the hill of the effective potential, 
the localized Higgs fields classically go out to the true vacuum 
and catastrophic Anti-de Sitter (AdS) domains are formed.
Although not all Higgs AdS domains threaten the existence of the Universe~\cite{Hook:2014uia, Espinosa:2015qea}, 
which highly depends on the evolution of the Higgs AdS domains 
(for the details see Ref.\cite{Espinosa:2015qea,Tetradis:2016vqb}),
some AdS domains expand eating other regions of the electroweak vacuum, and consume the entire Universe.
Thus, the existence of AdS domains in the Universe is still serious and 
the creation of the Higgs AdS domains/bubbles should not happen in our Universe.

Let us consider the conditions not to generate the AdS domains/bubbles.
The probability of the Higgs fluctuations 
can be expressed as the Gaussian distribution function
\begin{align}
P\left( \phi  \right) 
= \frac { 1 }{ \sqrt {2{ \pi  }\left< {  \delta \phi  }^{ 2 } \right>_{\rm ren} } } 
\exp \left( -\frac {{  \phi  }^{ 2 } }{ 2\left< {  \delta \phi  }^{ 2 } \right>_{\rm ren} }  \right)\label{eq:hqqqqdg}.
\end{align}
By using Eq.~(\ref{eq:hqqqqdg}), 
the probability not to produce  AdS domains/bubbles is given by
\begin{align}
{ P }\left(  \phi<{ \phi }_{ \rm max }\right) &\equiv \int _{ -{ \phi  }_{\rm  max } }^{ { \phi }_{ \rm max } }
{ P\left( \phi \right) d\phi } \\ &= {\rm erf}\left( \frac{{ \phi }_{ \rm max }  }
{ \sqrt { 2\left< {  \delta \phi  }^{ 2 } \right>_{\rm ren}  }}  \right)\label{eq:hswegg}.
\end{align}
where we define ${ \phi }_{ \rm max }$ to be the effective Higgs potential of Eq.~(\ref{eq:gkljkkdg}) takes its maximal value
\footnote{ The effective Higgs potential with the large effective mass $m_{\rm eff}$
can be approximated as
\begin{align*}
V_{\rm eff }\left(  \phi \right) \simeq \frac { 1 }{ 2 }m^{2}_{\rm eff}{ \phi }^{ 2 }
\left( 1-\frac { 1 }{ 2 } { \left( \frac {  \phi }{ { \phi}_{\rm  max } }  \right)  }^{ 2 } \right), 
\end{align*}
where ${  \phi }_{\rm  max }$ is approximately given by
\begin{align*}
{\phi }_{ \rm max }=\sqrt { -\frac { m^{2}_{\rm eff}}{ { \lambda }\left( \mu  \right)} }.
\end{align*}
In numerical approximation, we can approximate the maximal field value as
$\phi_{\rm max}\simeq 10\cdot m_{\rm eff}$ for the effective Higgs potential.}.
Thus, the probability that the localized Higgs fields roll down into the true vacuum is given by
\begin{align}
{ P }\left(  \phi>{ \phi }_{ \rm max }\right)&=1- {\rm erf}\left( \frac{{ \phi }_{ \rm max }  }
{ \sqrt { 2\left< {  \delta \phi  }^{ 2 } \right>_{\rm ren}  }}  \right)\label{eq:afklsjiijgg} \\ &\simeq 
\frac {\sqrt{2\left< {  \delta \phi  }^{ 2 } \right>_{\rm ren}  }}{\pi{ \phi }_{ \rm max }}
\exp \left( -\frac {{  \phi  }^{ 2 } }{ 2\left< {  \delta \phi  }^{ 2 } \right>_{\rm ren} }  \right) \nonumber.
\end{align}
The vacuum decay probability of the inflationary Universe can be expressed as
\begin{align}
{ e }^{ 3{ N }_{ \rm hor } }{ P }\left(  \phi>{ \phi }_{ \rm max }\right)<1,\label{eq:aaawegg}
\end{align}
where ${ e }^{ 3{ N }_{ \rm hor } }$ corresponds to the physical volume of the Universe at the
end of the inflation, and we can take the e-folding number $N_{\rm hor }\simeq N_{\rm CMB }\simeq60$.
By substituting Eq.~(\ref{eq:aaawegg}) into Eq.~(\ref{eq:afklsjiijgg}),
we obtain the following relation of the electroweak vacuum stability 
\begin{align}
\frac{\left< {  \delta \phi  }^{ 2 } \right>_{\rm ren}}{{ \phi }_{ \rm max }^{2}}<\frac{1}{6N_{\rm hor} }\label{eq:hswedg}.
\end{align}
The above condition can be determined by the effective Higgs potential of Eq.~(\ref{eq:gkljkkdg}) 
and the Higgs vacuum fluctuations of Eqs.~(\ref{eq:sffdusg}),
~(\ref{eq:kfsdfhdsg}) and ~(\ref{eq:ssddsdsg}).
The inflationary Universe restricts ${ \xi (\mu) }\gtrsim O\left(10^{-2}\right)$ 
or $\lambda_{\phi S} \gtrsim O\left(10^{-12}\right)$ not to generate 
the AdS domains/bubbles.
These obtained constraints are somewhat tighter 
than the destabilization conditions of the effective Higgs potential.
If the relatively large coupling 
${ \xi }$ or $\lambda_{\phi S}$ are introduced,
the false Higgs vacuum can be safe during inflation.
But after inflation the large coupling ${ \xi }$ or $\lambda_{\phi S}$ 
generate large Higgs fluctuations via the parametric/tachyonic resonance.

After all the Higgs fluctuations in the non-adiabatic case as discussed in Section~\ref{sec:non-adiabatic} 
generally destabilize the false electroweak vacuum.
On the other hand, the Higgs fluctuations in the adiabatic case as discussed in Section~\ref{sec:adiabatic} have little effect on
the electroweak vacuum stability.
However, if there are large inflaton field 
or some scalar fields $S$ satisfying both relations
$\lambda (\mu) m_{S}^{2}/48\pi^{2} \gtrsim  \Lambda_{I}^2$ and $\lambda(\mu) m_{S}^{2}/48\pi^{2} \gtrsim   \lambda_{\phi S}S^{2}$,
the Higgs fluctuations destabilize the effective Higgs potential or generate the AdS domains/bubbles.
The cosmological stability of the electroweak vacuum 
highly unstable due to vacuum fluctuations of the Higgs field
and imposes severe cosmological constraints.

\section{Conclusion and summary} 
\label{sec:conclusion}
In this paper, we have thoroughly
investigated the stability of the electroweak vacuum in 
the FLRW background.
Adopting the adiabatic (WKB) approximation or adiabatic
regularization methods, 
we have clearly shown that the Higgs vacuum fluctuations
depend on the curvature scale and also the masses of the Higgs field 
or other scalar field.
Next, we have discussed 
the renormalization issues of the vacuum field 
fluctuations and shown that the standard effective potential 
is modified by the gravitational backreaction effects.
Furthermore in Section~\ref{sec:instability} we have shown how the vacuum fluctuations of the Higgs field
influence the stability of the electroweak vacuum in a rigid manner of the QFT in curved spacetime.
The Higgs fluctuations in the non-adiabatic case as discussed in Section~\ref{sec:non-adiabatic} 
generally destabilize the effective Higgs potential, or generate the Higgs AdS domains or bubbles.
On the other hand, the Higgs fluctuations in the adiabatic case as discussed in Section~\ref{sec:adiabatic}
does not generally cause the collapse of the Higgs vacuum.
However, if there are large background scalar fields as discussed in Section~\ref{sec:background}, the vacuum fluctuations of the Higgs field 
can destabilize the effective Higgs potential and give the upper bound 
on the scalar mass to be $m_{S} \lesssim  10^{13}\ {\rm GeV}$.
We have provided new cosmological constraints and comprehensive description about the Higgs vacuum stability in the FLRW background

\acknowledgments
We would like to thank Satoshi Iso,
Holger B. Nielsen, Mihoko Nojiri and 
Kin-ya Oda for helpful comments and discussions.
This work is supported in part by MEXT KAKENHI 
Nos.15H05889, JP16H00877 and JP18H04594
(K.K.), and JSPS KAKENHI No. 26247042  and JP1701131 (K.K.).

\appendix*
\section{Adiabatic (WKB) Approximation Method}
In this appendix we introduce
a detailed description of the adiabatic (WKB) approximation method 
following literature~\cite{Ringwald:1987ui,Parker:1974qw}. In order to give the 
renormalized vacuum field fluctuations we must solve Eq.~(\ref{eq:fufudsdhg}) with the suitable in-vacuum as follows,
\begin{align}
{ n  }_{ k }'=\frac{{\Omega  }_{ k }'}{{\Omega  }_{ k }}{\rm Re}z_{k}, \ \
{ z  }_{ k }'=\frac{{\Omega  }_{ k }'}{{\Omega  }_{ k }}\left(n_{k}+\frac{1}{2}\right)-2i{\Omega  }_{ k }z_{k}.\label{eq:dsdgsdhg}
\end{align}
For simplicity we assume ${ z  }_{ k }={ u  }_{ k }+i{ v  }_{ k }$, i.e ${ u  }_{ k }={\rm Re}z_{k}$ and ${ v  }_{ k }={\rm Im}z_{k}$.
By using these relations we can rewrite Eq.~(\ref{eq:dsdgsdhg}) as folllows
\begin{align}
{ n  }_{ k }'&=\frac{{\Omega  }_{ k }'}{{\Omega  }_{ k }}{ u  }_{ k }, \label{eq:dehrhsdhg} \\
{ u  }_{ k }'&=\frac{{\Omega  }_{ k }'}{{\Omega  }_{ k }}\left(n_{k}+\frac{1}{2}\right) + 2{\Omega  }_{ k }{ v  }_{ k }, \label{eq:weygsdhg} \\
{ v  }_{ k }'&=-2{\Omega  }_{ k }{ u  }_{ k } \label{eq:dfkjfrsdhg}.
\end{align}
Here, we introduce a single formal adiabatic parameter $T$ and 
a rescaling time variable $\tau \equiv \eta /T $.
The adiabatic (WKB) condition of Eq.~(\ref{eq:adgkdhjhf}) can be restated by
\begin{align}
\frac { d }{ d\eta  } \Omega \left( \eta /T \right) =\frac { 1 }{ T } \frac { d }{ d\tau  } \Omega \left( \tau  \right),  \label{eq:ddssjdhg}
\end{align}
with $T\rightarrow \infty$. By using this procedure we can rewrite Eqs.~(\ref{eq:dehrhsdhg}),
~(\ref{eq:weygsdhg}) and ~(\ref{eq:dfkjfrsdhg}) as follows,
\begin{align}
\frac { 1 }{ T } { n  }_{ k }'&=\frac { 1 }{ T } \frac{{\Omega  }_{ k }'}{{\Omega  }_{ k }}{ u  }_{ k }, \label{eq:defjkjkhg} \\
\frac { 1 }{ T } { u  }_{ k }'&=\frac { 1 }{ T } \frac{{\Omega  }_{ k }'}{{\Omega  }_{ k }}
\left(n_{k}+\frac{1}{2}\right) + 2{\Omega  }_{ k }{ v  }_{ k }, \label{eq:wdhjsdsdhg} \\
\frac { 1 }{ T } { v  }_{ k }'&=-2{\Omega  }_{ k }{ u  }_{ k } \label{eq:sdhjhdsdhg}.
\end{align}
Next we expand ${ n  }_{ k }$, ${ u  }_{ k }$ and ${ v  }_{ k }$ in inverse powers of $T$ as 
\begin{align}
{ n  }_{ k }&={ n  }_{ k }^{(0)}+\frac { 1 }{ T }{ n  }_{ k }^{(1)}+\frac { 1 }{ T^{2} }{ n  }_{ k }^{(2)}+\cdots, \label{eq:fsdhfdjkhg} \\
{ u  }_{ k }&={ u  }_{ k }^{(0)}+\frac { 1 }{ T }{ u  }_{ k }^{(1)}+\frac { 1 }{ T^{2} }{ u  }_{ k }^{(2)}+\cdots, \label{eq:wfsfjdhdhg} \\
{ v  }_{ k }&={ v  }_{ k }^{(0)}+\frac { 1 }{ T }{ v  }_{ k }^{(1)}+\frac { 1 }{ T^{2} }{ v  }_{ k }^{(2)}+\cdots\label{eq:fdfjddsdhg},
\end{align}
where superscripts $(i)$ express the adiabatic order, and the zeroth order expressions are given by 
\begin{align}
{ n  }_{ k }^{(0)}={\rm const},\quad { u  }_{ k }^{(0)}=0,\quad { v  }_{ k }^{(0)}=0,
\end{align}
where we solve Eqs.~(\ref{eq:defjkjkhg}), ~(\ref{eq:wdhjsdsdhg}) 
and ~(\ref{eq:sdhjhdsdhg}) with an iterative procedure.
The above integration constant can be determined by the initial conditions for ${ n  }_{ k }\left(\eta_{0}\right)$,
and ${ z  }_{ k }\left(\eta_{0}\right)$, which 
corresponds  to the choice of the in-vacuum.
For the conformal vacuum ${ n  }_{ k }\left(\eta_{0}\right)={ z  }_{ k }\left(\eta_{0}\right)=0$,
the zeroth-order adiabatic number density ${ n  }_{ k }^{(0)}$ is zero.
For the first adiabatic order, we can obtain the following expression
\begin{align}
{ n  }_{ k }^{(1)}=0,\quad { u  }_{ k }^{(1)}=0,\quad { v  }_{ k }^{(1)}=-\frac { 1 }{ 2 } \frac{{\Omega  }_{ k }'}{{\Omega  }_{ k }^{2}}
\left(n_{k}^{(0)}+\frac{1}{2}\right) ,
\end{align}
where the odd-order adiabatic number density is zero. 
Next we can obtain the second order adiabatic expressions as follows 
\begin{align}
{ n }_{ k }^{ (2) }=\frac{1}{16}\frac{{ \Omega ' }_{ k }^{ 2 }}{{ \Omega  }_{ k }^{ 4}},\quad 
{ u }_{k}^{(2)}=\frac{1}{8}\frac{{ \Omega ''  }_{ k }}{{ \Omega  }_{ k }^{ 3}}
-\frac{1}{4}\frac{{ \Omega ' }_{ k }^{ 2 }}{{ \Omega  }_{ k }^{ 4}},\quad { v  }_{ k }^{(2)}=0,
\end{align}
In the same way, the third order adiabatic expressions can be given by
\begin{align}
{ n  }_{ k }^{(3)}=&0,\quad { u  }_{ k }^{(3)}=0,\\ 
{ v  }_{ k }^{(3)}=&\frac{1}{16{\Omega  }_{ k }^{4}}
\left({ \Omega'''  }_{ k }-7\frac{{ \Omega' }_{ k }{ \Omega''}_{ k }}{{ \Omega }_{ k }}
+\frac{15}{2}\frac{{ \Omega' }_{ k }^{3}}{{ \Omega }_{ k }^{2}}\right) ,
\end{align}
Finally, the forth order adiabatic expressions are given by
\begin{align}
{ n }_{ k }^{ (4) }=&
-\frac{{ \Omega ' }_{ k }{ \Omega '''}_{ k }}{32{ \Omega  }_{ k }^{6}}
+\frac{{ \Omega '' }_{ k }^{2}}{64{ \Omega  }_{ k }^{6}}
+\frac{5{ \Omega ' }_{ k }^{2}{ \Omega ''}_{ k }}{32{ \Omega  }_{ k }^{7}}
-\frac{45{ \Omega ' }_{ k }^{4}}{256{ \Omega  }_{ k }^{8}},\\ 
{ u }_{k}^{(4)}=&
-\frac{{ \Omega ''''}_{ k }}{32{ \Omega  }_{ k }^{5}}
+\frac{11{ \Omega ' }_{ k }{ \Omega ''' }_{ k }  }{32{ \Omega  }_{ k }^{6}}
-\frac{115{ \Omega ' }_{ k }^{2}{ \Omega ''}_{ k }}{64{ \Omega  }_{ k }^{7}}\nonumber  \\& 
+\frac{7{ \Omega '' }_{ k }^{2}}{32{ \Omega  }_{ k }^{6}}
+\frac{45{ \Omega ' }_{ k }^{4}}{32{ \Omega  }_{ k }^{8}}, \\
{ v }_{k}^{(4)}=& 0.
\end{align}

\bibliography{electroweak}

\providecommand{\noopsort}[1]{}\providecommand{\singleletter}[1]{#1}%
\begin{thebibliography}{86}%
\makeatletter
\providecommand \@ifxundefined [1]{%
 \@ifx{#1\undefined}
}%
\providecommand \@ifnum [1]{%
 \ifnum #1\expandafter \@firstoftwo
 \else \expandafter \@secondoftwo
 \fi
}%
\providecommand \@ifx [1]{%
 \ifx #1\expandafter \@firstoftwo
 \else \expandafter \@secondoftwo
 \fi
}%
\providecommand \natexlab [1]{#1}%
\providecommand \enquote  [1]{``#1''}%
\providecommand \bibnamefont  [1]{#1}%
\providecommand \bibfnamefont [1]{#1}%
\providecommand \citenamefont [1]{#1}%
\providecommand \href@noop [0]{\@secondoftwo}%
\providecommand \href [0]{\begingroup \@sanitize@url \@href}%
\providecommand \@href[1]{\@@startlink{#1}\@@href}%
\providecommand \@@href[1]{\endgroup#1\@@endlink}%
\providecommand \@sanitize@url [0]{\catcode `\\12\catcode `\$12\catcode
  `\&12\catcode `\#12\catcode `\^12\catcode `\_12\catcode `\%12\relax}%
\providecommand \@@startlink[1]{}%
\providecommand \@@endlink[0]{}%
\providecommand \url  [0]{\begingroup\@sanitize@url \@url }%
\providecommand \@url [1]{\endgroup\@href {#1}{\urlprefix }}%
\providecommand \urlprefix  [0]{URL }%
\providecommand \Eprint [0]{\href }%
\providecommand \doibase [0]{http://dx.doi.org/}%
\providecommand \selectlanguage [0]{\@gobble}%
\providecommand \bibinfo  [0]{\@secondoftwo}%
\providecommand \bibfield  [0]{\@secondoftwo}%
\providecommand \translation [1]{[#1]}%
\providecommand \BibitemOpen [0]{}%
\providecommand \bibitemStop [0]{}%
\providecommand \bibitemNoStop [0]{.\EOS\space}%
\providecommand \EOS [0]{\spacefactor3000\relax}%
\providecommand \BibitemShut  [1]{\csname bibitem#1\endcsname}%
\let\auto@bib@innerbib\@empty
\bibitem [{\citenamefont {Aad}\ \emph {et~al.}(2015)\citenamefont {Aad} \emph
  {et~al.}}]{Aad:2015zhl}%
  \BibitemOpen
  \bibfield  {author} {\bibinfo {author} {\bibfnamefont {G.}~\bibnamefont
  {Aad}} \emph {et~al.} (\bibinfo {collaboration} {ATLAS, CMS}),\ }\bibfield
  {booktitle} {\emph {\bibinfo {booktitle} {{Proceedings, Meeting of the APS
  Division of Particles and Fields (DPF 2015)}}},\ }\href {\doibase
  10.1103/PhysRevLett.114.191803} {\bibfield  {journal} {\bibinfo  {journal}
  {Phys. Rev. Lett.}\ }\textbf {\bibinfo {volume} {114}},\ \bibinfo {pages}
  {191803} (\bibinfo {year} {2015})},\ \Eprint
  {http://arxiv.org/abs/1503.07589} {arXiv:1503.07589 [hep-ex]} \BibitemShut
  {NoStop}%
\bibitem [{\citenamefont {Aad}\ \emph {et~al.}(2013)\citenamefont {Aad} \emph
  {et~al.}}]{Aad:2013wqa}%
  \BibitemOpen
  \bibfield  {author} {\bibinfo {author} {\bibfnamefont {G.}~\bibnamefont
  {Aad}} \emph {et~al.} (\bibinfo {collaboration} {ATLAS}),\ }\href {\doibase
  10.1016/j.physletb.2014.05.011, 10.1016/j.physletb.2013.08.010} {\bibfield
  {journal} {\bibinfo  {journal} {Phys.Lett.}\ }\textbf {\bibinfo {volume}
  {B726}},\ \bibinfo {pages} {88} (\bibinfo {year} {2013})},\ \Eprint
  {http://arxiv.org/abs/1307.1427} {arXiv:1307.1427 [hep-ex]} \BibitemShut
  {NoStop}%
\bibitem [{\citenamefont {Chatrchyan}\ \emph {et~al.}(2014)\citenamefont
  {Chatrchyan} \emph {et~al.}}]{Chatrchyan:2013mxa}%
  \BibitemOpen
  \bibfield  {author} {\bibinfo {author} {\bibfnamefont {S.}~\bibnamefont
  {Chatrchyan}} \emph {et~al.} (\bibinfo {collaboration} {CMS}),\ }\href
  {\doibase 10.1103/PhysRevD.89.092007} {\bibfield  {journal} {\bibinfo
  {journal} {Phys.Rev.}\ }\textbf {\bibinfo {volume} {D89}},\ \bibinfo {pages}
  {092007} (\bibinfo {year} {2014})},\ \Eprint {http://arxiv.org/abs/1312.5353}
  {arXiv:1312.5353 [hep-ex]} \BibitemShut {NoStop}%
\bibitem [{\citenamefont {Giardino}\ \emph {et~al.}(2014)\citenamefont
  {Giardino}, \citenamefont {Kannike}, \citenamefont {Masina}, \citenamefont
  {Raidal},\ and\ \citenamefont {Strumia}}]{Giardino:2013bma}%
  \BibitemOpen
  \bibfield  {author} {\bibinfo {author} {\bibfnamefont {P.~P.}\ \bibnamefont
  {Giardino}}, \bibinfo {author} {\bibfnamefont {K.}~\bibnamefont {Kannike}},
  \bibinfo {author} {\bibfnamefont {I.}~\bibnamefont {Masina}}, \bibinfo
  {author} {\bibfnamefont {M.}~\bibnamefont {Raidal}}, \ and\ \bibinfo {author}
  {\bibfnamefont {A.}~\bibnamefont {Strumia}},\ }\href {\doibase
  10.1007/JHEP05(2014)046} {\bibfield  {journal} {\bibinfo  {journal} {JHEP}\
  }\textbf {\bibinfo {volume} {1405}},\ \bibinfo {pages} {046} (\bibinfo {year}
  {2014})},\ \Eprint {http://arxiv.org/abs/1303.3570} {arXiv:1303.3570
  [hep-ph]} \BibitemShut {NoStop}%
\bibitem [{\citenamefont {Khachatryan}\ \emph {et~al.}(2016)\citenamefont
  {Khachatryan} \emph {et~al.}}]{Khachatryan:2015hba}%
  \BibitemOpen
  \bibfield  {author} {\bibinfo {author} {\bibfnamefont {V.}~\bibnamefont
  {Khachatryan}} \emph {et~al.} (\bibinfo {collaboration} {CMS}),\ }\href
  {\doibase 10.1103/PhysRevD.93.072004} {\bibfield  {journal} {\bibinfo
  {journal} {Phys. Rev.}\ }\textbf {\bibinfo {volume} {D93}},\ \bibinfo {pages}
  {072004} (\bibinfo {year} {2016})},\ \Eprint
  {http://arxiv.org/abs/1509.04044} {arXiv:1509.04044 [hep-ex]} \BibitemShut
  {NoStop}%
\bibitem [{\citenamefont {Kobzarev}\ \emph {et~al.}(1975)\citenamefont
  {Kobzarev}, \citenamefont {Okun},\ and\ \citenamefont
  {Voloshin}}]{Kobzarev:1974cp}%
  \BibitemOpen
  \bibfield  {author} {\bibinfo {author} {\bibfnamefont {I.~{\relax Yu}.}\
  \bibnamefont {Kobzarev}}, \bibinfo {author} {\bibfnamefont {L.~B.}\
  \bibnamefont {Okun}}, \ and\ \bibinfo {author} {\bibfnamefont {M.~B.}\
  \bibnamefont {Voloshin}},\ }\href@noop {} {\bibfield  {journal} {\bibinfo
  {journal} {Sov. J. Nucl. Phys.}\ }\textbf {\bibinfo {volume} {20}},\ \bibinfo
  {pages} {644} (\bibinfo {year} {1975})},\ \bibinfo {note} {[Yad.
  Fiz.20,1229(1974)]}\BibitemShut {NoStop}%
\bibitem [{\citenamefont {Coleman}(1977)}]{Coleman:1977py}%
  \BibitemOpen
  \bibfield  {author} {\bibinfo {author} {\bibfnamefont {S.~R.}\ \bibnamefont
  {Coleman}},\ }\href {\doibase 10.1103/PhysRevD.15.2929,
  10.1103/PhysRevD.16.1248} {\bibfield  {journal} {\bibinfo  {journal} {Phys.
  Rev.}\ }\textbf {\bibinfo {volume} {D15}},\ \bibinfo {pages} {2929} (\bibinfo
  {year} {1977})},\ \bibinfo {note} {[Erratum: Phys.
  Rev.D16,1248(1977)]}\BibitemShut {NoStop}%
\bibitem [{\citenamefont {Callan}\ and\ \citenamefont
  {Coleman}(1977)}]{Callan:1977pt}%
  \BibitemOpen
  \bibfield  {author} {\bibinfo {author} {\bibfnamefont {C.~G.}\ \bibnamefont
  {Callan}, \bibfnamefont {Jr.}}\ and\ \bibinfo {author} {\bibfnamefont
  {S.~R.}\ \bibnamefont {Coleman}},\ }\href {\doibase 10.1103/PhysRevD.16.1762}
  {\bibfield  {journal} {\bibinfo  {journal} {Phys. Rev.}\ }\textbf {\bibinfo
  {volume} {D16}},\ \bibinfo {pages} {1762} (\bibinfo {year}
  {1977})}\BibitemShut {NoStop}%
\bibitem [{\citenamefont {Degrassi}\ \emph {et~al.}(2012)\citenamefont
  {Degrassi}, \citenamefont {Di~Vita}, \citenamefont {Elias-Miro},
  \citenamefont {Espinosa}, \citenamefont {Giudice} \emph
  {et~al.}}]{Degrassi:2012ry}%
  \BibitemOpen
  \bibfield  {author} {\bibinfo {author} {\bibfnamefont {G.}~\bibnamefont
  {Degrassi}}, \bibinfo {author} {\bibfnamefont {S.}~\bibnamefont {Di~Vita}},
  \bibinfo {author} {\bibfnamefont {J.}~\bibnamefont {Elias-Miro}}, \bibinfo
  {author} {\bibfnamefont {J.~R.}\ \bibnamefont {Espinosa}}, \bibinfo {author}
  {\bibfnamefont {G.~F.}\ \bibnamefont {Giudice}},  \emph {et~al.},\ }\href
  {\doibase 10.1007/JHEP08(2012)098} {\bibfield  {journal} {\bibinfo  {journal}
  {JHEP}\ }\textbf {\bibinfo {volume} {1208}},\ \bibinfo {pages} {098}
  (\bibinfo {year} {2012})},\ \Eprint {http://arxiv.org/abs/1205.6497}
  {arXiv:1205.6497 [hep-ph]} \BibitemShut {NoStop}%
\bibitem [{\citenamefont {Isidori}\ \emph {et~al.}(2001)\citenamefont
  {Isidori}, \citenamefont {Ridolfi},\ and\ \citenamefont
  {Strumia}}]{Isidori:2001bm}%
  \BibitemOpen
  \bibfield  {author} {\bibinfo {author} {\bibfnamefont {G.}~\bibnamefont
  {Isidori}}, \bibinfo {author} {\bibfnamefont {G.}~\bibnamefont {Ridolfi}}, \
  and\ \bibinfo {author} {\bibfnamefont {A.}~\bibnamefont {Strumia}},\ }\href
  {\doibase 10.1016/S0550-3213(01)00302-9} {\bibfield  {journal} {\bibinfo
  {journal} {Nucl.Phys.}\ }\textbf {\bibinfo {volume} {B609}},\ \bibinfo
  {pages} {387} (\bibinfo {year} {2001})},\ \Eprint
  {http://arxiv.org/abs/hep-ph/0104016} {arXiv:hep-ph/0104016 [hep-ph]}
  \BibitemShut {NoStop}%
\bibitem [{\citenamefont {Ellis}\ \emph {et~al.}(2009)\citenamefont {Ellis},
  \citenamefont {Espinosa}, \citenamefont {Giudice}, \citenamefont {Hoecker},\
  and\ \citenamefont {Riotto}}]{Ellis:2009tp}%
  \BibitemOpen
  \bibfield  {author} {\bibinfo {author} {\bibfnamefont {J.}~\bibnamefont
  {Ellis}}, \bibinfo {author} {\bibfnamefont {J.}~\bibnamefont {Espinosa}},
  \bibinfo {author} {\bibfnamefont {G.}~\bibnamefont {Giudice}}, \bibinfo
  {author} {\bibfnamefont {A.}~\bibnamefont {Hoecker}}, \ and\ \bibinfo
  {author} {\bibfnamefont {A.}~\bibnamefont {Riotto}},\ }\href {\doibase
  10.1016/j.physletb.2009.07.054} {\bibfield  {journal} {\bibinfo  {journal}
  {Phys.Lett.}\ }\textbf {\bibinfo {volume} {B679}},\ \bibinfo {pages} {369}
  (\bibinfo {year} {2009})},\ \Eprint {http://arxiv.org/abs/0906.0954}
  {arXiv:0906.0954 [hep-ph]} \BibitemShut {NoStop}%
\bibitem [{\citenamefont {Elias-Miro}\ \emph {et~al.}(2012)\citenamefont
  {Elias-Miro}, \citenamefont {Espinosa}, \citenamefont {Giudice},
  \citenamefont {Isidori}, \citenamefont {Riotto} \emph
  {et~al.}}]{EliasMiro:2011aa}%
  \BibitemOpen
  \bibfield  {author} {\bibinfo {author} {\bibfnamefont {J.}~\bibnamefont
  {Elias-Miro}}, \bibinfo {author} {\bibfnamefont {J.~R.}\ \bibnamefont
  {Espinosa}}, \bibinfo {author} {\bibfnamefont {G.~F.}\ \bibnamefont
  {Giudice}}, \bibinfo {author} {\bibfnamefont {G.}~\bibnamefont {Isidori}},
  \bibinfo {author} {\bibfnamefont {A.}~\bibnamefont {Riotto}},  \emph
  {et~al.},\ }\href {\doibase 10.1016/j.physletb.2012.02.013} {\bibfield
  {journal} {\bibinfo  {journal} {Phys.Lett.}\ }\textbf {\bibinfo {volume}
  {B709}},\ \bibinfo {pages} {222} (\bibinfo {year} {2012})},\ \Eprint
  {http://arxiv.org/abs/1112.3022} {arXiv:1112.3022 [hep-ph]} \BibitemShut
  {NoStop}%
\bibitem [{\citenamefont {Espinosa}\ \emph {et~al.}(2008)\citenamefont
  {Espinosa}, \citenamefont {Giudice},\ and\ \citenamefont
  {Riotto}}]{Espinosa:2007qp}%
  \BibitemOpen
  \bibfield  {author} {\bibinfo {author} {\bibfnamefont {J.}~\bibnamefont
  {Espinosa}}, \bibinfo {author} {\bibfnamefont {G.}~\bibnamefont {Giudice}}, \
  and\ \bibinfo {author} {\bibfnamefont {A.}~\bibnamefont {Riotto}},\ }\href
  {\doibase 10.1088/1475-7516/2008/05/002} {\bibfield  {journal} {\bibinfo
  {journal} {JCAP}\ }\textbf {\bibinfo {volume} {0805}},\ \bibinfo {pages}
  {002} (\bibinfo {year} {2008})},\ \Eprint {http://arxiv.org/abs/0710.2484}
  {arXiv:0710.2484 [hep-ph]} \BibitemShut {NoStop}%
\bibitem [{\citenamefont {Fairbairn}\ and\ \citenamefont
  {Hogan}(2014)}]{Fairbairn:2014zia}%
  \BibitemOpen
  \bibfield  {author} {\bibinfo {author} {\bibfnamefont {M.}~\bibnamefont
  {Fairbairn}}\ and\ \bibinfo {author} {\bibfnamefont {R.}~\bibnamefont
  {Hogan}},\ }\href {\doibase 10.1103/PhysRevLett.112.201801} {\bibfield
  {journal} {\bibinfo  {journal} {Phys.Rev.Lett.}\ }\textbf {\bibinfo {volume}
  {112}},\ \bibinfo {pages} {201801} (\bibinfo {year} {2014})},\ \Eprint
  {http://arxiv.org/abs/1403.6786} {arXiv:1403.6786 [hep-ph]} \BibitemShut
  {NoStop}%
\bibitem [{\citenamefont {Lebedev}\ and\ \citenamefont
  {Westphal}(2013)}]{Lebedev:2012sy}%
  \BibitemOpen
  \bibfield  {author} {\bibinfo {author} {\bibfnamefont {O.}~\bibnamefont
  {Lebedev}}\ and\ \bibinfo {author} {\bibfnamefont {A.}~\bibnamefont
  {Westphal}},\ }\href {\doibase 10.1016/j.physletb.2012.12.069} {\bibfield
  {journal} {\bibinfo  {journal} {Phys.Lett.}\ }\textbf {\bibinfo {volume}
  {B719}},\ \bibinfo {pages} {415} (\bibinfo {year} {2013})},\ \Eprint
  {http://arxiv.org/abs/1210.6987} {arXiv:1210.6987 [hep-ph]} \BibitemShut
  {NoStop}%
\bibitem [{\citenamefont {Kobakhidze}\ and\ \citenamefont
  {Spencer-Smith}(2013)}]{Kobakhidze:2013tn}%
  \BibitemOpen
  \bibfield  {author} {\bibinfo {author} {\bibfnamefont {A.}~\bibnamefont
  {Kobakhidze}}\ and\ \bibinfo {author} {\bibfnamefont {A.}~\bibnamefont
  {Spencer-Smith}},\ }\href {\doibase 10.1016/j.physletb.2013.04.013}
  {\bibfield  {journal} {\bibinfo  {journal} {Phys.Lett.}\ }\textbf {\bibinfo
  {volume} {B722}},\ \bibinfo {pages} {130} (\bibinfo {year} {2013})},\ \Eprint
  {http://arxiv.org/abs/1301.2846} {arXiv:1301.2846 [hep-ph]} \BibitemShut
  {NoStop}%
\bibitem [{\citenamefont {Enqvist}\ \emph {et~al.}(2013)\citenamefont
  {Enqvist}, \citenamefont {Meriniemi},\ and\ \citenamefont
  {Nurmi}}]{Enqvist:2013kaa}%
  \BibitemOpen
  \bibfield  {author} {\bibinfo {author} {\bibfnamefont {K.}~\bibnamefont
  {Enqvist}}, \bibinfo {author} {\bibfnamefont {T.}~\bibnamefont {Meriniemi}},
  \ and\ \bibinfo {author} {\bibfnamefont {S.}~\bibnamefont {Nurmi}},\ }\href
  {\doibase 10.1088/1475-7516/2013/10/057} {\bibfield  {journal} {\bibinfo
  {journal} {JCAP}\ }\textbf {\bibinfo {volume} {1310}},\ \bibinfo {pages}
  {057} (\bibinfo {year} {2013})},\ \Eprint {http://arxiv.org/abs/1306.4511}
  {arXiv:1306.4511 [hep-ph]} \BibitemShut {NoStop}%
\bibitem [{\citenamefont {Herranen}\ \emph {et~al.}(2014)\citenamefont
  {Herranen}, \citenamefont {Markkanen}, \citenamefont {Nurmi},\ and\
  \citenamefont {Rajantie}}]{Herranen:2014cua}%
  \BibitemOpen
  \bibfield  {author} {\bibinfo {author} {\bibfnamefont {M.}~\bibnamefont
  {Herranen}}, \bibinfo {author} {\bibfnamefont {T.}~\bibnamefont {Markkanen}},
  \bibinfo {author} {\bibfnamefont {S.}~\bibnamefont {Nurmi}}, \ and\ \bibinfo
  {author} {\bibfnamefont {A.}~\bibnamefont {Rajantie}},\ }\href {\doibase
  10.1103/PhysRevLett.113.211102} {\bibfield  {journal} {\bibinfo  {journal}
  {Phys.Rev.Lett.}\ }\textbf {\bibinfo {volume} {113}},\ \bibinfo {pages}
  {211102} (\bibinfo {year} {2014})},\ \Eprint {http://arxiv.org/abs/1407.3141}
  {arXiv:1407.3141 [hep-ph]} \BibitemShut {NoStop}%
\bibitem [{\citenamefont {Kobakhidze}\ and\ \citenamefont
  {Spencer-Smith}(2014)}]{Kobakhidze:2014xda}%
  \BibitemOpen
  \bibfield  {author} {\bibinfo {author} {\bibfnamefont {A.}~\bibnamefont
  {Kobakhidze}}\ and\ \bibinfo {author} {\bibfnamefont {A.}~\bibnamefont
  {Spencer-Smith}},\ }\href@noop {} {\  (\bibinfo {year} {2014})},\ \Eprint
  {http://arxiv.org/abs/1404.4709} {arXiv:1404.4709 [hep-ph]} \BibitemShut
  {NoStop}%
\bibitem [{\citenamefont {Kamada}(2015)}]{Kamada:2014ufa}%
  \BibitemOpen
  \bibfield  {author} {\bibinfo {author} {\bibfnamefont {K.}~\bibnamefont
  {Kamada}},\ }\href {\doibase 10.1016/j.physletb.2015.01.024} {\bibfield
  {journal} {\bibinfo  {journal} {Phys. Lett.}\ }\textbf {\bibinfo {volume}
  {B742}},\ \bibinfo {pages} {126} (\bibinfo {year} {2015})},\ \Eprint
  {http://arxiv.org/abs/1409.5078} {arXiv:1409.5078 [hep-ph]} \BibitemShut
  {NoStop}%
\bibitem [{\citenamefont {Enqvist}\ \emph {et~al.}(2014)\citenamefont
  {Enqvist}, \citenamefont {Meriniemi},\ and\ \citenamefont
  {Nurmi}}]{Enqvist:2014bua}%
  \BibitemOpen
  \bibfield  {author} {\bibinfo {author} {\bibfnamefont {K.}~\bibnamefont
  {Enqvist}}, \bibinfo {author} {\bibfnamefont {T.}~\bibnamefont {Meriniemi}},
  \ and\ \bibinfo {author} {\bibfnamefont {S.}~\bibnamefont {Nurmi}},\ }\href
  {\doibase 10.1088/1475-7516/2014/07/025} {\bibfield  {journal} {\bibinfo
  {journal} {JCAP}\ }\textbf {\bibinfo {volume} {1407}},\ \bibinfo {pages}
  {025} (\bibinfo {year} {2014})},\ \Eprint {http://arxiv.org/abs/1404.3699}
  {arXiv:1404.3699 [hep-ph]} \BibitemShut {NoStop}%
\bibitem [{\citenamefont {Hook}\ \emph {et~al.}(2015)\citenamefont {Hook},
  \citenamefont {Kearney}, \citenamefont {Shakya},\ and\ \citenamefont
  {Zurek}}]{Hook:2014uia}%
  \BibitemOpen
  \bibfield  {author} {\bibinfo {author} {\bibfnamefont {A.}~\bibnamefont
  {Hook}}, \bibinfo {author} {\bibfnamefont {J.}~\bibnamefont {Kearney}},
  \bibinfo {author} {\bibfnamefont {B.}~\bibnamefont {Shakya}}, \ and\ \bibinfo
  {author} {\bibfnamefont {K.~M.}\ \bibnamefont {Zurek}},\ }\href {\doibase
  10.1007/JHEP01(2015)061} {\bibfield  {journal} {\bibinfo  {journal} {JHEP}\
  }\textbf {\bibinfo {volume} {1501}},\ \bibinfo {pages} {061} (\bibinfo {year}
  {2015})},\ \Eprint {http://arxiv.org/abs/1404.5953} {arXiv:1404.5953
  [hep-ph]} \BibitemShut {NoStop}%
\bibitem [{\citenamefont {Kearney}\ \emph {et~al.}(2015)\citenamefont
  {Kearney}, \citenamefont {Yoo},\ and\ \citenamefont
  {Zurek}}]{Kearney:2015vba}%
  \BibitemOpen
  \bibfield  {author} {\bibinfo {author} {\bibfnamefont {J.}~\bibnamefont
  {Kearney}}, \bibinfo {author} {\bibfnamefont {H.}~\bibnamefont {Yoo}}, \ and\
  \bibinfo {author} {\bibfnamefont {K.~M.}\ \bibnamefont {Zurek}},\ }\href@noop
  {} {\  (\bibinfo {year} {2015})},\ \Eprint {http://arxiv.org/abs/1503.05193}
  {arXiv:1503.05193 [hep-th]} \BibitemShut {NoStop}%
\bibitem [{\citenamefont {Espinosa}\ \emph {et~al.}(2015)\citenamefont
  {Espinosa}, \citenamefont {Giudice}, \citenamefont {Morgante}, \citenamefont
  {Riotto}, \citenamefont {Senatore} \emph {et~al.}}]{Espinosa:2015qea}%
  \BibitemOpen
  \bibfield  {author} {\bibinfo {author} {\bibfnamefont {J.~R.}\ \bibnamefont
  {Espinosa}}, \bibinfo {author} {\bibfnamefont {G.~F.}\ \bibnamefont
  {Giudice}}, \bibinfo {author} {\bibfnamefont {E.}~\bibnamefont {Morgante}},
  \bibinfo {author} {\bibfnamefont {A.}~\bibnamefont {Riotto}}, \bibinfo
  {author} {\bibfnamefont {L.}~\bibnamefont {Senatore}},  \emph {et~al.},\
  }\href@noop {} {\  (\bibinfo {year} {2015})},\ \Eprint
  {http://arxiv.org/abs/1505.04825} {arXiv:1505.04825 [hep-ph]} \BibitemShut
  {NoStop}%
\bibitem [{\citenamefont {Kohri}\ and\ \citenamefont
  {Matsui}(2016{\natexlab{a}})}]{Kohri:2016qqv}%
  \BibitemOpen
  \bibfield  {author} {\bibinfo {author} {\bibfnamefont {K.}~\bibnamefont
  {Kohri}}\ and\ \bibinfo {author} {\bibfnamefont {H.}~\bibnamefont {Matsui}},\
  }\href@noop {} {\  (\bibinfo {year} {2016}{\natexlab{a}})},\ \Eprint
  {http://arxiv.org/abs/1607.08133} {arXiv:1607.08133 [hep-ph]} \BibitemShut
  {NoStop}%
\bibitem [{\citenamefont {Czerwinska}\ \emph {et~al.}(2016)\citenamefont
  {Czerwinska}, \citenamefont {Lalak}, \citenamefont {Lewicki},\ and\
  \citenamefont {Olszewski}}]{Czerwinska:2016fky}%
  \BibitemOpen
  \bibfield  {author} {\bibinfo {author} {\bibfnamefont {O.}~\bibnamefont
  {Czerwinska}}, \bibinfo {author} {\bibfnamefont {Z.}~\bibnamefont {Lalak}},
  \bibinfo {author} {\bibfnamefont {M.}~\bibnamefont {Lewicki}}, \ and\
  \bibinfo {author} {\bibfnamefont {P.}~\bibnamefont {Olszewski}},\ }\href
  {\doibase 10.1007/JHEP10(2016)004} {\bibfield  {journal} {\bibinfo  {journal}
  {JHEP}\ }\textbf {\bibinfo {volume} {10}},\ \bibinfo {pages} {004} (\bibinfo
  {year} {2016})},\ \Eprint {http://arxiv.org/abs/1606.07808} {arXiv:1606.07808
  [hep-ph]} \BibitemShut {NoStop}%
\bibitem [{\citenamefont {East}\ \emph {et~al.}(2017)\citenamefont {East},
  \citenamefont {Kearney}, \citenamefont {Shakya}, \citenamefont {Yoo},\ and\
  \citenamefont {Zurek}}]{East:2016anr}%
  \BibitemOpen
  \bibfield  {author} {\bibinfo {author} {\bibfnamefont {W.~E.}\ \bibnamefont
  {East}}, \bibinfo {author} {\bibfnamefont {J.}~\bibnamefont {Kearney}},
  \bibinfo {author} {\bibfnamefont {B.}~\bibnamefont {Shakya}}, \bibinfo
  {author} {\bibfnamefont {H.}~\bibnamefont {Yoo}}, \ and\ \bibinfo {author}
  {\bibfnamefont {K.~M.}\ \bibnamefont {Zurek}},\ }\href {\doibase
  10.1103/PhysRevD.95.023526} {\bibfield  {journal} {\bibinfo  {journal} {Phys.
  Rev.}\ }\textbf {\bibinfo {volume} {D95}},\ \bibinfo {pages} {023526}
  (\bibinfo {year} {2017})},\ \bibinfo {note} {[Phys. Rev.D95,023526(2017)]},\
  \Eprint {http://arxiv.org/abs/1607.00381} {arXiv:1607.00381 [hep-ph]}
  \BibitemShut {NoStop}%
\bibitem [{\citenamefont {Herranen}\ \emph {et~al.}(2015)\citenamefont
  {Herranen}, \citenamefont {Markkanen}, \citenamefont {Nurmi},\ and\
  \citenamefont {Rajantie}}]{Herranen:2015ima}%
  \BibitemOpen
  \bibfield  {author} {\bibinfo {author} {\bibfnamefont {M.}~\bibnamefont
  {Herranen}}, \bibinfo {author} {\bibfnamefont {T.}~\bibnamefont {Markkanen}},
  \bibinfo {author} {\bibfnamefont {S.}~\bibnamefont {Nurmi}}, \ and\ \bibinfo
  {author} {\bibfnamefont {A.}~\bibnamefont {Rajantie}},\ }\href@noop {} {\
  (\bibinfo {year} {2015})},\ \Eprint {http://arxiv.org/abs/1506.04065}
  {arXiv:1506.04065 [hep-ph]} \BibitemShut {NoStop}%
\bibitem [{\citenamefont {Kohri}\ and\ \citenamefont
  {Matsui}(2016{\natexlab{b}})}]{Kohri:2016wof}%
  \BibitemOpen
  \bibfield  {author} {\bibinfo {author} {\bibfnamefont {K.}~\bibnamefont
  {Kohri}}\ and\ \bibinfo {author} {\bibfnamefont {H.}~\bibnamefont {Matsui}},\
  }\href {\doibase 10.1103/PhysRevD.94.103509} {\bibfield  {journal} {\bibinfo
  {journal} {Phys. Rev.}\ }\textbf {\bibinfo {volume} {D94}},\ \bibinfo {pages}
  {103509} (\bibinfo {year} {2016}{\natexlab{b}})},\ \Eprint
  {http://arxiv.org/abs/1602.02100} {arXiv:1602.02100 [hep-ph]} \BibitemShut
  {NoStop}%
\bibitem [{\citenamefont {Ema}\ \emph {et~al.}(2016)\citenamefont {Ema},
  \citenamefont {Mukaida},\ and\ \citenamefont {Nakayama}}]{Ema:2016kpf}%
  \BibitemOpen
  \bibfield  {author} {\bibinfo {author} {\bibfnamefont {Y.}~\bibnamefont
  {Ema}}, \bibinfo {author} {\bibfnamefont {K.}~\bibnamefont {Mukaida}}, \ and\
  \bibinfo {author} {\bibfnamefont {K.}~\bibnamefont {Nakayama}},\ }\href@noop
  {} {\  (\bibinfo {year} {2016})},\ \Eprint {http://arxiv.org/abs/1602.00483}
  {arXiv:1602.00483 [hep-ph]} \BibitemShut {NoStop}%
\bibitem [{\citenamefont {Enqvist}\ \emph {et~al.}(2016)\citenamefont
  {Enqvist}, \citenamefont {Karciauskas}, \citenamefont {Lebedev},
  \citenamefont {Rusak},\ and\ \citenamefont {Zatta}}]{Enqvist:2016mqj}%
  \BibitemOpen
  \bibfield  {author} {\bibinfo {author} {\bibfnamefont {K.}~\bibnamefont
  {Enqvist}}, \bibinfo {author} {\bibfnamefont {M.}~\bibnamefont
  {Karciauskas}}, \bibinfo {author} {\bibfnamefont {O.}~\bibnamefont
  {Lebedev}}, \bibinfo {author} {\bibfnamefont {S.}~\bibnamefont {Rusak}}, \
  and\ \bibinfo {author} {\bibfnamefont {M.}~\bibnamefont {Zatta}},\ }\href
  {\doibase 10.1088/1475-7516/2016/11/025} {\bibfield  {journal} {\bibinfo
  {journal} {JCAP}\ }\textbf {\bibinfo {volume} {1611}},\ \bibinfo {pages}
  {025} (\bibinfo {year} {2016})},\ \Eprint {http://arxiv.org/abs/1608.08848}
  {arXiv:1608.08848 [hep-ph]} \BibitemShut {NoStop}%
\bibitem [{\citenamefont {Postma}\ and\ \citenamefont {van~de
  Vis}(2017)}]{Postma:2017hbk}%
  \BibitemOpen
  \bibfield  {author} {\bibinfo {author} {\bibfnamefont {M.}~\bibnamefont
  {Postma}}\ and\ \bibinfo {author} {\bibfnamefont {J.}~\bibnamefont {van~de
  Vis}},\ }\href@noop {} {\  (\bibinfo {year} {2017})},\ \Eprint
  {http://arxiv.org/abs/1702.07636} {arXiv:1702.07636 [hep-ph]} \BibitemShut
  {NoStop}%
\bibitem [{\citenamefont {Ema}\ \emph {et~al.}(2017)\citenamefont {Ema},
  \citenamefont {Karciauskas}, \citenamefont {Lebedev},\ and\ \citenamefont
  {Zatta}}]{Ema:2017loe}%
  \BibitemOpen
  \bibfield  {author} {\bibinfo {author} {\bibfnamefont {Y.}~\bibnamefont
  {Ema}}, \bibinfo {author} {\bibfnamefont {M.}~\bibnamefont {Karciauskas}},
  \bibinfo {author} {\bibfnamefont {O.}~\bibnamefont {Lebedev}}, \ and\
  \bibinfo {author} {\bibfnamefont {M.}~\bibnamefont {Zatta}},\ }\href@noop {}
  {\  (\bibinfo {year} {2017})},\ \Eprint {http://arxiv.org/abs/1703.04681}
  {arXiv:1703.04681 [hep-ph]} \BibitemShut {NoStop}%
\bibitem [{\citenamefont {Burda}\ \emph
  {et~al.}(2015{\natexlab{a}})\citenamefont {Burda}, \citenamefont {Gregory},\
  and\ \citenamefont {Moss}}]{Burda:2015isa}%
  \BibitemOpen
  \bibfield  {author} {\bibinfo {author} {\bibfnamefont {P.}~\bibnamefont
  {Burda}}, \bibinfo {author} {\bibfnamefont {R.}~\bibnamefont {Gregory}}, \
  and\ \bibinfo {author} {\bibfnamefont {I.}~\bibnamefont {Moss}},\ }\href
  {\doibase 10.1103/PhysRevLett.115.071303} {\bibfield  {journal} {\bibinfo
  {journal} {Phys. Rev. Lett.}\ }\textbf {\bibinfo {volume} {115}},\ \bibinfo
  {pages} {071303} (\bibinfo {year} {2015}{\natexlab{a}})},\ \Eprint
  {http://arxiv.org/abs/1501.04937} {arXiv:1501.04937 [hep-th]} \BibitemShut
  {NoStop}%
\bibitem [{\citenamefont {Burda}\ \emph
  {et~al.}(2015{\natexlab{b}})\citenamefont {Burda}, \citenamefont {Gregory},\
  and\ \citenamefont {Moss}}]{Burda:2015yfa}%
  \BibitemOpen
  \bibfield  {author} {\bibinfo {author} {\bibfnamefont {P.}~\bibnamefont
  {Burda}}, \bibinfo {author} {\bibfnamefont {R.}~\bibnamefont {Gregory}}, \
  and\ \bibinfo {author} {\bibfnamefont {I.}~\bibnamefont {Moss}},\ }\href
  {\doibase 10.1007/JHEP08(2015)114} {\bibfield  {journal} {\bibinfo  {journal}
  {JHEP}\ }\textbf {\bibinfo {volume} {08}},\ \bibinfo {pages} {114} (\bibinfo
  {year} {2015}{\natexlab{b}})},\ \Eprint {http://arxiv.org/abs/1503.07331}
  {arXiv:1503.07331 [hep-th]} \BibitemShut {NoStop}%
\bibitem [{\citenamefont {Burda}\ \emph {et~al.}(2016)\citenamefont {Burda},
  \citenamefont {Gregory},\ and\ \citenamefont {Moss}}]{Burda:2016mou}%
  \BibitemOpen
  \bibfield  {author} {\bibinfo {author} {\bibfnamefont {P.}~\bibnamefont
  {Burda}}, \bibinfo {author} {\bibfnamefont {R.}~\bibnamefont {Gregory}}, \
  and\ \bibinfo {author} {\bibfnamefont {I.}~\bibnamefont {Moss}},\ }\href
  {\doibase 10.1007/JHEP06(2016)025} {\bibfield  {journal} {\bibinfo  {journal}
  {JHEP}\ }\textbf {\bibinfo {volume} {06}},\ \bibinfo {pages} {025} (\bibinfo
  {year} {2016})},\ \Eprint {http://arxiv.org/abs/1601.02152} {arXiv:1601.02152
  [hep-th]} \BibitemShut {NoStop}%
\bibitem [{\citenamefont {Grinstein}\ and\ \citenamefont
  {Murphy}(2015)}]{Grinstein:2015jda}%
  \BibitemOpen
  \bibfield  {author} {\bibinfo {author} {\bibfnamefont {B.}~\bibnamefont
  {Grinstein}}\ and\ \bibinfo {author} {\bibfnamefont {C.~W.}\ \bibnamefont
  {Murphy}},\ }\href {\doibase 10.1007/JHEP12(2015)063} {\bibfield  {journal}
  {\bibinfo  {journal} {JHEP}\ }\textbf {\bibinfo {volume} {12}},\ \bibinfo
  {pages} {063} (\bibinfo {year} {2015})},\ \Eprint
  {http://arxiv.org/abs/1509.05405} {arXiv:1509.05405 [hep-ph]} \BibitemShut
  {NoStop}%
\bibitem [{\citenamefont {Tetradis}(2016)}]{Tetradis:2016vqb}%
  \BibitemOpen
  \bibfield  {author} {\bibinfo {author} {\bibfnamefont {N.}~\bibnamefont
  {Tetradis}},\ }\href {\doibase 10.1088/1475-7516/2016/09/036} {\bibfield
  {journal} {\bibinfo  {journal} {JCAP}\ }\textbf {\bibinfo {volume} {1609}},\
  \bibinfo {pages} {036} (\bibinfo {year} {2016})},\ \Eprint
  {http://arxiv.org/abs/1606.04018} {arXiv:1606.04018 [hep-ph]} \BibitemShut
  {NoStop}%
\bibitem [{\citenamefont {Cheung}\ and\ \citenamefont
  {Leichenauer}(2014)}]{Cheung:2013sxa}%
  \BibitemOpen
  \bibfield  {author} {\bibinfo {author} {\bibfnamefont {C.}~\bibnamefont
  {Cheung}}\ and\ \bibinfo {author} {\bibfnamefont {S.}~\bibnamefont
  {Leichenauer}},\ }\href {\doibase 10.1103/PhysRevD.89.104035} {\bibfield
  {journal} {\bibinfo  {journal} {Phys. Rev.}\ }\textbf {\bibinfo {volume}
  {D89}},\ \bibinfo {pages} {104035} (\bibinfo {year} {2014})},\ \Eprint
  {http://arxiv.org/abs/1309.0530} {arXiv:1309.0530 [hep-ph]} \BibitemShut
  {NoStop}%
\bibitem [{\citenamefont {Canko}\ \emph {et~al.}(2017)\citenamefont {Canko},
  \citenamefont {Gialamas}, \citenamefont {Jelic-Cizmek}, \citenamefont
  {Riotto},\ and\ \citenamefont {Tetradis}}]{Canko:2017ebb}%
  \BibitemOpen
  \bibfield  {author} {\bibinfo {author} {\bibfnamefont {D.}~\bibnamefont
  {Canko}}, \bibinfo {author} {\bibfnamefont {I.}~\bibnamefont {Gialamas}},
  \bibinfo {author} {\bibfnamefont {G.}~\bibnamefont {Jelic-Cizmek}}, \bibinfo
  {author} {\bibfnamefont {A.}~\bibnamefont {Riotto}}, \ and\ \bibinfo {author}
  {\bibfnamefont {N.}~\bibnamefont {Tetradis}},\ }\href@noop {} {\  (\bibinfo
  {year} {2017})},\ \Eprint {http://arxiv.org/abs/1706.01364} {arXiv:1706.01364
  [hep-th]} \BibitemShut {NoStop}%
\bibitem [{\citenamefont {Gorbunov}\ \emph {et~al.}(2017)\citenamefont
  {Gorbunov}, \citenamefont {Levkov},\ and\ \citenamefont
  {Panin}}]{Gorbunov:2017fhq}%
  \BibitemOpen
  \bibfield  {author} {\bibinfo {author} {\bibfnamefont {D.}~\bibnamefont
  {Gorbunov}}, \bibinfo {author} {\bibfnamefont {D.}~\bibnamefont {Levkov}}, \
  and\ \bibinfo {author} {\bibfnamefont {A.}~\bibnamefont {Panin}},\
  }\href@noop {} {\  (\bibinfo {year} {2017})},\ \Eprint
  {http://arxiv.org/abs/1704.05399} {arXiv:1704.05399 [astro-ph.CO]}
  \BibitemShut {NoStop}%
\bibitem [{\citenamefont {Kohri}\ and\ \citenamefont
  {Matsui}(2017{\natexlab{a}})}]{Kohri:2017ybt}%
  \BibitemOpen
  \bibfield  {author} {\bibinfo {author} {\bibfnamefont {K.}~\bibnamefont
  {Kohri}}\ and\ \bibinfo {author} {\bibfnamefont {H.}~\bibnamefont {Matsui}},\
  }\href@noop {} {\  (\bibinfo {year} {2017}{\natexlab{a}})},\ \Eprint
  {http://arxiv.org/abs/1708.02138} {arXiv:1708.02138 [hep-ph]} \BibitemShut
  {NoStop}%
\bibitem [{\citenamefont {Ringwald}(1987)}]{Ringwald:1987ui}%
  \BibitemOpen
  \bibfield  {author} {\bibinfo {author} {\bibfnamefont {A.}~\bibnamefont
  {Ringwald}},\ }\href {\doibase 10.1016/S0003-4916(87)80027-1} {\bibfield
  {journal} {\bibinfo  {journal} {Annals Phys.}\ }\textbf {\bibinfo {volume}
  {177}},\ \bibinfo {pages} {129} (\bibinfo {year} {1987})}\BibitemShut
  {NoStop}%
\bibitem [{\citenamefont {Shore}(1980)}]{Shore:1979as}%
  \BibitemOpen
  \bibfield  {author} {\bibinfo {author} {\bibfnamefont {G.~M.}\ \bibnamefont
  {Shore}},\ }\href {\doibase 10.1016/0003-4916(80)90326-7} {\bibfield
  {journal} {\bibinfo  {journal} {Annals Phys.}\ }\textbf {\bibinfo {volume}
  {128}},\ \bibinfo {pages} {376} (\bibinfo {year} {1980})}\BibitemShut
  {NoStop}%
\bibitem [{\citenamefont {Toms}(1982)}]{Toms:1982af}%
  \BibitemOpen
  \bibfield  {author} {\bibinfo {author} {\bibfnamefont {D.~J.}\ \bibnamefont
  {Toms}},\ }\href {\doibase 10.1103/PhysRevD.26.2713} {\bibfield  {journal}
  {\bibinfo  {journal} {Phys. Rev.}\ }\textbf {\bibinfo {volume} {D26}},\
  \bibinfo {pages} {2713} (\bibinfo {year} {1982})}\BibitemShut {NoStop}%
\bibitem [{\citenamefont {Toms}(1983)}]{Toms:1983qr}%
  \BibitemOpen
  \bibfield  {author} {\bibinfo {author} {\bibfnamefont {D.~J.}\ \bibnamefont
  {Toms}},\ }\href {\doibase 10.1016/0370-2693(83)90011-4} {\bibfield
  {journal} {\bibinfo  {journal} {Phys. Lett.}\ }\textbf {\bibinfo {volume}
  {B126}},\ \bibinfo {pages} {37} (\bibinfo {year} {1983})}\BibitemShut
  {NoStop}%
\bibitem [{\citenamefont {Hu}\ and\ \citenamefont
  {O'Connor}(1984)}]{Hu:1984js}%
  \BibitemOpen
  \bibfield  {author} {\bibinfo {author} {\bibfnamefont {B.~L.}\ \bibnamefont
  {Hu}}\ and\ \bibinfo {author} {\bibfnamefont {D.~J.}\ \bibnamefont
  {O'Connor}},\ }\href {\doibase 10.1103/PhysRevD.30.743} {\bibfield  {journal}
  {\bibinfo  {journal} {Phys. Rev.}\ }\textbf {\bibinfo {volume} {D30}},\
  \bibinfo {pages} {743} (\bibinfo {year} {1984})}\BibitemShut {NoStop}%
\bibitem [{\citenamefont {Buchbinder}\ and\ \citenamefont
  {Odintsov}(1984)}]{Buchbinder:1985js}%
  \BibitemOpen
  \bibfield  {author} {\bibinfo {author} {\bibfnamefont {I.~L.}\ \bibnamefont
  {Buchbinder}}\ and\ \bibinfo {author} {\bibfnamefont {S.~D.}\ \bibnamefont
  {Odintsov}},\ }\href {\doibase 10.1007/BF00897445} {\bibfield  {journal}
  {\bibinfo  {journal} {Sov. Phys. J.}\ }\textbf {\bibinfo {volume} {27}},\
  \bibinfo {pages} {554} (\bibinfo {year} {1984})}\BibitemShut {NoStop}%
\bibitem [{\citenamefont {Balakrishnan}\ and\ \citenamefont
  {Toms}(1992)}]{Balakrishnan:1991pm}%
  \BibitemOpen
  \bibfield  {author} {\bibinfo {author} {\bibfnamefont {J.}~\bibnamefont
  {Balakrishnan}}\ and\ \bibinfo {author} {\bibfnamefont {D.~J.}\ \bibnamefont
  {Toms}},\ }\href {\doibase 10.1103/PhysRevD.46.4413} {\bibfield  {journal}
  {\bibinfo  {journal} {Phys. Rev.}\ }\textbf {\bibinfo {volume} {D46}},\
  \bibinfo {pages} {4413} (\bibinfo {year} {1992})}\BibitemShut {NoStop}%
\bibitem [{\citenamefont {Muta}\ and\ \citenamefont
  {Odintsov}(1991)}]{Muta:1991mw}%
  \BibitemOpen
  \bibfield  {author} {\bibinfo {author} {\bibfnamefont {T.}~\bibnamefont
  {Muta}}\ and\ \bibinfo {author} {\bibfnamefont {S.~D.}\ \bibnamefont
  {Odintsov}},\ }\href {\doibase 10.1142/S0217732391004206} {\bibfield
  {journal} {\bibinfo  {journal} {Mod. Phys. Lett.}\ }\textbf {\bibinfo
  {volume} {A6}},\ \bibinfo {pages} {3641} (\bibinfo {year}
  {1991})}\BibitemShut {NoStop}%
\bibitem [{\citenamefont {Kirsten}\ \emph {et~al.}(1993)\citenamefont
  {Kirsten}, \citenamefont {Cognola},\ and\ \citenamefont
  {Vanzo}}]{Kirsten:1993jn}%
  \BibitemOpen
  \bibfield  {author} {\bibinfo {author} {\bibfnamefont {K.}~\bibnamefont
  {Kirsten}}, \bibinfo {author} {\bibfnamefont {G.}~\bibnamefont {Cognola}}, \
  and\ \bibinfo {author} {\bibfnamefont {L.}~\bibnamefont {Vanzo}},\ }\href
  {\doibase 10.1103/PhysRevD.48.2813} {\bibfield  {journal} {\bibinfo
  {journal} {Phys. Rev.}\ }\textbf {\bibinfo {volume} {D48}},\ \bibinfo {pages}
  {2813} (\bibinfo {year} {1993})},\ \Eprint
  {http://arxiv.org/abs/hep-th/9304092} {arXiv:hep-th/9304092 [hep-th]}
  \BibitemShut {NoStop}%
\bibitem [{\citenamefont {Elizalde}\ and\ \citenamefont
  {Odintsov}(1993)}]{Elizalde:1993ee}%
  \BibitemOpen
  \bibfield  {author} {\bibinfo {author} {\bibfnamefont {E.}~\bibnamefont
  {Elizalde}}\ and\ \bibinfo {author} {\bibfnamefont {S.~D.}\ \bibnamefont
  {Odintsov}},\ }\href {\doibase 10.1016/0370-2693(93)91427-O} {\bibfield
  {journal} {\bibinfo  {journal} {Phys. Lett.}\ }\textbf {\bibinfo {volume}
  {B303}},\ \bibinfo {pages} {240} (\bibinfo {year} {1993})},\ \bibinfo {note}
  {[Russ. Phys. J.37,25(1994)]},\ \Eprint {http://arxiv.org/abs/hep-th/9302074}
  {arXiv:hep-th/9302074 [hep-th]} \BibitemShut {NoStop}%
\bibitem [{\citenamefont {Elizalde}\ and\ \citenamefont
  {Odintsov}(1994{\natexlab{a}})}]{Elizalde:1993ew}%
  \BibitemOpen
  \bibfield  {author} {\bibinfo {author} {\bibfnamefont {E.}~\bibnamefont
  {Elizalde}}\ and\ \bibinfo {author} {\bibfnamefont {S.~D.}\ \bibnamefont
  {Odintsov}},\ }\href {\doibase 10.1016/0370-2693(94)90464-2} {\bibfield
  {journal} {\bibinfo  {journal} {Phys. Lett.}\ }\textbf {\bibinfo {volume}
  {B321}},\ \bibinfo {pages} {199} (\bibinfo {year} {1994}{\natexlab{a}})},\
  \Eprint {http://arxiv.org/abs/hep-th/9311087} {arXiv:hep-th/9311087 [hep-th]}
  \BibitemShut {NoStop}%
\bibitem [{\citenamefont {Elizalde}\ and\ \citenamefont
  {Odintsov}(1994{\natexlab{b}})}]{Elizalde:1994im}%
  \BibitemOpen
  \bibfield  {author} {\bibinfo {author} {\bibfnamefont {E.}~\bibnamefont
  {Elizalde}}\ and\ \bibinfo {author} {\bibfnamefont {S.~D.}\ \bibnamefont
  {Odintsov}},\ }\href {\doibase 10.1016/0370-2693(94)90151-1} {\bibfield
  {journal} {\bibinfo  {journal} {Phys. Lett.}\ }\textbf {\bibinfo {volume}
  {B333}},\ \bibinfo {pages} {331} (\bibinfo {year} {1994}{\natexlab{b}})},\
  \Eprint {http://arxiv.org/abs/hep-th/9403132} {arXiv:hep-th/9403132 [hep-th]}
  \BibitemShut {NoStop}%
\bibitem [{\citenamefont {Elizalde}\ \emph {et~al.}(1994)\citenamefont
  {Elizalde}, \citenamefont {Kirsten},\ and\ \citenamefont
  {Odintsov}}]{Elizalde:1994ds}%
  \BibitemOpen
  \bibfield  {author} {\bibinfo {author} {\bibfnamefont {E.}~\bibnamefont
  {Elizalde}}, \bibinfo {author} {\bibfnamefont {K.}~\bibnamefont {Kirsten}}, \
  and\ \bibinfo {author} {\bibfnamefont {S.~D.}\ \bibnamefont {Odintsov}},\
  }\href {\doibase 10.1103/PhysRevD.50.5137} {\bibfield  {journal} {\bibinfo
  {journal} {Phys. Rev.}\ }\textbf {\bibinfo {volume} {D50}},\ \bibinfo {pages}
  {5137} (\bibinfo {year} {1994})},\ \Eprint
  {http://arxiv.org/abs/hep-th/9404084} {arXiv:hep-th/9404084 [hep-th]}
  \BibitemShut {NoStop}%
\bibitem [{\citenamefont {Elizalde}\ \emph {et~al.}(1995)\citenamefont
  {Elizalde}, \citenamefont {Odintsov},\ and\ \citenamefont
  {Romeo}}]{Elizalde:1994gv}%
  \BibitemOpen
  \bibfield  {author} {\bibinfo {author} {\bibfnamefont {E.}~\bibnamefont
  {Elizalde}}, \bibinfo {author} {\bibfnamefont {S.~D.}\ \bibnamefont
  {Odintsov}}, \ and\ \bibinfo {author} {\bibfnamefont {A.}~\bibnamefont
  {Romeo}},\ }\href {\doibase 10.1103/PhysRevD.51.1680} {\bibfield  {journal}
  {\bibinfo  {journal} {Phys. Rev.}\ }\textbf {\bibinfo {volume} {D51}},\
  \bibinfo {pages} {1680} (\bibinfo {year} {1995})},\ \Eprint
  {http://arxiv.org/abs/hep-th/9410113} {arXiv:hep-th/9410113 [hep-th]}
  \BibitemShut {NoStop}%
\bibitem [{\citenamefont {Gorbar}\ and\ \citenamefont
  {Shapiro}(2003{\natexlab{a}})}]{Gorbar:2002pw}%
  \BibitemOpen
  \bibfield  {author} {\bibinfo {author} {\bibfnamefont {E.~V.}\ \bibnamefont
  {Gorbar}}\ and\ \bibinfo {author} {\bibfnamefont {I.~L.}\ \bibnamefont
  {Shapiro}},\ }\href {\doibase 10.1088/1126-6708/2003/02/021} {\bibfield
  {journal} {\bibinfo  {journal} {JHEP}\ }\textbf {\bibinfo {volume} {02}},\
  \bibinfo {pages} {021} (\bibinfo {year} {2003}{\natexlab{a}})},\ \Eprint
  {http://arxiv.org/abs/hep-ph/0210388} {arXiv:hep-ph/0210388 [hep-ph]}
  \BibitemShut {NoStop}%
\bibitem [{\citenamefont {Gorbar}\ and\ \citenamefont
  {Shapiro}(2003{\natexlab{b}})}]{Gorbar:2003yt}%
  \BibitemOpen
  \bibfield  {author} {\bibinfo {author} {\bibfnamefont {E.~V.}\ \bibnamefont
  {Gorbar}}\ and\ \bibinfo {author} {\bibfnamefont {I.~L.}\ \bibnamefont
  {Shapiro}},\ }\href {\doibase 10.1088/1126-6708/2003/06/004} {\bibfield
  {journal} {\bibinfo  {journal} {JHEP}\ }\textbf {\bibinfo {volume} {06}},\
  \bibinfo {pages} {004} (\bibinfo {year} {2003}{\natexlab{b}})},\ \Eprint
  {http://arxiv.org/abs/hep-ph/0303124} {arXiv:hep-ph/0303124 [hep-ph]}
  \BibitemShut {NoStop}%
\bibitem [{\citenamefont {Gorbar}\ and\ \citenamefont
  {Shapiro}(2004)}]{Gorbar:2003yp}%
  \BibitemOpen
  \bibfield  {author} {\bibinfo {author} {\bibfnamefont {E.~V.}\ \bibnamefont
  {Gorbar}}\ and\ \bibinfo {author} {\bibfnamefont {I.~L.}\ \bibnamefont
  {Shapiro}},\ }\href {\doibase 10.1088/1126-6708/2004/02/060} {\bibfield
  {journal} {\bibinfo  {journal} {JHEP}\ }\textbf {\bibinfo {volume} {02}},\
  \bibinfo {pages} {060} (\bibinfo {year} {2004})},\ \Eprint
  {http://arxiv.org/abs/hep-ph/0311190} {arXiv:hep-ph/0311190 [hep-ph]}
  \BibitemShut {NoStop}%
\bibitem [{\citenamefont {Maroto}\ and\ \citenamefont
  {Prada}(2014)}]{Maroto:2014oca}%
  \BibitemOpen
  \bibfield  {author} {\bibinfo {author} {\bibfnamefont {A.~L.}\ \bibnamefont
  {Maroto}}\ and\ \bibinfo {author} {\bibfnamefont {F.}~\bibnamefont {Prada}},\
  }\href {\doibase 10.1103/PhysRevD.90.123541, 10.1103/PhysRevD.93.069904}
  {\bibfield  {journal} {\bibinfo  {journal} {Phys. Rev.}\ }\textbf {\bibinfo
  {volume} {D90}},\ \bibinfo {pages} {123541} (\bibinfo {year} {2014})},\
  \bibinfo {note} {[Erratum: Phys. Rev.D93,no.6,069904(2016)]},\ \Eprint
  {http://arxiv.org/abs/1410.0302} {arXiv:1410.0302 [hep-ph]} \BibitemShut
  {NoStop}%
\bibitem [{\citenamefont {Czerwi?ska}\ \emph {et~al.}(2015)\citenamefont
  {Czerwi?ska}, \citenamefont {Lalak},\ and\ \citenamefont
  {Nakonieczny}}]{Czerwinska:2015xwa}%
  \BibitemOpen
  \bibfield  {author} {\bibinfo {author} {\bibfnamefont {O.}~\bibnamefont
  {Czerwi?ska}}, \bibinfo {author} {\bibfnamefont {Z.}~\bibnamefont {Lalak}}, \
  and\ \bibinfo {author} {\bibfnamefont {u.}~\bibnamefont {Nakonieczny}},\
  }\href {\doibase 10.1007/JHEP11(2015)207} {\bibfield  {journal} {\bibinfo
  {journal} {JHEP}\ }\textbf {\bibinfo {volume} {11}},\ \bibinfo {pages} {207}
  (\bibinfo {year} {2015})},\ \Eprint {http://arxiv.org/abs/1508.03297}
  {arXiv:1508.03297 [hep-th]} \BibitemShut {NoStop}%
\bibitem [{\citenamefont {Albareti}\ \emph {et~al.}(2016)\citenamefont
  {Albareti}, \citenamefont {Maroto},\ and\ \citenamefont
  {Prada}}]{Albareti:2016cvx}%
  \BibitemOpen
  \bibfield  {author} {\bibinfo {author} {\bibfnamefont {F.~D.}\ \bibnamefont
  {Albareti}}, \bibinfo {author} {\bibfnamefont {A.~L.}\ \bibnamefont
  {Maroto}}, \ and\ \bibinfo {author} {\bibfnamefont {F.}~\bibnamefont
  {Prada}},\ }\href@noop {} {\  (\bibinfo {year} {2016})},\ \Eprint
  {http://arxiv.org/abs/1602.02776} {arXiv:1602.02776 [hep-ph]} \BibitemShut
  {NoStop}%
\bibitem [{\citenamefont {Ford}(1985)}]{Ford:1984hs}%
  \BibitemOpen
  \bibfield  {author} {\bibinfo {author} {\bibfnamefont {L.~H.}\ \bibnamefont
  {Ford}},\ }\href {\doibase 10.1103/PhysRevD.31.710} {\bibfield  {journal}
  {\bibinfo  {journal} {Phys. Rev.}\ }\textbf {\bibinfo {volume} {D31}},\
  \bibinfo {pages} {710} (\bibinfo {year} {1985})}\BibitemShut {NoStop}%
\bibitem [{\citenamefont {Mottola}(1985)}]{Mottola:1984ar}%
  \BibitemOpen
  \bibfield  {author} {\bibinfo {author} {\bibfnamefont {E.}~\bibnamefont
  {Mottola}},\ }\href {\doibase 10.1103/PhysRevD.31.754} {\bibfield  {journal}
  {\bibinfo  {journal} {Phys. Rev.}\ }\textbf {\bibinfo {volume} {D31}},\
  \bibinfo {pages} {754} (\bibinfo {year} {1985})}\BibitemShut {NoStop}%
\bibitem [{\citenamefont {Mottola}(1986)}]{Mottola:1985qt}%
  \BibitemOpen
  \bibfield  {author} {\bibinfo {author} {\bibfnamefont {E.}~\bibnamefont
  {Mottola}},\ }\href {\doibase 10.1103/PhysRevD.33.1616} {\bibfield  {journal}
  {\bibinfo  {journal} {Phys. Rev.}\ }\textbf {\bibinfo {volume} {D33}},\
  \bibinfo {pages} {1616} (\bibinfo {year} {1986})}\BibitemShut {NoStop}%
\bibitem [{\citenamefont {Antoniadis}\ \emph {et~al.}(1986)\citenamefont
  {Antoniadis}, \citenamefont {Iliopoulos},\ and\ \citenamefont
  {Tomaras}}]{Antoniadis:1985pj}%
  \BibitemOpen
  \bibfield  {author} {\bibinfo {author} {\bibfnamefont {I.}~\bibnamefont
  {Antoniadis}}, \bibinfo {author} {\bibfnamefont {J.}~\bibnamefont
  {Iliopoulos}}, \ and\ \bibinfo {author} {\bibfnamefont {T.~N.}\ \bibnamefont
  {Tomaras}},\ }\href {\doibase 10.1103/PhysRevLett.56.1319} {\bibfield
  {journal} {\bibinfo  {journal} {Phys. Rev. Lett.}\ }\textbf {\bibinfo
  {volume} {56}},\ \bibinfo {pages} {1319} (\bibinfo {year}
  {1986})}\BibitemShut {NoStop}%
\bibitem [{\citenamefont {Kofman}\ \emph {et~al.}(1997)\citenamefont {Kofman},
  \citenamefont {Linde},\ and\ \citenamefont {Starobinsky}}]{Kofman:1997yn}%
  \BibitemOpen
  \bibfield  {author} {\bibinfo {author} {\bibfnamefont {L.}~\bibnamefont
  {Kofman}}, \bibinfo {author} {\bibfnamefont {A.~D.}\ \bibnamefont {Linde}}, \
  and\ \bibinfo {author} {\bibfnamefont {A.~A.}\ \bibnamefont {Starobinsky}},\
  }\href {\doibase 10.1103/PhysRevD.56.3258} {\bibfield  {journal} {\bibinfo
  {journal} {Phys. Rev.}\ }\textbf {\bibinfo {volume} {D56}},\ \bibinfo {pages}
  {3258} (\bibinfo {year} {1997})},\ \Eprint
  {http://arxiv.org/abs/hep-ph/9704452} {arXiv:hep-ph/9704452 [hep-ph]}
  \BibitemShut {NoStop}%
\bibitem [{\citenamefont {Bunch}(1980)}]{bunch1980adiabatic}%
  \BibitemOpen
  \bibfield  {author} {\bibinfo {author} {\bibfnamefont {T.}~\bibnamefont
  {Bunch}},\ }\href@noop {} {\bibfield  {journal} {\bibinfo  {journal} {Journal
  of Physics A: Mathematical and General}\ }\textbf {\bibinfo {volume} {13}},\
  \bibinfo {pages} {1297} (\bibinfo {year} {1980})}\BibitemShut {NoStop}%
\bibitem [{\citenamefont {Parker}\ and\ \citenamefont
  {Fulling}(1974)}]{Parker:1974qw}%
  \BibitemOpen
  \bibfield  {author} {\bibinfo {author} {\bibfnamefont {L.}~\bibnamefont
  {Parker}}\ and\ \bibinfo {author} {\bibfnamefont {S.~A.}\ \bibnamefont
  {Fulling}},\ }\href {\doibase 10.1103/PhysRevD.9.341} {\bibfield  {journal}
  {\bibinfo  {journal} {Phys. Rev.}\ }\textbf {\bibinfo {volume} {D9}},\
  \bibinfo {pages} {341} (\bibinfo {year} {1974})}\BibitemShut {NoStop}%
\bibitem [{\citenamefont {Fulling}\ \emph {et~al.}(1974)\citenamefont
  {Fulling}, \citenamefont {Parker},\ and\ \citenamefont
  {Hu}}]{Fulling:1974pu}%
  \BibitemOpen
  \bibfield  {author} {\bibinfo {author} {\bibfnamefont {S.~A.}\ \bibnamefont
  {Fulling}}, \bibinfo {author} {\bibfnamefont {L.}~\bibnamefont {Parker}}, \
  and\ \bibinfo {author} {\bibfnamefont {B.~L.}\ \bibnamefont {Hu}},\ }\href
  {\doibase 10.1103/PhysRevD.10.3905} {\bibfield  {journal} {\bibinfo
  {journal} {Phys. Rev.}\ }\textbf {\bibinfo {volume} {D10}},\ \bibinfo {pages}
  {3905} (\bibinfo {year} {1974})}\BibitemShut {NoStop}%
\bibitem [{\citenamefont {Fulling}\ and\ \citenamefont
  {Parker}(1974)}]{Fulling:1974zr}%
  \BibitemOpen
  \bibfield  {author} {\bibinfo {author} {\bibfnamefont {S.~A.}\ \bibnamefont
  {Fulling}}\ and\ \bibinfo {author} {\bibfnamefont {L.}~\bibnamefont
  {Parker}},\ }\href {\doibase 10.1016/0003-4916(74)90451-5} {\bibfield
  {journal} {\bibinfo  {journal} {Annals Phys.}\ }\textbf {\bibinfo {volume}
  {87}},\ \bibinfo {pages} {176} (\bibinfo {year} {1974})}\BibitemShut
  {NoStop}%
\bibitem [{\citenamefont {Anderson}\ and\ \citenamefont
  {Parker}(1987)}]{Anderson:1987yt}%
  \BibitemOpen
  \bibfield  {author} {\bibinfo {author} {\bibfnamefont {P.~R.}\ \bibnamefont
  {Anderson}}\ and\ \bibinfo {author} {\bibfnamefont {L.}~\bibnamefont
  {Parker}},\ }\href {\doibase 10.1103/PhysRevD.36.2963} {\bibfield  {journal}
  {\bibinfo  {journal} {Phys. Rev.}\ }\textbf {\bibinfo {volume} {D36}},\
  \bibinfo {pages} {2963} (\bibinfo {year} {1987})}\BibitemShut {NoStop}%
\bibitem [{\citenamefont {Haro}(2010{\natexlab{a}})}]{Haro:2010zz}%
  \BibitemOpen
  \bibfield  {author} {\bibinfo {author} {\bibfnamefont {J.}~\bibnamefont
  {Haro}},\ }\href {\doibase 10.1007/s11232-010-0123-2} {\bibfield  {journal}
  {\bibinfo  {journal} {Theor. Math. Phys.}\ }\textbf {\bibinfo {volume}
  {165}},\ \bibinfo {pages} {1490} (\bibinfo {year}
  {2010}{\natexlab{a}})}\BibitemShut {NoStop}%
\bibitem [{\citenamefont {Haro}(2010{\natexlab{b}})}]{Haro:2010mx}%
  \BibitemOpen
  \bibfield  {author} {\bibinfo {author} {\bibfnamefont {J.}~\bibnamefont
  {Haro}},\ }\href@noop {} {\  (\bibinfo {year} {2010}{\natexlab{b}})},\
  \Eprint {http://arxiv.org/abs/1011.4772} {arXiv:1011.4772 [gr-qc]}
  \BibitemShut {NoStop}%
\bibitem [{\citenamefont {Birrell}(1978)}]{birrell1978application}%
  \BibitemOpen
  \bibfield  {author} {\bibinfo {author} {\bibfnamefont {N.}~\bibnamefont
  {Birrell}},\ }in\ \href@noop {} {\emph {\bibinfo {booktitle} {Proceedings of
  the Royal Society of London A: Mathematical, Physical and Engineering
  Sciences}}},\ Vol.\ \bibinfo {volume} {361}\ (\bibinfo {organization} {The
  Royal Society},\ \bibinfo {year} {1978})\ pp.\ \bibinfo {pages}
  {513--526}\BibitemShut {NoStop}%
\bibitem [{\citenamefont {Kohri}\ and\ \citenamefont
  {Matsui}(2017{\natexlab{b}})}]{Kohri:2016lsj}%
  \BibitemOpen
  \bibfield  {author} {\bibinfo {author} {\bibfnamefont {K.}~\bibnamefont
  {Kohri}}\ and\ \bibinfo {author} {\bibfnamefont {H.}~\bibnamefont {Matsui}},\
  }\href {\doibase 10.1088/1475-7516/2017/06/006} {\bibfield  {journal}
  {\bibinfo  {journal} {JCAP}\ }\textbf {\bibinfo {volume} {1706}},\ \bibinfo
  {pages} {006} (\bibinfo {year} {2017}{\natexlab{b}})},\ \Eprint
  {http://arxiv.org/abs/1612.08818} {arXiv:1612.08818 [hep-th]} \BibitemShut
  {NoStop}%
\bibitem [{\citenamefont {Linde}(1982)}]{Linde:1982uu}%
  \BibitemOpen
  \bibfield  {author} {\bibinfo {author} {\bibfnamefont {A.~D.}\ \bibnamefont
  {Linde}},\ }\href {\doibase 10.1016/0370-2693(82)90293-3} {\bibfield
  {journal} {\bibinfo  {journal} {Phys. Lett.}\ }\textbf {\bibinfo {volume}
  {B116}},\ \bibinfo {pages} {335} (\bibinfo {year} {1982})}\BibitemShut
  {NoStop}%
\bibitem [{\citenamefont {Vilenkin}(1983)}]{Vilenkin:1983xq}%
  \BibitemOpen
  \bibfield  {author} {\bibinfo {author} {\bibfnamefont {A.}~\bibnamefont
  {Vilenkin}},\ }\href {\doibase 10.1103/PhysRevD.27.2848} {\bibfield
  {journal} {\bibinfo  {journal} {Phys. Rev.}\ }\textbf {\bibinfo {volume}
  {D27}},\ \bibinfo {pages} {2848} (\bibinfo {year} {1983})}\BibitemShut
  {NoStop}%
\bibitem [{\citenamefont {Davies}\ and\ \citenamefont
  {Unruh}(1979)}]{Davies:1979se}%
  \BibitemOpen
  \bibfield  {author} {\bibinfo {author} {\bibfnamefont {P.~C.~W.}\
  \bibnamefont {Davies}}\ and\ \bibinfo {author} {\bibfnamefont {W.~G.}\
  \bibnamefont {Unruh}},\ }\href {\doibase 10.1103/PhysRevD.20.388} {\bibfield
  {journal} {\bibinfo  {journal} {Phys. Rev.}\ }\textbf {\bibinfo {volume}
  {D20}},\ \bibinfo {pages} {388} (\bibinfo {year} {1979})}\BibitemShut
  {NoStop}%
\bibitem [{\citenamefont {Buttazzo}\ \emph {et~al.}(2013)\citenamefont
  {Buttazzo}, \citenamefont {Degrassi}, \citenamefont {Giardino}, \citenamefont
  {Giudice}, \citenamefont {Sala} \emph {et~al.}}]{Buttazzo:2013uya}%
  \BibitemOpen
  \bibfield  {author} {\bibinfo {author} {\bibfnamefont {D.}~\bibnamefont
  {Buttazzo}}, \bibinfo {author} {\bibfnamefont {G.}~\bibnamefont {Degrassi}},
  \bibinfo {author} {\bibfnamefont {P.~P.}\ \bibnamefont {Giardino}}, \bibinfo
  {author} {\bibfnamefont {G.~F.}\ \bibnamefont {Giudice}}, \bibinfo {author}
  {\bibfnamefont {F.}~\bibnamefont {Sala}},  \emph {et~al.},\ }\href {\doibase
  10.1007/JHEP12(2013)089} {\bibfield  {journal} {\bibinfo  {journal} {JHEP}\
  }\textbf {\bibinfo {volume} {1312}},\ \bibinfo {pages} {089} (\bibinfo {year}
  {2013})},\ \Eprint {http://arxiv.org/abs/1307.3536} {arXiv:1307.3536
  [hep-ph]} \BibitemShut {NoStop}%
\bibitem [{\citenamefont {Di~Luzio}\ and\ \citenamefont
  {Mihaila}(2014)}]{DiLuzio:2014bua}%
  \BibitemOpen
  \bibfield  {author} {\bibinfo {author} {\bibfnamefont {L.}~\bibnamefont
  {Di~Luzio}}\ and\ \bibinfo {author} {\bibfnamefont {L.}~\bibnamefont
  {Mihaila}},\ }\href {\doibase 10.1007/JHEP06(2014)079} {\bibfield  {journal}
  {\bibinfo  {journal} {JHEP}\ }\textbf {\bibinfo {volume} {1406}},\ \bibinfo
  {pages} {079} (\bibinfo {year} {2014})},\ \Eprint
  {http://arxiv.org/abs/1404.7450} {arXiv:1404.7450 [hep-ph]} \BibitemShut
  {NoStop}%
\bibitem [{\citenamefont {Andreassen}\ \emph {et~al.}(2015)\citenamefont
  {Andreassen}, \citenamefont {Frost},\ and\ \citenamefont
  {Schwartz}}]{Andreassen:2014eha}%
  \BibitemOpen
  \bibfield  {author} {\bibinfo {author} {\bibfnamefont {A.}~\bibnamefont
  {Andreassen}}, \bibinfo {author} {\bibfnamefont {W.}~\bibnamefont {Frost}}, \
  and\ \bibinfo {author} {\bibfnamefont {M.~D.}\ \bibnamefont {Schwartz}},\
  }\href {\doibase 10.1103/PhysRevD.91.016009} {\bibfield  {journal} {\bibinfo
  {journal} {Phys.Rev.}\ }\textbf {\bibinfo {volume} {D91}},\ \bibinfo {pages}
  {016009} (\bibinfo {year} {2015})},\ \Eprint {http://arxiv.org/abs/1408.0287}
  {arXiv:1408.0287 [hep-ph]} \BibitemShut {NoStop}%
\bibitem [{\citenamefont {Andreassen}\ \emph {et~al.}(2014)\citenamefont
  {Andreassen}, \citenamefont {Frost},\ and\ \citenamefont
  {Schwartz}}]{Andreassen:2014gha}%
  \BibitemOpen
  \bibfield  {author} {\bibinfo {author} {\bibfnamefont {A.}~\bibnamefont
  {Andreassen}}, \bibinfo {author} {\bibfnamefont {W.}~\bibnamefont {Frost}}, \
  and\ \bibinfo {author} {\bibfnamefont {M.~D.}\ \bibnamefont {Schwartz}},\
  }\href {\doibase 10.1103/PhysRevLett.113.241801} {\bibfield  {journal}
  {\bibinfo  {journal} {Phys.Rev.Lett.}\ }\textbf {\bibinfo {volume} {113}},\
  \bibinfo {pages} {241801} (\bibinfo {year} {2014})},\ \Eprint
  {http://arxiv.org/abs/1408.0292} {arXiv:1408.0292 [hep-ph]} \BibitemShut
  {NoStop}%
\bibitem [{\citenamefont {Lalak}\ \emph {et~al.}(2016)\citenamefont {Lalak},
  \citenamefont {Lewicki},\ and\ \citenamefont {Olszewski}}]{Lalak:2016zlv}%
  \BibitemOpen
  \bibfield  {author} {\bibinfo {author} {\bibfnamefont {Z.}~\bibnamefont
  {Lalak}}, \bibinfo {author} {\bibfnamefont {M.}~\bibnamefont {Lewicki}}, \
  and\ \bibinfo {author} {\bibfnamefont {P.}~\bibnamefont {Olszewski}},\ }\href
  {\doibase 10.1103/PhysRevD.94.085028} {\bibfield  {journal} {\bibinfo
  {journal} {Phys. Rev.}\ }\textbf {\bibinfo {volume} {D94}},\ \bibinfo {pages}
  {085028} (\bibinfo {year} {2016})},\ \Eprint
  {http://arxiv.org/abs/1605.06713} {arXiv:1605.06713 [hep-ph]} \BibitemShut
  {NoStop}%
\bibitem [{\citenamefont {Espinosa}\ \emph
  {et~al.}(2016{\natexlab{a}})\citenamefont {Espinosa}, \citenamefont {Garny},\
  and\ \citenamefont {Konstandin}}]{Espinosa:2016uaw}%
  \BibitemOpen
  \bibfield  {author} {\bibinfo {author} {\bibfnamefont {J.~R.}\ \bibnamefont
  {Espinosa}}, \bibinfo {author} {\bibfnamefont {M.}~\bibnamefont {Garny}}, \
  and\ \bibinfo {author} {\bibfnamefont {T.}~\bibnamefont {Konstandin}},\
  }\href {\doibase 10.1103/PhysRevD.94.055026} {\bibfield  {journal} {\bibinfo
  {journal} {Phys. Rev.}\ }\textbf {\bibinfo {volume} {D94}},\ \bibinfo {pages}
  {055026} (\bibinfo {year} {2016}{\natexlab{a}})},\ \Eprint
  {http://arxiv.org/abs/1607.08432} {arXiv:1607.08432 [hep-ph]} \BibitemShut
  {NoStop}%
\bibitem [{\citenamefont {Espinosa}\ \emph
  {et~al.}(2016{\natexlab{b}})\citenamefont {Espinosa}, \citenamefont {Garny},
  \citenamefont {Konstandin},\ and\ \citenamefont {Riotto}}]{Espinosa:2016nld}%
  \BibitemOpen
  \bibfield  {author} {\bibinfo {author} {\bibfnamefont {J.~R.}\ \bibnamefont
  {Espinosa}}, \bibinfo {author} {\bibfnamefont {M.}~\bibnamefont {Garny}},
  \bibinfo {author} {\bibfnamefont {T.}~\bibnamefont {Konstandin}}, \ and\
  \bibinfo {author} {\bibfnamefont {A.}~\bibnamefont {Riotto}},\ }\href@noop {}
  {\  (\bibinfo {year} {2016}{\natexlab{b}})},\ \Eprint
  {http://arxiv.org/abs/1608.06765} {arXiv:1608.06765 [hep-ph]} \BibitemShut
  {NoStop}%
\end{thebibliography}%
\end{document}